\documentclass[a4paper,11pt]{article}

\usepackage[utf8]{inputenc}
\usepackage[numbers,sort&compress]{natbib}

\usepackage[margin=1.0in]{geometry}

\usepackage[T1]{fontenc}
\usepackage{graphicx}
\usepackage{amsmath}
\usepackage{hyperref}
\usepackage{placeins}
\usepackage{slashed}
\usepackage[colorinlistoftodos, shadow]{todonotes}
\usepackage{titlesec}
\usepackage{gensymb}
\usepackage{textcomp}
\usepackage{amssymb}
\usepackage{comment}
\usepackage{subfigure} % subfigure with nice labeling (a), (b), ...
\usepackage{multirow} % multirow for tables

\allowdisplaybreaks

\def\refeq#1{\mbox{(\ref{#1})}}
\def\reffi#1{\mbox{Fig.~\ref{#1}}}
\def\reffis#1{\mbox{Figs.~\ref{#1}}}
\def\refta#1{\mbox{Tab.~\ref{#1}}}

\def\refse#1{\mbox{Sec.~\ref{#1}}}

\def\citere#1{\mbox{Ref.~\cite{#1}}}
\def\citeres#1{\mbox{Refs.~\cite{#1}}}

\newcommand{\im}{\mathrm{i}}

\newcommand{\GeV}{\unskip\,\mathrm{GeV}}
\newcommand{\MeV}{\unskip\,\mathrm{MeV}}
\newcommand{\TeV}{\unskip\,\mathrm{TeV}}

\def\mathswitch#1{\relax\ifmmode#1\else$#1$\fi}
\def\mathswitchr#1{\relax\ifmmode{\mathrm{#1}}\else$\mathrm{#1}$\fi}
\def\mathswitchit#1{\relax\ifmmode{#1}\else$#1$\fi}

\newcommand{\lsim}
{\;\raisebox{-.3em}{$\stackrel{\displaystyle <}{\sim}$}\;}

\newcommand{\order}[1]{\ensuremath{ \mathcal{O}( #1 ) }}

\newcommand{\Pf}{\mathswitchr f}
\newcommand{\Ph}{\mathswitchr h}
\newcommand{\PH}{\mathswitchr H}

\newcommand{\PAO}{\mathswitchr {A_0}}
\newcommand{\PHP}{\mathswitchr {H^+}}

\newcommand{\Pu}{\mathswitchr u}
\newcommand{\Pd}{\mathswitchr d}
\newcommand{\Ps}{\mathswitchr s}
\newcommand{\Pc}{\mathswitchr c}
\newcommand{\Pt}{\mathswitchr t}

\newcommand{\Pe}{\mathswitchr e}
\newcommand{\Pep}{\mathswitchr {e^+}}
\newcommand{\Pem}{\mathswitchr {e^-}}

\newcommand{\PW}{\mathswitchr W}
\newcommand{\PZ}{\mathswitchr Z}

\newcommand{\Pg}{g}

\newcommand{\MW}{\mathswitch {M_\PW}}

\newcommand{\MZ}{\mathswitch {M_\PZ}}
\newcommand{\MH}{\mathswitch {M_\PH}}
\newcommand{\Mh}{\mathswitch {M_\Ph}}
\newcommand{\MAO}{\mathswitch {M_\PAO}}
\newcommand{\MHP}{\mathswitch {M_\PHP}}

\newcommand{\scrs}{\scriptscriptstyle}
\newcommand{\sw}{\mathswitch {s_{\scrs\PW}}}
\newcommand{\cw}{\mathswitch {c_{\scrs\PW}}}

\newcommand{\cb}{\mathswitch {c_\beta}}
\renewcommand{\sb}{\mathswitch {s_\beta}}

\newcommand{\GF}{\mathswitch {G_\mu}}

\newcommand{\ri}{{\mathrm{i}}}

\newcommand{\rT}{{\mathrm{T}}}

\newcommand{\EW}{{\mathrm{EW}}}
\newcommand{\QCD}{{\mathrm{QCD}}}

\newcommand{\OS}{{\mathrm{OS}}}

\newcommand{\SM}{{\mathrm{SM}}}

\newcommand{\NLO}{{\mathrm{NLO}}}

\newcommand{\MSbar}{\mathswitch {\overline{\mr{MS}}}}

\newcommand{\Pb}{\mathswitchr b}

\newcommand{\mr}{\mathrm}

\def\bfi{\begin{figure}}
\def\efi{\end{figure}}

\def\draftdate{\relax}
\def\mda{\relax}
\def\mua{\relax}
\def\mla{\relax}
\def\draft{
\def\thtystars{******************************}
\def\sixtystars{\thtystars\thtystars}
\typeout{}
\typeout{\sixtystars**}
\typeout{* Draft mode!
         For final version remove \protect\draft\space in source file *}
\typeout{\sixtystars**}
\typeout{}
\def\draftdate{\today}
\def\mua{\marginpar[\boldmath\hfil$\uparrow$]%
                   {\boldmath$\uparrow$\hfil}%
                    \typeout{marginpar: $\uparrow$}\ignorespaces}
\def\mda{\marginpar[\boldmath\hfil$\downarrow$]%
                   {\boldmath$\downarrow$\hfil}%
                    \typeout{marginpar: $\downarrow$}\ignorespaces}
\def\mla{\marginpar[\boldmath\hfil$\rightarrow$]%
                   {\boldmath$\leftarrow $\hfil}%
                    \typeout{marginpar: $\leftrightarrow$}\ignorespaces}
\def\Mua{\marginpar[\boldmath\hfil$\Uparrow$]%
                   {\boldmath$\Uparrow$\hfil}%
                    \typeout{marginpar: $\uparrow$}\ignorespaces}
\def\Mda{\marginpar[\boldmath\hfil$\Downarrow$]%
                   {\boldmath$\Downarrow$\hfil}%
                    \typeout{marginpar: $\downarrow$}\ignorespaces}
\def\Mla{\marginpar[\boldmath\hfil$\Rightarrow$]%
                   {\boldmath$\Leftarrow $\hfil}%
                    \typeout{marginpar: $\leftrightarrow$}\ignorespaces}
\def\muua{\marginpar[\boldmath\hfil$\upuparrows$]%
                   {\boldmath$\upuparrows$\hfil}%
                    \typeout{marginpar: $\upuparrows$}\ignorespaces}
\def\mdda{\marginpar[\boldmath\hfil$\downdownarrows$]%
                   {\boldmath$\downdownarrows$\hfil}%
                    \typeout{marginpar: $\downdownarrows$}\ignorespaces}
\def\mlla{\marginpar[\boldmath\hfil$\leftleftarrows$]%
                   {\boldmath$\leftleftarrows $\hfil}%
                    \typeout{marginpar: $\leftleftarrows$}\ignorespaces}                    
\overfullrule 5pt
\oddsidemargin -15mm
\marginparwidth 29mm
}

\numberwithin{equation}{section}

\begin{document}

\thispagestyle{empty}
\def\thefootnote{\fnsymbol{footnote}}
\setcounter{footnote}{1}
\null
\draftdate\hfill FR-PHENO-2017-013, CP3-Origins-2017-045 DNRF90
\vfill
\begin{center}
  {\Large {{\boldmath\bf {Precision calculations for $\Ph \to \PW\PW/\PZ\PZ \to 4$ fermions 
\\[0.5em]
in the Two-Higgs-Doublet Model with} \textsc{Prophecy4f}}
\par} \vskip 2.5em
{\large
{\sc Lukas Altenkamp$^{1}$, Stefan Dittmaier$^{1}$, Heidi Rzehak$^{2}$
     }\\[2ex]
{\normalsize \it 
$^1$Albert-Ludwigs-Universit\"at Freiburg, Physikalisches Institut, \\
79104 Freiburg, Germany
}\\[2ex]
{\normalsize \it
$^2$University of Southern Denmark, $\text{CP}^3$-Origins \\
Campusvej 55, DK-5230 Odense M, Denmark}
}}
\par \vskip 1em
\end{center}\par
\vskip .0cm \vfill {\bf Abstract:} 
\par 
We have calculated the next-to-leading-order electroweak and QCD
corrections to the decay
processes $\Ph\to\PW\PW/\PZ\PZ\to4\,$fermions of the light CP-even Higgs boson~h
of various types of
Two-Higgs-Doublet Models (Types~I and II, ``lepton-specific'' and ``flipped'' models).
The input parameters are defined in four different renormalization schemes,
where parameters that are not directly accessible by experiments are defined
in the \MSbar{} scheme. Numerical results are presented for the corrections to partial 
decay widths for various benchmark scenarios previously motivated in the literature,
where we investigate the dependence on the \MSbar{} renormalization
scale and on the choice of the
renormalization scheme in detail. We find that it is crucial to be precise with
these issues in parameter analyses, since parameter conversions between different schemes
can involve sizeable or large corrections, especially in scenarios that are
close to experimental exclusion limits or theoretical bounds.
It even turns out that some renormalization schemes are not applicable in specific
regions of parameter space.
Our investigation of differential distributions shows that corrections beyond the 
Standard Model are mostly constant offsets induced by the mixing between the light and
heavy CP-even Higgs bosons, so that differential analyses of $\Ph\to4f$ decay observables do not
help to identify Two-Higgs-Doublet Models. Moreover, the decay widths do not
significantly depend on the specific type of those models.
The calculations are implemented in the public Monte Carlo generator 
\textsc{Prophecy4f} and ready for application.
\par
\vskip 1cm
\noindent
October 2017
\par
\null
\setcounter{page}{0}
\clearpage
\def\thefootnote{\arabic{footnote}}
\setcounter{footnote}{0}

\section{Introduction}
\label{se:intro}

The CERN Large Hadron Collider (LHC) was built to explore the validity of the
Standard Model (SM) of particle physics at energy scales ranging from the
electroweak scale $\sim$100~GeV up to energies of some TeV and to search for new
phenomena and new particles in this energy domain. The discovery of a Higgs
particle at LHC Run~1 in 2012~\cite{Aad:2012tfa,Chatrchyan:2012ufa}
was a first big achievement in this enterprise.
Since first studies of the properties of this Higgs particle (spin, CP parity,
couplings to the heaviest SM particles) show good agreement between measurements
and SM predictions, the SM is in better shape than ever to describe all known
particle phenomena up to very few exceptions.
Assuming the absence of spectacular new-physics signals in LHC data, this means
that any deviation from the SM hides in small and subtle effects. To extract
those differences from data, both experimental analyses and theoretical
predictions have to be performed with the highest possible precision.
On the other hand, assuming
that a new signal materializes at the $5\sigma$ level,
the properties of the newly discovered particle
have to be investigated with precision, in order to tell different models apart that can
accommodate the new phenomenon. 

Most of the promising candidates for models beyond the SM
modify or extend the scalar sector of electroweak (EW) symmetry breaking,
which introduces the Higgs boson in the SM. Lacking clear evidence of the realization
of a specific model extension, it is well motivated to prepare experimentally
testable predictions within generic SM extensions which are building blocks in 
larger models.
Two-Higgs-Doublet Models (THDMs)~\cite{Lee:1973iz,HHGuide:1990},
where a second Higgs doublet is added to the SM field content, provide an interesting class
of such generic models. While the gauge structure of the SM is kept, THDMs contain
five physical Higgs bosons in contrast to the one of the SM.
Three out of the five are neutral and two are charged. In the CP-conserving case, 
considered in this paper, one of the neutral Higgs bosons is CP-odd and two are CP-even.
An important issue in identifying or constraining
a THDM, thus, consists in telling the SM Higgs boson from an SM-like CP-even
Higgs boson of THDMs.
To this end, several phenomenological studies in THDMs have been carried out recently 
by various groups~\cite{Celis:2013rcs,Celis:2013ixa,Chang:2013ona,Eberhardt:2013uba,Harlander:2013mla,%
Hespel:2014sla,Baglio:2014nea,Broggio:2014mna,%
Haber2015,%
Goncalves:2016qhh,Dorsch:2016tab,%
Cacciapaglia:2016tlr,Cacchio:2016qyh,Aggleton:2016tdd,Han:2017pfo,
%Karmakar:2017yek,
Bhatia:2017ttp,Arbey:2017gmh}.

In this paper we investigate the decay observables of the SM-like neutral, light, 
CP-even Higgs boson~h decaying into four fermions,
$\Ph\to \PW\PW/\PZ\PZ \to 4f$,
in the THDM, including next-to-leading-order (NLO) corrections of the EW and strong interactions.
The fermions in the final state of these processes can either be quarks or leptons, 
and especially the latter can be resolved very well in the detector.
These four-body decays were already crucial in the Higgs-boson discovery, but also play
a major role in precision studies of the Higgs boson, in particular, to determine the 
couplings to the EW gauge bosons W and Z.
They also provide a window to physics beyond the 
SM~\cite{Nelson:1986ki,Soni:1993jc,Chang:1993jy,Skjold:1993jd,Buszello:2002uu,Arens:1994wd,Choi:2002jk,Boselli:2017pef},
as, due to its high precision, small deviations 
from the SM can be measured, and differential distributions can be investigated and tested 
against the SM. 
The Monte Carlo program 
\textsc{Prophecy4f}~\cite{Bredenstein:2006rh,Bredenstein:2006nk,Bredenstein:2006ha} 
performs the calculation of the full NLO EW and QCD corrections 
for all $\Ph\to \PW\PW/\PZ\PZ \to 4f$ channels
in the complex-mass scheme \cite{Denner:1999gp,Denner:2005fg,Denner:2006ic} to
describe the intermediate W- and Z-boson resonances.
It provides differential distributions as well as unweighted events for leptonic final states. 
The corrections to $\Ph\to \PZ\PZ \to 4\,$leptons were also calculated and matched to a QED parton
shower in \citere{Boselli:2015aha}.
In the following, we describe results obtained with an updated version of \textsc{Prophecy4f},
extended to the computation in the THDM in such a way that the usage 
of the program and its applicability as event generator basically remain the same. 

It is our goal to analyze the Higgs decay in the context of the most relevant THDM scenarios. 
To compute phenomenologically relevant results, we need to take into account current 
constraints which also restrict the large parameter space. 
The constraints come from direct LHC searches for heavy Higgs bosons
and from theoretical aspects like vacuum stability, perturbative unitarity, 
or perturbativity of the couplings, which are required for a meaningful perturbative evaluation. 
The recent report of the LHC Higgs Cross Section Working 
Group~\cite{deFlorian:2016spz} summarizes a selection of relevant benchmark scenarios 
proposed in other papers among which we study the most relevant.
The results are compared with the SM prediction, and deviations are quantified. 
In addition to the usual investigation of residual scale uncertainties, 
we compare the results of different renormalization schemes recently
presented and discussed in the literature~\cite{Krause:2016oke,Denner:2016etu,Altenkamp:2017ldc,Denner:2017vms}.
Specifically, we employ the four different schemes described in \citere{Altenkamp:2017ldc}
and vary the renormalization scale to investigate the perturbative stability of the 
predictions in the benchmark scenarios. 
Similar to what has already been found in the Minimal Supersymmetric SM~\cite{Freitas:2002um}, 
the different renormalization schemes may suffer from problems like gauge dependence, 
singularities in relations between parameters, or unnaturally large corrections. 
The comparison of the results obtained with different renormalization schemes 
allows us to determine regions where they behave well and yield reliable results.
Electroweak corrections to other Higgs-boson decay channels in the THDM were
investigated in \citeres{Krause:2016oke,Krause:2016xku}.
Generic tools to calculate Higgs decay widths in the THDM, such as
{\sc Hdecay}~\cite{Djouadi:1997yw}
and {\sc THDMC}~\cite{Eriksson:2009ws}  
are currently restricted to QCD corrections (see, e.g., \citeres{Heinemeyer:2013tqa,Harlander:2013qxa} for more details).

This paper is structured as follows: 
In \refse{sec:Prophecy4f} we briefly describe
the program \textsc{Prophecy4f}, on which our implementation is based, 
and give some details on our NLO calculation of corrections to the 
$\Ph{\to}4f$ decays in the THDM, including a survey of Feynman diagrams, the salient features
of the calculation, and a short outline of the implementation into \textsc{Prophecy4f}. 
In \refse{sec:setup}, we describe the setup of our numerical analysis and the chosen THDM scenarios.
The numerical results are presented and discussed in detail in \refse{sec:results}. 
We conclude in \refse{se:conclusion} and provide some supplementary results in the appendix.

\section{\boldmath{Predicting $\Ph\to\PW\PW/\PZ\PZ\to4f$ in the THDM
with the Monte Carlo program \textsc{Prophecy4f}}}
\label{sec:Prophecy4f}

\subsection{Preliminaries and functionality of \textsc{Prophecy4f}}

The Monte Carlo program 
\textsc{Prophecy4f}~\cite{Bredenstein:2006rh,Bredenstein:2006nk,Bredenstein:2006ha} 
provides a ``\textbf{PROP}er description of the \textbf{H}iggs d\textbf{EC}a\textbf{Y} 
into \textbf{4 F}ermions'' by calculating the decay observables of the process 
$\Ph\to\PW\PW/\PZ\PZ\to4f$ at NLO EW+QCD accuracy in the SM.
The original \textsc{Prophecy4f} code contains the matrix elements of all 19 possible 
$4f$ final states, which are listed in \refta{tab:finalstates}, in a generic way. 
It takes into account the full off-shell effects of the intermediate W and Z bosons 
and treats the W- and Z-boson resonances in the complex-mass 
scheme~\cite{Denner:1999gp,Denner:2005fg,Denner:2006ic}, which maintains gauge invariance
and NLO precision both in resonant and non-resonant phase-space regions.
For the evaluation of the one-loop integrals in the virtual corrections we have 
replaced the original internal integral library by the publicly available
Fortran library \textsc{Collier}~\cite{Denner:2016kdg}.
Ultraviolet (UV)~divergences are treated in dimensional regularization, while the (soft and collinear)
infrared (IR) 
divergences of the loop integrals and in the real photon or gluon emission 
are regularized by small photon, gluon, and external fermion masses. 
The final-state fermions are considered in the massless limit, i.e.\ small fermion masses
are only kept as regulators in the singular logarithms.%
\footnote{Mass effects are mostly negligible for the decays via W or Z~bosons. 
For leptonic final states those effects were discussed at leading order in \citere{Berge:2015jra}.}
However, in diagrams with a closed fermion loop the full mass dependence of those fermions 
is kept which allows to extend the calculation to include a heavy fourth fermion 
generation, as done in \citere{Denner:2011vt}.
The cancellation of the IR divergences can be performed via phase-space 
slicing~\cite{Harris:2001sx} or dipole 
subtraction~\cite{Catani1997,Dittmaier:1999mb,Dittmaier:2008md}.
\begin{table}
  \centering
   \renewcommand{\arraystretch}{1.1}
\begin{tabular}{|c|ccc|}\hline
Final states & leptonic & semi-leptonic & hadronic\\\hline
\multirow{4}{*}{neutral current}& $\nu_\Pe \bar{\nu}_\Pe \nu_\mu \bar{\nu}_\mu$~(3) &$\nu_\Pe \bar{\nu}_\Pe \Pu \bar{\Pu}$~(6)&$\Pu \bar{\Pu} \Pc \bar{\Pc}$~(1)\\
& $ \Pe^- \Pe^+ \mu^- \mu^+$~(3) &$\nu_\Pe \bar{\nu}_\Pe \Pd \bar{\Pd}$~(9) & $\Pd \bar{\Pd} \Ps \bar{\Ps}$~(3)\\
& $\nu_\Pe \bar{\nu}_\Pe  \mu^-\mu^+$~(6) & $ \Pe^-\Pe^+ \Pu \bar{\Pu}$~(6) & $\Pu \bar{\Pu} \Ps \bar{\Ps}$~(4)\\
&&  $\Pe^- \Pe^+  \Pd \bar{\Pd}$~(9)&\\\hline
\multirow{2}{*}{neutral current with interference}& $ \Pe^- \Pe^+ \Pe^-\Pe^+$~(3) &&$\Pu \bar{\Pu} \Pu \bar{\Pu}$~(2)\\
& $\nu_\Pe \bar{\nu}_\Pe \nu_\Pe \bar{\nu}_\Pe$~(3) &&$\Pd \bar{\Pd} \Pd \bar{\Pd}$~(3)\\
\hline
charged current & $ \nu_\Pe \Pe^+ \mu^- \bar{\nu}_\mu$~(6) & $  \nu_\Pe \Pe^+ \Pd \bar{\Pu}$~(12)& $\Pu \bar{\Pd} \Ps \bar{\Pc}$~(2)\\\hline
charged and neutral current & $\nu_\Pe \Pe^+  \Pe^- \bar{\nu}_\Pe$~(3) &&  $\Pu \bar{\Pd} \Pd \bar{\Pu}$~(2)\\\hline
\end{tabular}  
  \caption{The possible final states for the decay $\Ph \to \PW\PW/\PZ\PZ \to 4 f$. 
They can be distinguished by the intermediate gauge boson and the number of lepton pairs. 
Final states that differ only by generation indices, but have the same diagrams have 
identical matrix elements and are only stated once. 
The multiplicity of a final state obtained by changing the generation indices is 
given in parentheses.}
\label{tab:finalstates}
\end{table}

The integration over the phase space is done using an adaptive multi-channel
Monte Carlo integrator, where the integrand is evaluated at pseudo-random 
phase-space points, 
and the density of the points is adapted iteratively to the 
integrand to provide a better convergence.
The Monte Carlo generator can also be used to generate samples of unweighted events
for leptonic final states, which is particularly interesting for experimental analyses.
\textsc{Prophecy4f} automatically provides distributions for leptonic and semi-leptonic 
final states. Distributions for fully hadronic final states are not predefined, 
since this should be done in the hadronic production environment.

The $\Ph{\to}4f$ decay width is the sum of all partial widths of the 19 independent 
final states listed in \refta{tab:finalstates}. 
All other final states differ only by generation indices and yield the same result, 
since the external fermion masses are neglected. One can weight these independent 
final states with their multiplicity (given in parentheses in \refta{tab:finalstates}) 
instead of computing partial widths for all existing final states. 
However, it is also of interest to separate the contributions from ZZ or WW intermediate 
states and WW/ZZ interferences in the partial width, as, e.g., described in
\citeres{Dittmaier2011,Dittmaier2012},
\begin{align}
\label{eq:totwidth}
\Gamma_{\Ph{\to}4f}= \Gamma_{\Ph{\to}\PW\PW{\to}4f}
+\Gamma_{\Ph{\to}\PZ\PZ{\to}4f}+\Gamma_{\PW\PW/\PZ\PZ-\mr{int}}.
\end{align}
The decomposition is trivial for $4f$ states to which only WW or ZZ intermediate
states contribute; only one of the first two terms contributes in this case.
Both WW and ZZ intermediate states can only contribute if all four final-state fermions
are in the same generation (in the absence of quark mixing, which does not
play a role in these processes). In such cases the WW and ZZ parts can be extracted
by replacing the $4f$ state by $f\bar f'F'\bar F$ 
and $f\bar fF\bar F$ states
with the same flavours as in the original $f\bar f'f' \bar f$ 
state, but taking $f$ and $F$ from
different generations. The interference term is then obtained by subtracting the
WW and ZZ parts from the full squared matrix element.
Exemplarily for the $\nu_\Pe \Pe^+  \Pe^- \bar{\nu}_\Pe$ final state this reads
\begin{align}
\Gamma_{\Ph {\to} \PW\PW{\to} \nu_\Pe \Pe^+  \Pe^- \bar{\nu}_\Pe}
&=\Gamma_{\Ph {\to} \nu_\Pe \Pe^+ \mu^- \bar{\nu}_\mu},
\\
\Gamma_{\Ph {\to} \PZ\PZ{\to} \nu_\Pe \Pe^+  \Pe^- \bar{\nu}_\Pe}
&=\Gamma_{\Ph {\to} \nu_\Pe \bar{\nu}_\Pe  \mu^-\mu^+},
\\
\Gamma_{\PW\PW/\PZ\PZ-\mr{int}, \nu_\Pe \Pe^+  \Pe^- \bar{\nu}_\Pe}
&=\Gamma_{\Ph {\to} \nu_\Pe \Pe^+  \Pe^- \bar{\nu}_\Pe}
-\Gamma_{\Ph {\to} \nu_\Pe \Pe^+ \mu^- \bar{\nu}_\mu}
-\Gamma_{\Ph {\to} \nu_\Pe \bar{\nu}_\Pe  \mu^-\mu^+}.
\end{align}
With this procedure the contribution of all final states to the WW, ZZ 
partial widths, 
and the WW/ZZ interference contribution can be computed \cite{Dittmaier2011,Dittmaier2012}, 
\begin{align}
\Gamma_{\Ph {\to} \PW\PW{\to}4f}
={}& 9\Gamma_{\Ph {\to}  \nu_\Pe \Pe^+ \mu^- \bar{\nu}_\mu }+12\Gamma_{\Ph {\to}\nu_\Pe \Pe^+  \Pu \bar{\Pd}}+ 4 \Gamma_{\Ph {\to} \Pu \bar{\Pd} \Ps \bar{\Pc}},
\\
\Gamma_{\Ph {\to} \PZ\PZ{\to}4f}
={}& 3\Gamma_{\Ph {\to} \nu_\Pe \bar{\nu}_\Pe \nu_\mu \bar{\nu}_\mu}+3 \Gamma_{\Ph {\to}  \Pe^- \Pe^+ \mu^-  \mu^+}+9 \Gamma_{\Ph {\to} \nu_\Pe \bar{\nu}_\Pe  \mu^- \mu^+}+3 \Gamma_{\Ph {\to} \Pe^- \Pe^+ \Pe^- \Pe^+ }
\nonumber\\
&+3 \Gamma_{\Ph {\to} \nu_\Pe \bar{\nu}_\Pe \nu_\Pe \bar{\nu}_\Pe}+6 \Gamma_{\Ph {\to} \nu_\Pe \bar{\nu}_\Pe \Pu \bar{\Pu}}+9 \Gamma_{\Ph {\to} \nu_\Pe \bar{\nu}_\Pe \Pd \bar{\Pd}}+6 \Gamma_{\Ph {\to}  \Pe^- \Pe^+ \Pu \bar{\Pu}} +9 \Gamma_{\Ph {\to}  \Pe^- \Pe^+ \Pd \bar{\Pd}}
\nonumber\\
&+ \Gamma_{\Ph {\to} \Pu \bar{\Pu} \Pc \bar{\Pc}}+ 3 \Gamma_{\Ph {\to} \Pd \bar{\Pd} \Ps \bar{\Ps}} +6 \Gamma_{\Ph {\to} \Pu \bar{\Pu} \Ps \bar{\Ps}}+2 \Gamma_{\Ph {\to} \Pu \bar{\Pu} \Pu \bar{\Pu}}+3 \Gamma_{\Ph {\to} \Pd \bar{\Pd} \Pd \bar{\Pd}},
\\
\Gamma_{\PW\PW/\PZ\PZ-\mr{int}}
={}&3 \Gamma_{\Ph {\to} \nu_\Pe \Pe^+  \Pe^- \bar{\nu}_\Pe}- 3 \Gamma_{\Ph {\to} \nu_\Pe \bar{\nu}_\Pe \mu^-  \mu^+}-3  \Gamma_{\Ph {\to}  \nu_\Pe \Pe^+ \mu^- \bar{\nu}_\mu }
\nonumber\\
 &+2 \Gamma_{\Ph {\to} \Pu \bar{\Pd} \Pd \bar{\Pu}}-2 \Gamma_{\Ph {\to} \Pu \bar{\Pu} \Ps \bar{\Ps}}- 2 \Gamma_{\Ph {\to} \Pu \bar{\Pd} \Ps \bar{\Pc}}.
\end{align}

\subsection{Details of the NLO calculation and implementation into \textsc{Prophecy4f}}
\label{sec:implementation}

We extend \textsc{Prophecy4f} to the calculation of the corresponding decays of the
light CP-even Higgs boson~h in THDMs. 
Specifically, we consider THDMs of Type~I 
and II, as well as models of
``lepton-specific'' and ``flipped'' types.
For the calculation of the decay $\Ph \to \PW\PW/\PZ\PZ \to 4f$, 
we identify the decaying Higgs boson~h with the discovered Higgs boson of mass $125\GeV$. 
The calculation is similar to the one in the SM described in detail in 
Refs.~\cite{Bredenstein:2006rh,Bredenstein:2006ha}. 
As the particle content of the SM is extended in the THDM, 
all diagrams of the SM appear also in the THDM calculation. 
However, coupling factors of interactions involving scalar particles are modified in the 
THDM and have to be adapted. In addition, new diagrams including heavy Higgs bosons appear 
and need to be taken into account. In the following, 
we discuss the leading-order (LO)
matrix elements and the EW and QCD NLO corrections.

\subsubsection{Lowest order}

\begin{figure}
\centering
\subfigure[]{\includegraphics[scale=1]{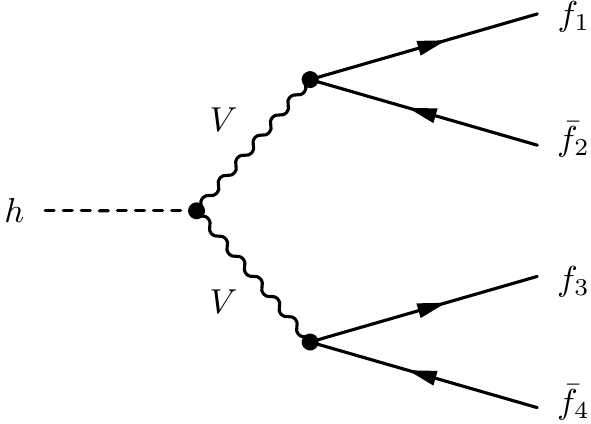}\label{fig:hdecayLO1}}
\hspace{10pt}
\subfigure[]{\includegraphics[scale=1]{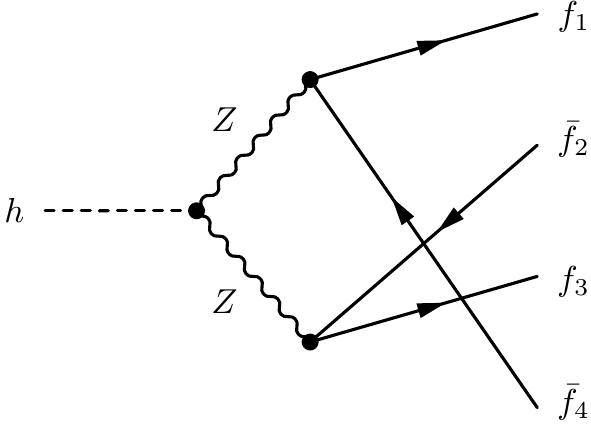}\label{fig:hdecayLO2}}
\caption{Tree-level diagrams of the decay $\Ph \to 4 f$ with $V=\PW,\PZ$. The diagram on the r.h.s.\
exists only if the fermion pairs are quarks or leptons of the same generation. 
The only couplings that change in the transition from the SM to the THDM
are the $\Ph\PW\PW$ and $\Ph\PZ\PZ$
couplings which involve an additional factor of $\sin{(\beta-\alpha)}$.}
\label{fig:hdecayLO}
\end{figure}

At LO, the decay of the Higgs boson proceeds via a pair of (off-shell) gauge bosons 
$V=\PW,\PZ$
which subsequently decay into fermions, as shown in Fig.~\ref{fig:hdecayLO}. 
The diagrams involving a coupling of scalars to external fermions vanish and 
can be omitted due to the neglect of the external fermion masses. 
The only change in the THDM w.r.t.\ the SM is that the $\Ph VV$ coupling 
acquires an additional factor of $\sin{(\beta-\alpha)}$, so that the LO matrix element becomes
\begin{align}
\label{eq:MEborn}
 \mathcal{M}^{VV}_{\mr{THDM, LO}}= \sin{(\beta-\alpha)}\,\mathcal{M}^{VV}_{\mr{SM, LO}},
\end{align}
where $\alpha$ is the mixing angle between the two neutral CP-even Higgs bosons~h and H,%
\footnote{In order to avoid a conflict in our notation, we define $\alpha_\mr{em}=e^2/(4\pi)$ 
as electromagnetic coupling constant and consistently keep the symbol $\alpha$ for the rotation angle.}
and $\beta$ is the mixing angle in both the neutral CP-odd as well as in the charged scalar
sector 
which is related to the ratio of the vacuum expectation values of the two
scalar doublets. We consistently follow the conventions of \citere{Altenkamp:2017ldc} 
for all quantities of the THDM.

\subsubsection{Electroweak corrections}
\label{se:ewrcs}

\begin{figure}
  \centering
    \subfigure[]{\includegraphics[scale=1]{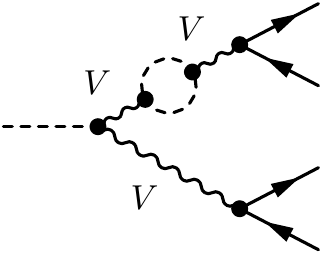}}\hspace{10pt}
\subfigure[]{\includegraphics[scale=1]{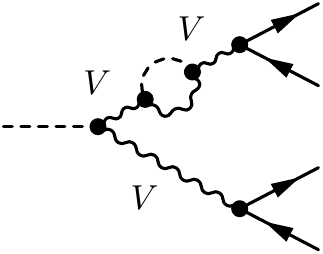}}\hspace{10pt}
  \subfigure[]{\includegraphics[scale=1]{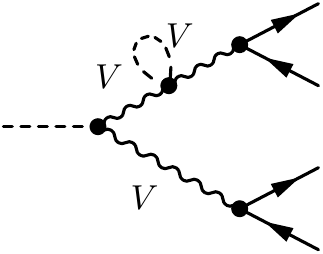}}\hspace{10pt}
  \subfigure[]{\includegraphics[scale=1]{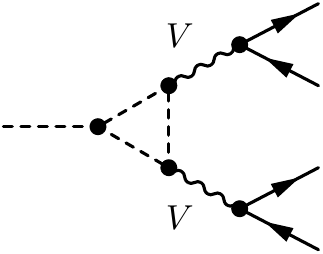}}\\[-1em]
\subfigure[]{\includegraphics[scale=1]{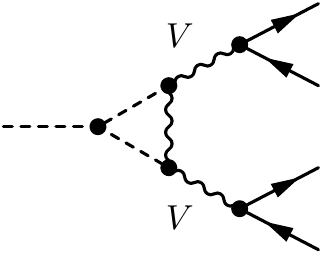}}\hspace{10pt}
  \subfigure[]{\includegraphics[scale=1]{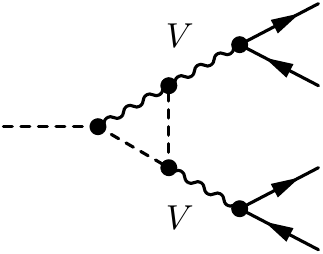}}\hspace{10pt}
\subfigure[]{\includegraphics[scale=1]{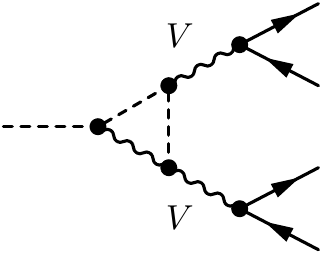}}\hspace{10pt}
  \subfigure[]{\includegraphics[scale=1]{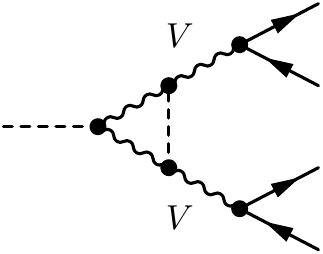}}\\[-1em]
\subfigure[]{\includegraphics[scale=1]{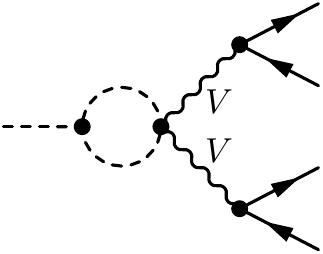}\vspace*{-5em}}\hspace{10pt}
\caption{Exemplary generic one-loop diagrams of the $\Ph VV$ vertex correction
with additional Higgs bosons $\PH,\PAO,\PHP$. The internal dashed lines represent any light or 
heavy Higgs boson;
the gauge boson $V$ can be $\PW,\PZ,\gamma$, depending on the charge flow and final state.}
\label{fig:EWvirtHVVbosTHDM}%
\end{figure}

In the EW corrections, heavy Higgs bosons appear in loop diagrams, viz.\ 
in self-energy and vertex corrections. 
Generic diagrams are shown in Fig.~\ref{fig:EWvirtHVVbosTHDM}. 
Four- and five-point diagrams do not contain heavy Higgs bosons, as this would require an
$\Ph f\bar{f}$ coupling which is proportional to the mass $m_f$ of an external fermion $f$. 
The one-loop diagrams that do not include these heavy particles are in direct 
correspondence to the SM diagrams described in detail in Ref.~\cite{Bredenstein:2006rh}.
However, the coupling factors of the internal (massive) fermions and the vector bosons to the 
light Higgs field need to be adapted in the THDM.

The counterterm contribution can be split into two parts. 
The first one, $\mathcal{M}^{\mr{CT}}_\mr{SM}$,
is analogous to the counterterm contribution in the SM, although all renormalization 
constants appearing in this part are defined within the THDM using the renormalization 
conditions described in \citere{Altenkamp:2017ldc} and in general receive contributions from 
the exchange of heavy Higgs bosons.
The second part is composed of the renormalization constants of the mixing angles 
$\alpha$, $\beta$, entering via the factor $\sin{(\beta-\alpha)}$
in $\mathcal{M}^{VV}_{\mr{SM, LO}}$,
and the field renormalization constant of the mixing of the neutral 
CP-even fields. 
The full counterterm can be written as
\begin{align}
\label{eq:CTampZ}
{\mathcal{M}}^{\mr{CT}}_{\mr{THDM}}
&=
c_{\beta-\alpha} \Big(\delta \beta -\delta \alpha + \frac{1}{2}\delta Z_{\PH\Ph}   
\Big)\mathcal{M}^{\mr{LO}}_{\mr{SM}}+ s_{\beta-\alpha}\,\mathcal{M}^{\mr{CT}}_\mr{SM},
\end{align}
where we introduced the abbreviations $s_x\equiv\sin x $, $c_x\equiv\cos x$, $t_x\equiv\tan x$.
Following \citere{Altenkamp:2017ldc},
we employ four different renormalization schemes in order to define the mixing angles
at NLO, i.e.\ to fix the renormalization constants $\delta \alpha $, $\delta \beta$:
\begin{itemize}
\item
$\MSbar(\alpha)$ scheme: \\
In this scheme $\alpha$ and $\beta$ are independent parameters and fixed in the 
\MSbar{} scheme. Tadpole parameters are renormalized in such a way that 
renormalized tadpole parameters vanish. As discussed in detail in
\citeres{Krause:2016oke,Denner:2016etu}, this treatment introduces gauge dependences
in the relation between bare parameters and, thus, the relation between renormalized
parameters and predicted observables inherit some gauge dependence.
Since we work in the 't~Hooft--Feynman gauge, all predictions (not only the ones
presented in this work) should be made in the same gauge to obtain a meaningful
confrontation of theory with data.
\item
$\MSbar(\lambda_3)$ scheme: \\
This scheme coincides with the $\MSbar(\alpha)$ scheme up to the point that
$\alpha$ is traded for the (dimensionless) Higgs self-coupling 
parameter $\lambda_3$ as independent parameter. The coupling $\lambda_3$ 
is fixed by an \MSbar{} condition, and $\alpha$ can be calculated from 
$\lambda_3$ and the other free parameters by tree-level relations.
Renormalized tadpoles are again forced to vanish, however, in this scheme
the relations between free parameters and predicted observables are 
gauge independent within the class of $R_\xi$ gauges at NLO, because $\lambda_3$
as basic coupling in the original Higgs potential does not introduce gauge dependences
and the \MSbar{} renormalization of $\beta$ is known to be gauge independent
in $R_\xi$ gauges at NLO~\cite{Krause:2016oke,Denner:2016etu}.
\item
FJ$(\alpha)$ scheme: \\
In this scheme, which is also described in \citeres{Krause:2016oke,Denner:2016etu}
in slightly different technical realizations, 
again $\alpha$ and $\beta$ are independent parameters,
but gauge dependences are avoided by treating tadpole contributions
differently, following a method proposed by Fleischer and 
Jegerlehner (FJ)~\cite{Fleischer:1980ub} a long time ago already in the SM.%
\footnote{A similar scheme, called $\beta_h$~scheme, was suggested in \citere{Actis:2006ra}.}
The basic idea is that bare tadpoles are defined to be zero, which preserves
gauge independence in the relations between bare parameters of the theory. 
As a consequence, explicit tadpole loop contributions have to be taken into account
in all loop calculations. This somewhat unpleasant feature can be avoided 
by
 introducing a new set of renormalization constants upon splitting off
constant contributions from those fields that develop vacuum expectation values
by field transformation in the functional integral (see \citeres{Krause:2016oke,Denner:2016etu}).
Equivalently, the whole procedure of the $\MSbar(\alpha)$ scheme,
i.e.\ the full counterterm Lagrangian including tadpole counterterms,
can be kept, but the renormalization constants $\delta \alpha $, $\delta \beta$,
which contain only pure UV~divergences in the $\MSbar(\alpha)$ scheme,
now receive some finite contributions from the different renormalization
prescription of the tadpoles. This procedure is described in \citere{Altenkamp:2017ldc}
in detail.
\item
FJ$(\lambda_3)$ scheme: \\
In this scheme $\beta$ and $\lambda_3$ are independent parameters, as in the
$\MSbar(\lambda_3)$ scheme, but tadpoles are treated following the gauge-independent FJ
prescription.
\end{itemize}
More details on the different schemes and explicit results for the renormalization constants
can be found in \citere{Altenkamp:2017ldc}.
In all four schemes the parameters $\alpha$, $\beta$,
and the Higgs-quartic-coupling parameter $\lambda_5$ are defined directly
in the \MSbar{} scheme or are indirectly connected to \MSbar{} parameters, i.e.\
all $\alpha$, $\beta$, and $\lambda_5$ depend on a renormalization scale $\mu_\mr{r}$.
In \citere{Altenkamp:2017ldc}
the $\mu_\mr{r}$~dependence of $\alpha$, $\beta$, and $\lambda_5$ was taken into account by numerically
solving the renormalization group equations in the four different renormalization schemes.
Using these results on the running of $\alpha$, $\beta$, and $\lambda_5$, we will investigate
the scale dependence of the NLO-corrected $\Ph{\to}4f$ decay widths.
In particular, we check whether the implicit $\mu_\mr{r}$~dependence of $\alpha$ and $\beta$,
which already enters the LO amplitude, is compensated by the explicit
$\mu_\mr{r}$~dependence contained in the loop corrections.

The diagrams of the real emission can be obtained from the LO diagrams by adding photon radiation. 
The photon couplings in the THDM and in the SM are identical, i.e.\
the real emission 
matrix element $\mathcal{M}^\mr{R,EW}_\mr{THDM}$ of the THDM results from the
matrix element $\mathcal{M}^\mr{R,EW}_\mr{SM}$ of the SM by multiplication with the
coupling factor $\sin{(\beta-\alpha)}$,
\begin{align}
\mathcal{M}^\mr{R,EW}_\mr{THDM}=s_{\beta-\alpha} \,\mathcal{M}^\mr{R,EW}_\mr{SM}. 
\label{eq:realEW}
\end{align}
The calculation of the SM amplitude $\mathcal{M}^\mr{R,EW}_\mr{SM}$
is described in detail in Ref.~\cite{Bredenstein:2006rh}.  
The IR-singular structure is not altered in the transition from the SM
to the THDM, so that the subtraction and 
slicing procedures can be applied straightforwardly in the same way as it was done in
the SM calculation~\cite{Bredenstein:2006rh}. 
 
\subsubsection{QCD corrections}
\label{sec:QCD}

\begin{figure}
  \centering
\subfigure[]{\includegraphics[scale=1]{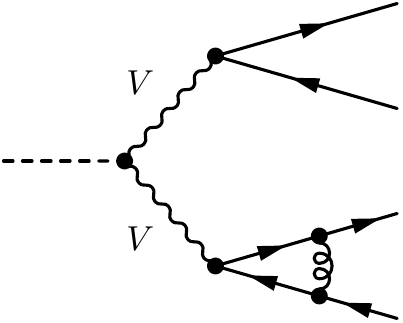}\label{fig:QCDvirt1}}\hspace{10pt}
\subfigure[]{\includegraphics[scale=1]{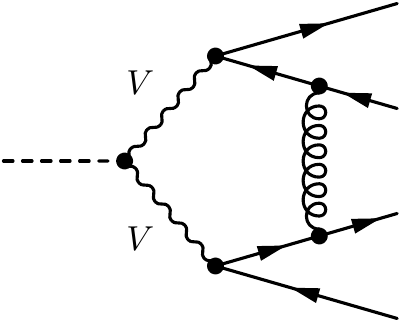}\label{fig:QCDvirt2}}\hspace{10pt}
\subfigure[]{\includegraphics[scale=1]{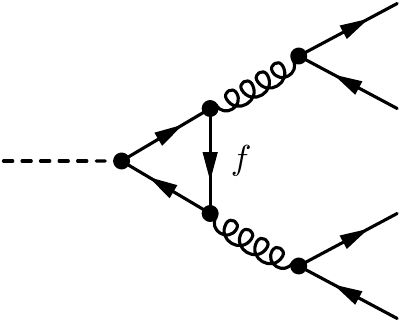}\label{fig:QCDvirt3}}\hspace{10pt}\\
\subfigure[]{\includegraphics[scale=1]{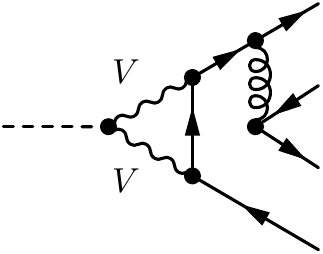}\label{fig:QCDvirt4}}\hspace{10pt}
\subfigure[]{\includegraphics[scale=1]{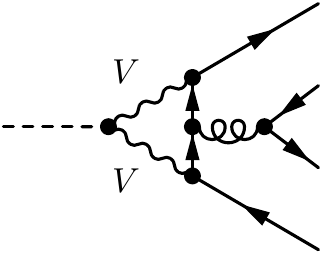}\label{fig:QCDvirt5}}\hspace{10pt}
\caption{Exemplary diagrams for the one-loop virtual QCD corrections. In the semi-leptonic case, only the first diagram type exists. 
Only the interference of the last four diagrams with the crossed LO diagram 
of \reffi{fig:hdecayLO}(b)
has a non-vanishing colour structure, demanding the quarks to be identical.}
\label{fig:QCDvirt}
\end{figure}

As the THDM does not change the strongly interacting part of the theory, 
the computation of the QCD corrections is much simpler than for the EW corrections. 
Some diagrams of the virtual QCD corrections are shown in Fig.~\ref{fig:QCDvirt}.
In the diagrams (a), (b), (d), (e) the only coupling that changes w.r.t.\ the SM is 
the $\Ph VV$ coupling with an additional factor of $s_{\beta-\alpha}$. 
In the diagrams represented by Fig.~\ref{fig:QCDvirt3}, $\Ph q\bar{q}$ couplings appear 
instead of the $\Ph VV$, where $q$ is any quark. 
The $\Ph q\bar{q}$ couplings 
depend on the type of THDM and are given in Tab.~\ref{tab:yukint}. 
\begin{table}
  \centering
\begin{tabular}{|c|c|c|c|c|}\hline
&Type I & Type II &  Lepton-specific &Flipped\\\hline
$\xi^{l}_\Ph$ & $\cos{\alpha}/\sin{\beta}$ & $-\sin{\alpha}/\cos{\beta}$ & $-\sin{\alpha}/\cos{\beta}$& $\cos{\alpha}/\sin{\beta}$  \\
$\xi^u_\Ph$ & $\cos{\alpha}/\sin{\beta}$ & $\cos{\alpha}/\sin{\beta}$ & $\cos{\alpha}/\sin{\beta}$ &$\cos{\alpha}/\sin{\beta}$ \\
$\xi^d_\Ph$& $\cos{\alpha}/\sin{\beta}$ & $-\sin{\alpha}/\cos{\beta}$ & $\cos{\alpha}/\sin{\beta}$ &$-\sin{\alpha}/\cos{\beta}$  \\\hline
\end{tabular}  
  \caption{The coupling strengths $\xi^f_\Ph$ of the light, CP-even Higgs boson~h 
to the fermions $f$ relative to the SM values for different types of THDM models.}
\label{tab:yukint}
\end{table}
The QCD counterterm contribution is identical to the one appearing in the SM calculation
up to the overall coupling factor $s_{\beta-\alpha}$ in the matrix elements.

The diagrams for the real QCD corrections can be obtained from Fig.~\ref{fig:hdecayLO} 
by adding gluon emission off quarks and antiquarks. Similar to the EW case, 
the THDM does not affect the additional gluon emission, so that 
\begin{align}
 \mathcal{M}^\mr{R,QCD}_\mr{THDM}=s_{\beta-\alpha}\, \mathcal{M}^\mr{R,QCD}_\mr{SM}.
\end{align}
The quark loop diagrams do not contain IR singularities, so that the singular structure 
of the one-loop matrix element matches the one of the corresponding SM amplitude multiplied by 
$s_{\beta-\alpha}$. As the LO amplitude contains the same factor, 
the SM subtraction and slicing algorithms can be applied without modification.

\subsubsection{Complex-mass scheme}
To treat the vector-boson resonances in a proper way, we employ the complex-mass scheme which is explained in detail in Refs.~\cite{Denner:1999gp,Denner:2005fg,Denner:2006ic}. This prescription consists of an analytic continuation of the masses of unstable particles into the complex plane which preserves gauge invariance as well as all algebraic relations between amplitudes or Green functions that do not involve complex conjugation (such as Ward and Slavnov Taylor identities). 
The complex mass $\mu_V$ of $V$ is directly connected to the real pole mass $M_V$ and the decay width $\Gamma_V$,
\begin{align}
 \mu_V^2 = M_V^2 - \im M_V \Gamma_V,
\end{align}
with $V=\PW,\PZ$. 
For our process at NLO, it is sufficient to treat only the W and Z boson in the complex-mass scheme even though the other scalar particles are not stable. We assume that in the THDM, the light Higgs boson has properties similar to the SM Higgs boson, i.e.\ its width is very small, $\order{\Gamma_\Ph/\Mh}<\order{10^{-4}}$. Effects of this order can be neglected, as they are smaller than the contributions from NLO and have the same size as the uncertainties due to the separation of $\Ph$~production and decay. The other unstable Higgs bosons of the THDM enter only in loop diagrams, and the corrections from the complex masses are negligible as $\Gamma_S\ll M_S$ where $\Gamma_S$ and $M_S$ are the decay width and  real pole mass of the considered Higgs boson. 
A fully consistent replacement of the real masses by its complex counterparts includes also a complex definition of the weak mixing angle $\theta_\PW$,
\begin{align}
 \cos^2 \theta_\PW = \cw^2=1-\sw^2=\frac{\mu_\PW^2}{\mu_\PZ^2}
= \frac{\MW^2- \ri\MW \Gamma_\PW}{\MZ^2- \ri\MZ \Gamma_\PZ}.
\end{align}
The generalization of this prescription to the one-loop level leads to complex renormalization constants~\cite{Denner:2005fg}, 
for instance to the complex mass renormalization constants
\begin{align}
 \delta \mu_\PW^2&=\Sigma^{\PW\PW}_\rT(\MW^2)+ (\mu_\PW^2-\MW^2)\Sigma'^{\PW\PW}_\rT(\MW^2)+\frac{\im \alpha_{\mathrm{em}}}{\pi} \MW \Gamma_\PW,\nonumber\\
 \delta \mu_\PZ^2&=\Sigma^{\PZ\PZ}_\rT(\MZ^2)+ (\mu_\PZ^2-\MZ^2)\Sigma'^{\PZ\PZ}_\rT(\MZ^2),
 \end{align}
where $\Sigma^{VV}_\rT(p^2)$ denotes the transverse parts of the $V$-boson self-energy with momentum transfer $p$ and
$\Sigma'^{VV}_\rT$ its derivative w.r.t.\ $p^2$.
As a consequence the renormalization constants of the weak mixing angle are
\begin{align}
 \frac{\delta \sw}{\sw}=-\frac{\cw^2}{\sw^2}\frac{\delta \cw}{\cw}=-\frac{\cw^2}{2\sw^2}\left( \frac{\delta \mu_\PW^2}{\mu_\PW^2}-\frac{\delta \mu_\PZ^2}{\mu_\PZ^2}\right).
\end{align}
The field renormalization constants of the vector bosons are given in Eq.~(4.30) of \citere{Denner:2005fg}.
In particular, they enter 
the calculation of the electric charge renormalization constant.
The field renormalization constants of the stable fermions and the scalars are also affected by treating W and Z~bosons in the complex-mass scheme as we do not take the real parts of the self-energies. Due to the appearing complex parameters, the self-energies and also the renormalization constants become complex. However, the  field renormalization constants of internal fields drop out and those of external fields factorize from the LO, so that the imaginary parts drop 
out after squaring the matrix element at NLO.

\subsubsection{Implementation and checks}

The implementation of our calculation is performed in two independent ways: 
In the first method, we use the \textsc{FeynArts}~\cite{Hahn2001} model generated as described in
\citere{Altenkamp:2017ldc} and adapt it to the specific demands of the \textsc{Prophecy4f} calculation, so that masses of fermions belonging to closed loops are treated with the full mass dependence and the complex vector-boson masses are implemented. The amplitudes are generated and processed
using \textsc{FeynArts}~\cite{Hahn2001} and \textsc{FormCalc}~\cite{Hahn1999,Hahn2000}
and implemented into the \textsc{Prophecy4f} code. Additionally, the coupling factors in the \textsc{Prophecy4f} code of the $\PH VV$ and $\PH f\bar{f}$ couplings are adapted to the THDM so that the LO, real photonic corrections, and the QCD corrections can be obtained by simple rescaling. 
In a second, independent calculation the amplitudes are generated via a 
\textsc{FeynArts}~1~\cite{Kublbeck:1990xc} model file, 
and the counterterms are inserted by hand. These two implementations allow us to check the results and ensure their correctness.

Apart from performing two independent loop calculations, we have verified
our one-loop matrix elements by numerically comparing our results to
the ones obtained in \citeres{Denner:2016etu,Denner:2017vms} for the related
$\PW\Ph/\PZ\Ph$ production channels (including $\PW/\PZ$ decays)
using crossing symmetry.

\section{Input parameters and scenarios for the THDM}
\label{sec:setup}

For the SM-like 
input parameters we take the values recommended by the LHC Higgs Cross Section Working Group \cite{deFlorian:2016spz} which essentially follow the Particle Data Group \cite{Agashe:2014kda}:
\begin{align}
 G_\mu ={} & 0.11663787\cdot 10^{-4} \GeV^{-2},& \alpha_\mr{s}={} & 0.118,
\nonumber \\
  \MZ^\OS={} &91.1876 \GeV, & \MW^\OS={} &80.385 \GeV, & \Mh={} & 125 \GeV,
\nonumber\\
 \Gamma^\OS_\PZ={}&2.4952 \GeV, &\Gamma^\OS_\PW={}&2.085 \GeV, 
\nonumber\\
  m_\Pe={}&510.998928 \text{ keV},& m_\mu={}&105.6583715 \MeV, & m_\tau={}&1.77682 \GeV,
\nonumber\\
  m_\Pu={} &100 \MeV,& m_\Pc={} &1.51 \GeV, & m_\Pt={} &172.5 \GeV,
\nonumber\\
 m_\Pd={} &100 \MeV,  &  m_\Ps={} &100  \text{ MeV},  & m_\Pb={} &4.92 \GeV. 
\end{align}
We employ the $G_\mu$ scheme 
where the electromagnetic coupling is derived from the Fermi constant~$\GF$, 
\begin{align}
\alpha_{G_\mu}=\frac{\sqrt{2} G_\mu \MW^2}{\pi}\left(1-\frac{\MW^2}{\MZ^2}\right),
\end{align}
which absorbs the running of the electromagnetic coupling $\alpha_{\text{em}}$
 from the Thomson limit to
the electroweak scale and accounts for universal corrections to the $\rho$-parameter.
The CKM matrix is consistently taken as the unit matrix, since all quark-mixing 
effects drop out in flavour sums, since we work with massless light quarks without
mixing to the third quark generation.

\textsc{Prophecy4f} performs its calculation in the complex-mass scheme and 
automatically converts the experimentally measured on-shell gauge
boson masses 
$M_{V}^\mr{OS}$ 
to pole masses $M_{V}^\mr{pole}$ of the propagators
according to 
\begin{align}
M_{V}^\mr{pole}=M_{V}^\mr{OS}/ \sqrt{1+(\Gamma_V^\mr{OS}/M_{V}^\mr{OS})^2}, && \Gamma_{V}^\mr{pole}=\Gamma_{V}^\mr{OS}/ \sqrt{1+(\Gamma_V^\mr{OS}/M_{V}^\mr{OS})^2}.
\end{align}
From these measured input values, the program recalculates the widths of the vector bosons in $\order{\alpha_{\mr{em}}}$ in the SM using real mass parameters everywhere. This recalculation ensures that the branching ratios of the vector bosons are correctly normalized and add up to one for the SM. In the THDM, the heavy Higgs bosons enter the width in the mass counterterms, however, as we are close to the alignment limit ($c_{\beta-\alpha}\to0$) the effects are negligible.
For an easier reproducibility of our results, we keep the SM values,
which are also compatible with the measured W/Z~widths.
The final-state fermions are considered massless.
The fermion masses are only inserted as regulator masses in soft and/or collinear divergent terms and 
in mass terms of closed fermion loops. 
The strong coupling constant $\alpha_\mr{s}$
appears only in the QCD corrections (see \refse{sec:QCD}), and we take its value at the Z-boson mass.

As central renormalization scale $\mu_0$ 
we use the average mass of all scalar degrees of freedom,
\begin{align}
\label{eq:centralscale}
 \mu_0=\frac{1}{5}(\Mh+\MH+\MAO+2\MHP).
\end{align}
This scale choice might seem surprising
at first glance, since the light Higgs-boson mass is the centre-of-mass energy of our process. However, the loop diagrams including heavy scalar particles $S=\PH,\PAO,\PH^\pm$
with mass $M_S$ introduce potentially large terms of $\ln{(M_S^2/\mu^2)}$ in the amplitudes as long as the mixing angle $\beta-\alpha$ stays away from the alignment limit. 
Therefore we adapt the choice of the scale to the arithmetic mean of the Higgs-boson masses, and the scale variations performed in \refse{sec:results} confirm this choice. The input values of the additional parameters of the THDM depend on the investigated scenario and are given in the following. The scale-dependent input parameters $c_{\beta-\alpha}$, $t_\beta$, $\lambda_5$ are defined at the central scale $\mu_0$ by 
default.

To provide collinear-safe differential observables, a photon recombination is performed in the real corrections. This procedure invokes the addition of the photon momentum to the 
one of the fermion in the histogram if the invariant mass of a photon and a charged fermion is smaller than $5\GeV$. When this is possible with more than one fermion, the photon is added to the fermion that yields the smallest invariant mass. We apply the photon recombination in all our calculations; further details about
its impact are discussed in Ref.~\cite{Bredenstein:2006rh}.

\begin{figure}
\centering
\includegraphics[width=0.45\columnwidth]{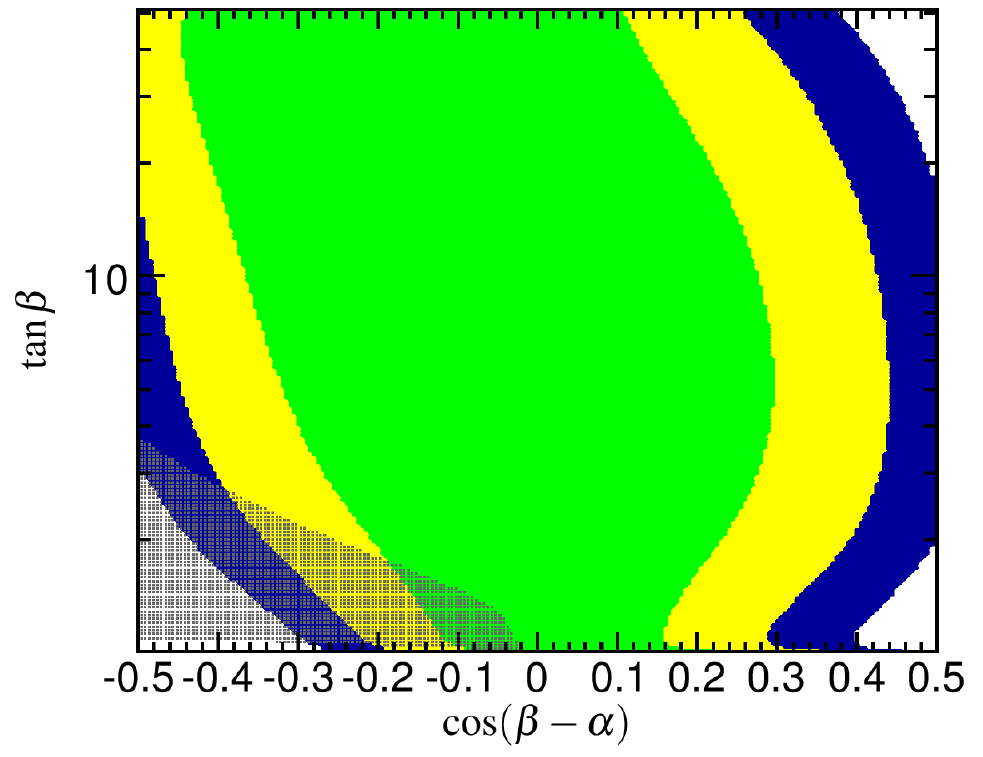}
\includegraphics[width=0.45\columnwidth]{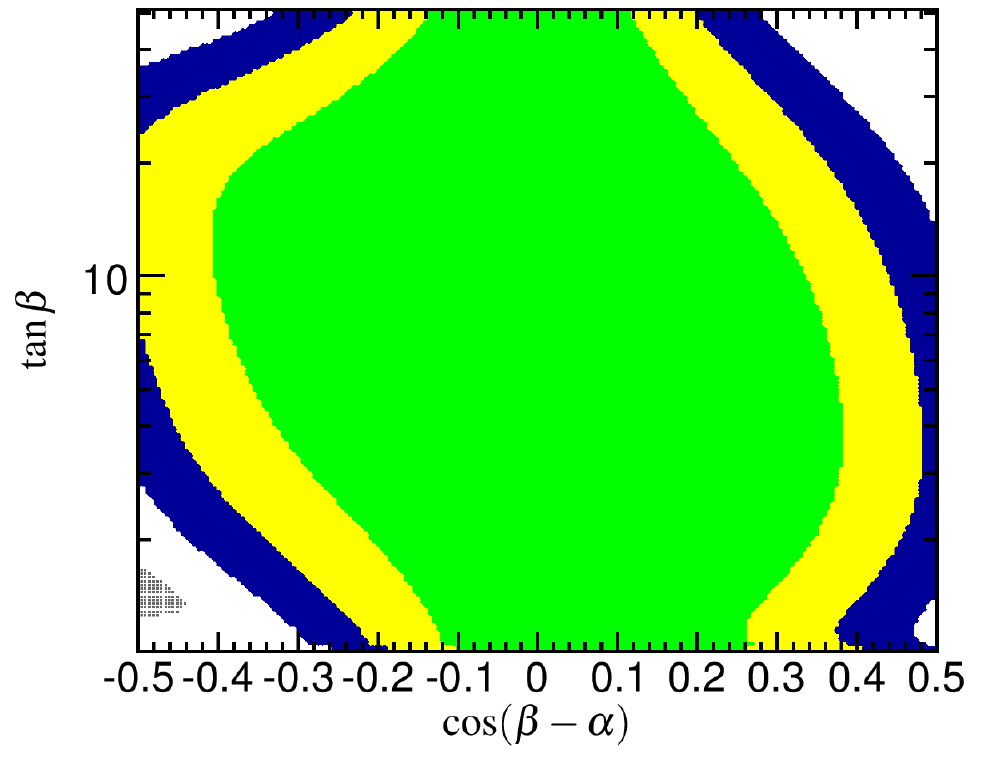}
\caption{Direct constraints from LHC Higgs searches on the parameter space for the THDM Type I with $\MH=300\GeV$ (left) and $\MH=600\GeV$ (right). In both cases $\Mh=125\GeV$, $Z_4=Z_5=-2$ and $Z_7=0$ are given in the hybrid basis (c.f. Ref.~\cite{Haber2015}). The colours indicate compatibility with the observed Higgs signal at $1\,\sigma$ (green), $2\,\sigma$ (yellow), and $3\,\sigma$ (blue). Exclusion bounds at $95\%$ C.L.~from the non-observation of the additional Higgs states are overlaid in gray. The graphics and description are taken from Ref.~\cite{Haber2015}.}
\label{fig:lhcconst}
\vspace*{1em}
\includegraphics[width=0.45\columnwidth]{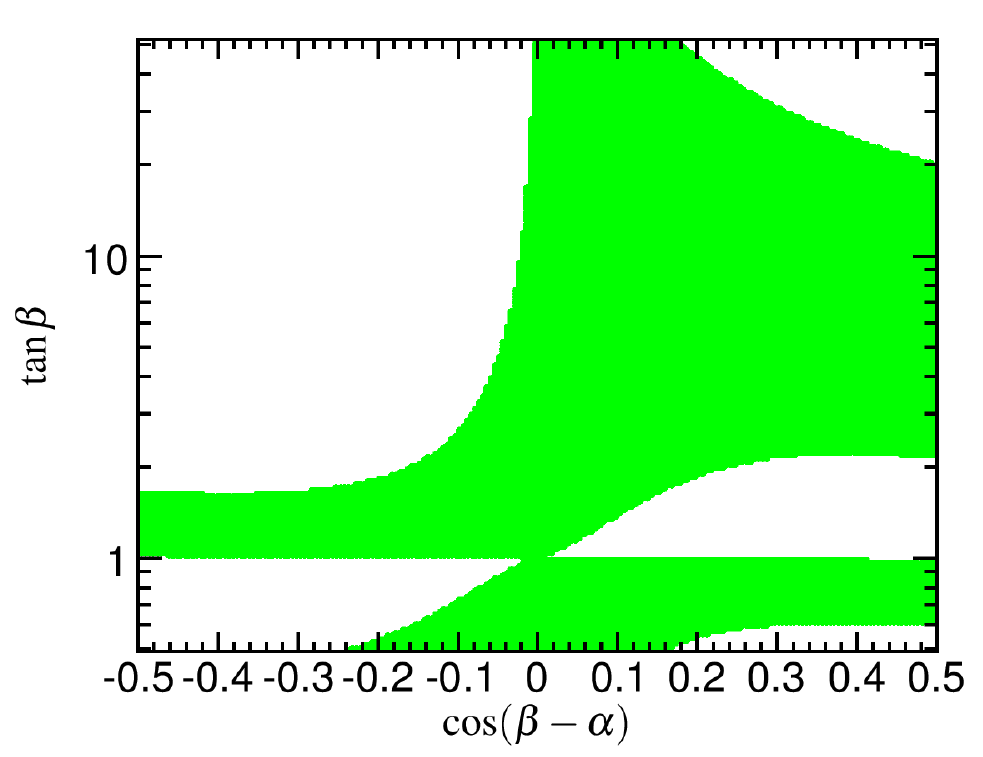}
\includegraphics[width=0.45\columnwidth]{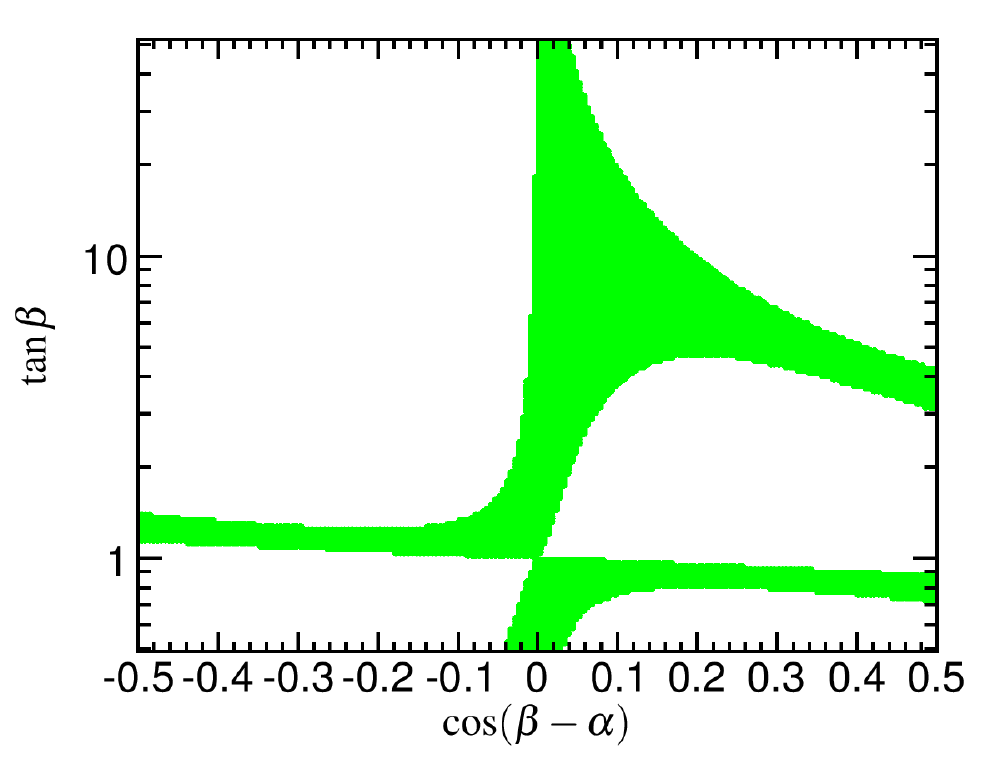}
\caption{Example THDM parameter regions respecting perturbative unitarity and stability constraints (green) for the scenario of
\reffi{fig:lhcconst}.  The graphics are taken from Ref.~\cite{Haber2015}.} 
%$\MH=300\GeV$ (left) and $\MH=600\GeV$ (right), $Z_4=Z_5=-2$ and $Z_7=0$. The graphics and descriptions are taken from Ref.~\cite{Haber2015}.}
\label{fig:stabconstr}
\vspace*{2em}
\hspace{-10pt}\includegraphics{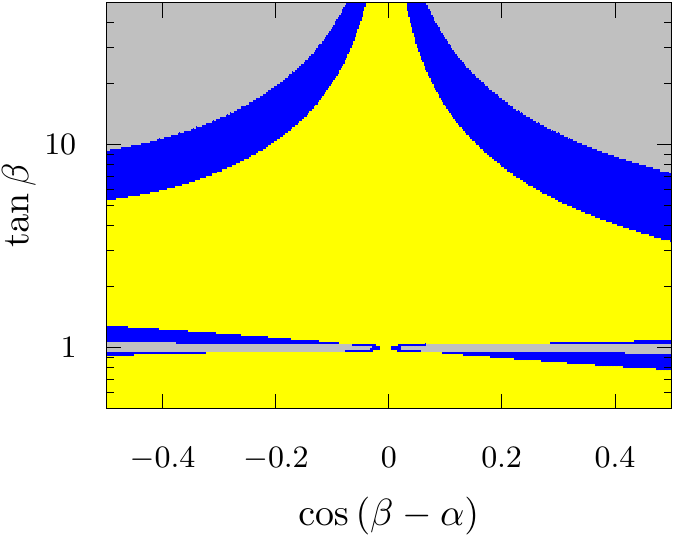}\hspace{7pt}
\includegraphics{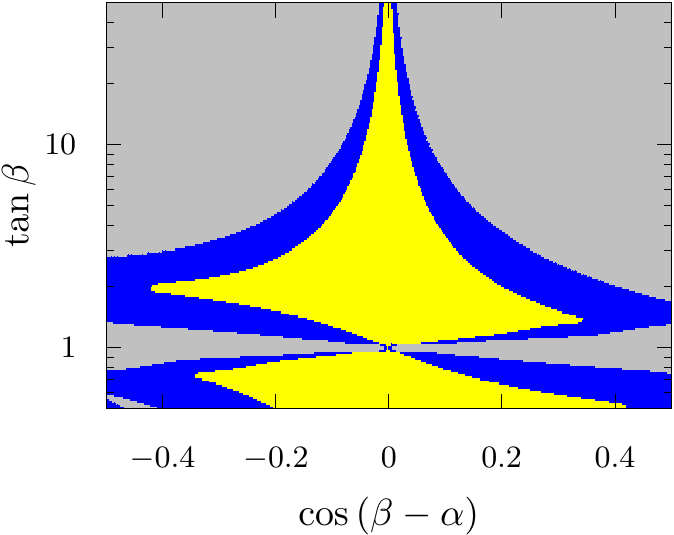}
\caption{The perturbativity measure for the scenario of
\reffi{fig:lhcconst}. 
Gray areas are ruled out, while the blue and yellow areas show the maximal Higgs self-coupling strengths $\lambda/(4\pi)$ between 0.5 and 1, and 0.3 and 0.5, respectively. Parameter sets with values smaller than 0.3 do not occur.}
\label{fig:perturbativityconstr}
\end{figure}

The recent report \cite{deFlorian:2016spz} of the LHC Higgs Cross Section Working Group summarizes a selection of 
benchmark scenarios proposed in other papers. 
We study the most relevant for our process. In particular, the scenarios proposed as BP1A  in Ref.~\cite{Haber2015} are relevant for our work. For these benchmark scenarios
 experimental constraints from direct LHC searches, shown in Fig.~\ref{fig:lhcconst}, as well as theoretical constraints from vacuum stability and perturbative unitarity, illustrated in Fig.~\ref{fig:stabconstr}, are taken into account. 
Additionally, we employ perturbativity constraints to improve this scenario. Large coupling factors can
lead to a breakdown of perturbation theory,
so that we demand sufficiently small coupling factors. 
To this end, we compute the size of each coupling factor $\lambda (S_1 S_2 S_3 S_4)$ of all the four-point Higgs-boson vertices at tree level, 
where $S_i = h, H, A, G, H^\pm, G^\pm$ for $i = 1,\dots, 4$, and use the largest coupling factor, 
$\lambda/(4 \pi)=\max|\lambda (S_1 S_2 S_3 S_4)|/(4 \pi)$, as a measure. 
We use \textsc{Mathema\-ti\-ca} and our \textsc{FeynArts} model files
 exploiting
  the hybrid basis (c.f. Ref.~\cite{Haber2015}). The parameters $Z_4$, $Z_5$, 
and $Z_7$ of the hybrid basis are related to our 
input parameters via
  \begin{align}
  \MAO^2 &= c_{\beta-\alpha}^2  \Mh^2 + \MH^2 s_{\beta-\alpha}^2 - v^2 Z_5,\\
   \MHP^2 &=  \MAO^2 - \frac{1}{2} v^2 (Z_4 - Z_5),\\ \label{eq:lam5}
\lambda_5 &= Z_5 + \frac{1}{2} t_{2\beta} \left[\frac{s_{2(\beta-\alpha)}}{2 v^2}(\Mh^2 - \MH^2) - Z_7 \right]
  \end{align}
  with $v^2 = 1/(\sqrt{2} \GF)$.

As the masses and mixing angles appear in the couplings, the perturbativity criterion 
gradually constrains the parameter space. 
Since the convergence of the perturbation series becomes worse with increasing coupling factors 
a clear discrimination of perturbative and non-perturbative parameter points is impossible. However, 
values of $\lambda/(4\pi)$
larger than~$1$ indicate that higher-order corrections do not systematically become smaller and perturbativity is not given anymore which rules out such parameter points.
Values between 0.5 and 1 usually still yield large higher-order corrections and need to be taken with care. 
The result of the perturbativity analysis is given in Fig.~\ref{fig:perturbativityconstr} for $\MH=300\GeV$ (left) and $\MH=600\GeV$ (right). Excluded areas are shown in gray, while blue ($0.5<\lambda/(4\pi)<1$) and yellow ($0.3<\lambda/(4\pi)<0.5$) indicate different sizes of the coupling factors. Parameter points where all couplings are smaller than 0.3 do not appear. The excluded trench at $\tan{\beta}=1$ is 
a singularity of the hybrid basis used in Ref.~\cite{Haber2015}, since  $t_{2\beta}$ in Eq.~\eqref{eq:lam5} and, hence, the coupling factors diverge at this point.
Overlaying these results with the previous experimental and theoretical constraints shows a significant reduction of the allowed parameter region. Nevertheless, we need to modify the scenarios proposed by Ref.~\cite{Haber2015} only slightly to obtain the low- and high-mass scenarios as well as the benchmark plane scenario described later.
\begin{figure}
\centering
\includegraphics[width=0.45\columnwidth]{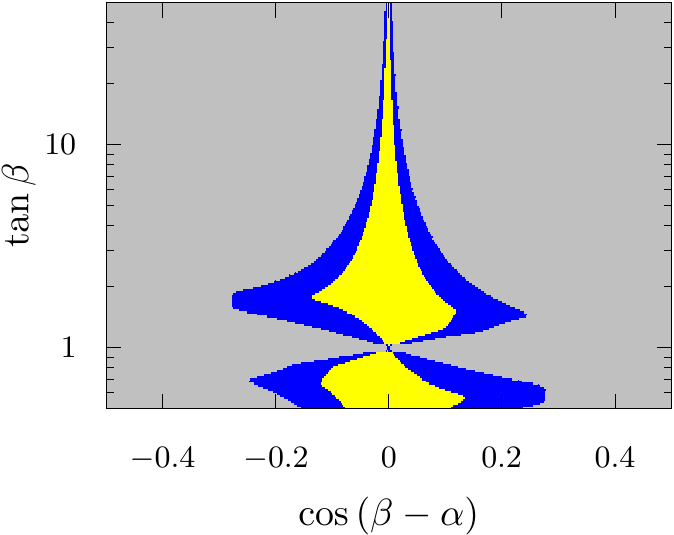}
\caption{The perturbativity measure for $\MH=1\TeV$ with the remaining parameters
as in the scenario of \reffi{fig:lhcconst}. 
Large areas are excluded by coupling factors $\lambda/(4\pi)>1$ (gray) whereas values between 0.5 and 1 
(0.3 and 0.5) are coloured blue (yellow).}
\label{fig:perturbativityconstr1000}
\end{figure}
We do not consider heavy Higgs masses in the TeV range, because the allowed parameter space is dramatically reduced in this region, which can be seen in Fig.~\ref{fig:perturbativityconstr1000}. Only parameters close to the alignment limit and with $t_\beta\approx 2$ and $t_\beta\approx0.5$ remain allowed
for $|c_{\beta-\alpha}|\sim0.1$.

\begin{enumerate}
 \item {\bf{Low-mass scenario:}} The low-mass scenario, which we already introduced in Ref.~\cite{Altenkamp:2017ldc}, consists of a heavy neutral CP-even Higgs boson $\PH$ of mass $\MH=300\GeV$.  
 The input values are based on a benchmark scenario of Ref.~\cite{Haber2015} and consist of a THDM of Type~I 
with  
\begin{align}
\Mh=125 \GeV, \quad
 \MH=300 \GeV, \quad \MAO=\MHP=460 \GeV, \quad \lambda_5=-1.9,\quad \tan \beta=2. 
\label{eq:scenarioA}
 \end{align}
Specifically, scenario~A contains a scan in $c_{\beta-\alpha}$, as this is the only parameter of the THDM appearing at LO. The range of the scan is limited by constraints from experiments and perturbative unitarity. These constraints indicate that values of $|c_{\beta-\alpha}|$ exceeding 0.1 are phenomenologically disfavoured \cite{Haber2015}. However, we perform our analysis with less stringent bounds to get a more complete picture. We take two distinguished points of the scan region named Aa and Ab with $c_{\beta-\alpha}=\pm0.1$ to perform scale variations:
\begin{subequations}
\begin{align}
\mbox{A:}  \quad \cos{(\beta-\alpha)} &= -0.2\ldots0.2, \\
\mbox{Aa:} \quad \cos{(\beta-\alpha)} &= +0.1, \\
\mbox{Ab:} \quad \cos{(\beta-\alpha)} &= -0.1. 
\end{align}
\label{eq:cba_A}
\end{subequations} 
\item {\bf{High-mass scenario:}} The high-mass scenario is again
based on a Type~I THDM, however, with heavier Higgs bosons,
\begin{align}
\Mh=125 \GeV, \quad
 \MH=600 \GeV, \quad \MAO=\MHP=690 \GeV.
\label{eq:scenarioB}
 \end{align}
Constraints from stability and perturbative unitarity (Fig.~\ref{fig:stabconstr}) reveal that positive and negative values of $c_{\beta-\alpha}$ are only allowed in different regions of $\tan{\beta}$. Therefore we define two parameter scans (B1, B2) which are applicable for positive (B1) and negative (B2) values of $c_{\beta-\alpha}$, and B1a, B2b are two distinguished points of the scan region:
\begin{subequations}
\begin{align}
\mbox{B1:}  \quad \cos{(\beta-\alpha)} &= -0\ldots0.15, \quad \lambda_5=-1.9,\quad \tan \beta=4.5,\\
\mbox{B1a:} \quad \cos{(\beta-\alpha)} &= +0.1, \\%\quad \lambda_5=-1.9,\quad \tan \beta=4.5,\\
\mbox{B2:}  \quad \cos{(\beta-\alpha)} &= -0.15\ldots 0, \quad \lambda_5=-2.4,\quad \tan \beta=1.5,\\
\mbox{B2b:} \quad \cos{(\beta-\alpha)} &= -0.1. %, \quad \lambda_5=-2.4,\quad \tan \beta=1.5,\\
\end{align}
\label{eq:cba_B}
\end{subequations} 
\item {\bf{Different THDM types:}} In this scenario, we compare different types of THDMs. Yukawa couplings appear in our process only in closed fermion loops in Higgs-boson two-point functions and in $\Ph VV$ and $\Ph\Pg\Pg$ vertex corrections,
so that the top-quark contribution is dominant. The couplings to up-type quarks is identical in all types of THDM, so that we expect negligible effects from changing the type.
The comparison is performed for scenarios~Aa and B1a.
\item {\bf{Benchmark plane:}} For this scenario, we analyze a large area of the $\MH{-}\tan{\beta}$ plane:
\begin{align}
 \MH&=300\ldots750 \GeV, \quad \tan \beta=1\ldots 50. 
\label{eq:scenarioTHDMs}
 \end{align}
The fixed parameters are based on the Type I non-alignment scenario of Ref.~\cite{Haber2015}, and are given in the hybrid basis (c.f. Ref.~\cite{Haber2015}) by
\begin{align}
\cos{(\beta-\alpha)}&=0.1,&
\Mh&=125 \GeV,& Z_4&=Z_5=-2,& Z_7&=0. 
\label{eq:scenarioBenchmark}
 \end{align}
\item {\bf{Baryogenesis:}} The $\mr{BP3}_{B}$ scenario of Ref.~\cite{deFlorian:2016spz} was initially proposed in Ref.~\cite{Dorsch:2014qja}. With a second Higgs doublet, a first-order electroweak phase transition is possible, which could explain the baryon asymmetry in the universe. The main signature for this model is the decay of a pseudoscalar Higgs boson. Nevertheless, the non-alignment of the benchmark points $\mr{BP3}_{B}$ renders these scenarios also interesting for our study. The parameterization in the original form uses  $m_{12}$ for the Higgs self-coupling parameter from which we compute $\lambda_5$ using 
 \begin{align}
  m^2_{12}&=\cb \sb (\MAO^2+\lambda_5 v^2).\label{eq:coeffrafom12}
 \end{align}
The input parameters are 
\begin{align}
\Mh=125 \GeV, \quad
 \MH=200 \GeV, \quad \MAO=\MHP=420 \GeV, \quad \lambda_5=-2.58,\quad \tan \beta=3, \nonumber
\end{align}
\begin{subequations}
and the two proposed scenarios differ by
\begin{align}
\mbox{$\mr{BP3}_{B1}$:}  \quad \cos{(\beta-\alpha)} &= 0.3, \quad \text{Type I}\\
\mbox{$\mr{BP3}_{B2}$:} \quad \cos{(\beta-\alpha)} &= 0.5. \quad \text{Type II}
\end{align}
\end{subequations} 
\item {\bf{Fermiophobic heavy Higgs:}} By choosing a Type~I THDM as well as a vanishing mixing angle $\alpha$, the heavy Higgs boson $\PH$ decouples from the fermions. Such a scenario was proposed in Ref.~\cite{Hespel:2014sla} with a direct detection of the heavy Higgs bosons as the leading signature. However, the alignment limit ($c_{\beta-\alpha}=0$) cannot be reached in this model as this would require large values of $\tan{\beta}$ which are ruled out by stability constraints. This gives rise to possibly sizable effects on the light Higgs-boson decay. Different $\tan{\beta}$ values
can be chosen, and with larger values the alignment limit is approached:
\begin{subequations}
\begin{align}
\mbox{$\mr{BP6}_{a}$:} \quad \tan{\beta} &= 40,\\
\mbox{$\mr{BP6}_{b}$:} \quad \tan{\beta} &= 20,\\
\mbox{$\mr{BP6}_{c}$:} \quad \tan{\beta} &= 10.
\end{align}
\end{subequations} 
We transform the input parameter $m_{12}$ to our convention using Eq.~\eqref{eq:coeffrafom12} and use the same fixed
$\lambda_5$ for all $\tan{\beta}$, and 
\begin{align}
\Mh=125 \GeV, \quad
 \MH=200 \GeV, \quad \MAO=\MHP=500 \GeV, \quad \lambda_5=-3.46 \quad s_\alpha=0. 
\end{align}
\end{enumerate}
 
\section{Numerical results}
\label{sec:results}

In this section we present our numerical results for the decay $\Ph\to4f$ of the light CP-even Higgs boson $\Ph$ in the THDM
for the different scenarios described in the previous section, beginning with the low-mass scenario. 
There, we investigate at first the conversion of the renormalized input parameters between different renormalization schemes and the running of the couplings.
Afterwards we discuss the scale dependence of the $\mr{h} {\to} 4f$ width and show the dependence on $c_{\beta-\alpha}$. 
First results of this study have already been published in Ref.~\cite{Altenkamp:2017ldc}. 
Finally, we study the partial widths and differential distributions in order to identify deviations from the SM expectations. The same procedure is performed for the high-mass scenario (split into two regions with positive or negative $c_{\beta-\alpha}$), while we do not perform such a detailed analysis for the other scenarios.

\subsection{Low-mass scenario}
\subsubsection{Conversion of the input parameters}
\label{sec:conversionA}

The numerical values of an
input parameter defined via different renormalization conditions in two renormalization schemes do in general 
not coincide in one and the same physics scenario,
but have to be properly converted from one scheme to the other.
This means that the mixing angles $\alpha$ and $\beta$ have to be converted in the transitions
between the four renormalization schemes described in \refse{se:ewrcs}, as already discussed in
\citere{Altenkamp:2017ldc}.
For a generic parameter $p$, the renormalized values $p^{(1)}$ and $p^{(2)}$ in two different renormalization 
schemes~1 and 2 are connected via 
\begin{align}
p_0=p^{(1)}+\delta p^{(1)} (p^{(1)})=p^{(2)} +\delta p^{(2)} (p^{(2)}),
\label{eq:pconversion_full}
\end{align}
where $p_0$ is the corresponding bare parameter 
and $\delta p^{(1,2)}$ are the NLO renormalization constants
of $\order{\alpha_\mr{em}}$.
In case of more parameters, this is a set of coupled equations.
For the conversion of $p^{(2)}$ to $p^{(1)}$, we can either use the linearized solution
\begin{align}
 p^{(1)}=p^{(2)} +\delta p^{(2)} (p^{(2)})-\delta p^{(1)} (p^{(2)}),
\end{align}
where $\delta p^{(1)} (p^{(1)})$ is approximated by $\delta p^{(1)} (p^{(2)})$,
or solve the set of implicit equations~\refeq{eq:pconversion_full} numerically.
The full solution of the implicit set of equations \refeq{eq:pconversion_full} has the
advantage that converting parameters from one scheme to another and back is an identity,
while this is only approximately the case in the linearized approach.
The error of the linearized approximation is beyond our desired NLO accuracy 
as long as the perturbation series behaves well and higher-order terms are small. 
The comparison of the results obtained with the two methods allows for a consistency check of the computation.
For the conversion of $\alpha$ and $\beta$,
we employ the \MSbar{}($\alpha$) scheme as one of the two schemes, 
so that we only have to deal with one set of finite counterterm contributions at a time. 

\begin{figure}
  \centering
  \subfigure[]{
\label{fig:plot_Umrechnung-von-alpha-LM}
\includegraphics{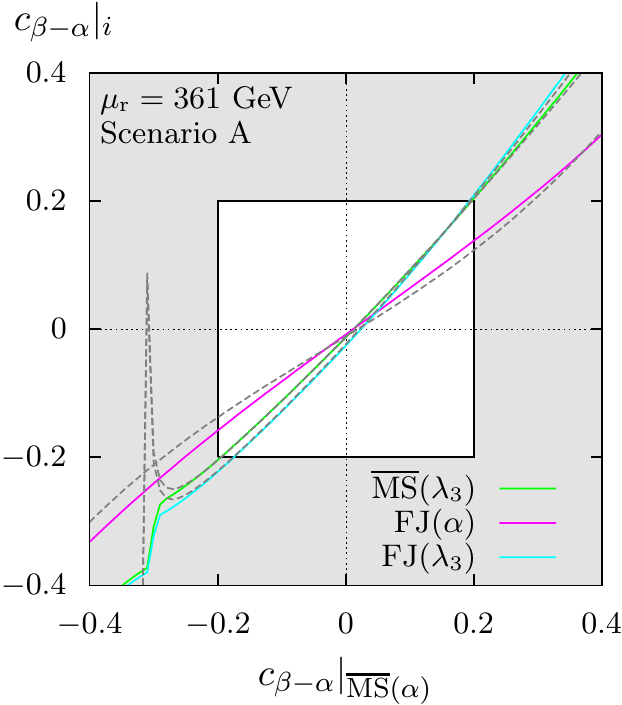}
}
\hspace{15pt}
\subfigure[]{
\label{fig:plot_Umrechnung-nach-alpha-LM}
\includegraphics{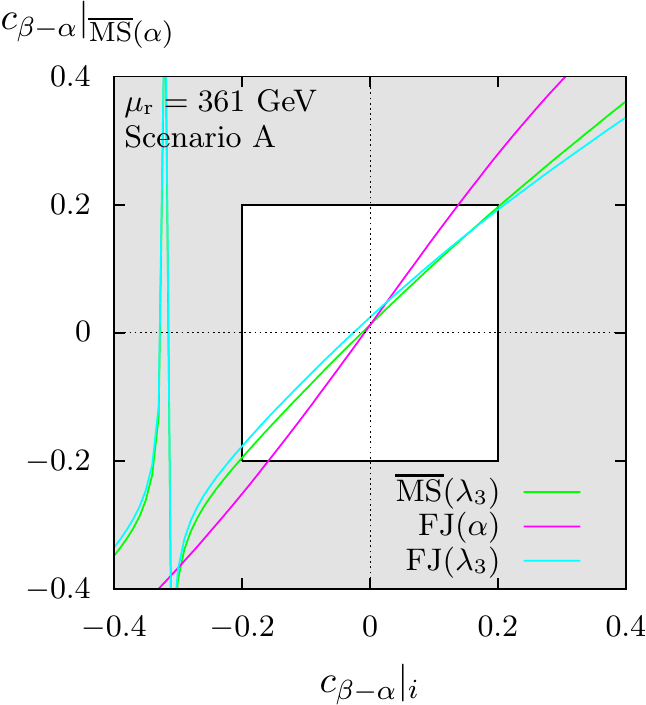}
}
\vspace*{-1em}
  \caption{Panel (a): Conversion of the value of $c_{\beta-\alpha}$ from \MSbar{}$(\alpha)$ to the \MSbar{}$(\lambda_3)$ (green), FJ($\alpha$) (pink), and FJ($\lambda_3$) schemes (turquoise) for scenario~A. Panel (b) shows the conversion to the \MSbar{}$(\alpha)$ scheme using the same colour coding. The solid lines are obtained by solving the implicit equations numerically, the dashed lines correspond to the linearized approximation. The phenomenologically relevant region is highlighted in the centre.}
\label{fig:plotconversionA}%
\end{figure}

In scenario~A, 
we extend the range of $c_{\beta-\alpha}$ values to $-0.4$  to $0.4$, 
so that we get a larger picture even though the regions with large $|c_{\beta-\alpha}|$ are ruled out by phenomenology.
The results are shown in Fig.~\ref{fig:plotconversionA} with a conversion from (l.h.s.) and to the \MSbar{}$(\alpha)$ scheme (r.h.s.), while the \MSbar{}$(\lambda_3)$ (green), FJ($\alpha$) (pink), and FJ$(\lambda_3)$ (turquoise) schemes are employed as the second scheme. 
All other conversions can be seen as a combination of the presented ones. On the left-hand side, the gray dashed lines are the result obtained using the linearized approximation\footnote{On the right-hand side, the conversion is exact, since the finite part of the $c_{\beta-\alpha}$ counterterm vanishes in the \MSbar{}$(\alpha)$ scheme.}.
 In both plots, we highlight the phenomenologically relevant region in the centre. 
 
All curves show only 
small changes in the parameter values, and the numerical solution agrees well with the linearized conversion, affirming that the finite higher-order contributions of the counterterms  are small, and perturbation theory is applicable. However, we would like to mention that a parameter set in the alignment limit 
$(c_{\beta-\alpha}\to0)$ is not preserved in the conversion to other renormalization schemes. 
The alignment limit, thus, depends on the choice of the renormalization scheme.
Outside the phenomenologically relevant region,
the curves for the transformation involving the schemes with $\lambda_3$ as an independent parameter have a small region where large effects occur. This is an artifact as these schemes become singular at $c_{2\alpha}=0$. In the vicinity of the singularity (corresponding here to $c_{\beta-\alpha}\approx -0.32$), the $\MSbar$ renormalization of $\lambda_3 $ introduces large finite contributions to the conversion equation resulting in a breakdown of perturbation theory%
\footnote{In this parameter-space region, one could choose $\lambda_1$ or $\lambda_2$ instead of $\lambda_3$ as input parameter in order to avoid this singularity.}. 
This occurs in the \MSbar{}$(\lambda_3)$ and the FJ($\lambda_3$) renormalization schemes which limits their use. 
For scenario~A, the phenomenologically relevant region is, however, not affected by this artifact.

\subsubsection{The running of \boldmath{$c_{\beta-\alpha}$}}

\begin{figure}
  \centering
  \subfigure[]{
\label{fig:running_cba0.1-LM}
\includegraphics{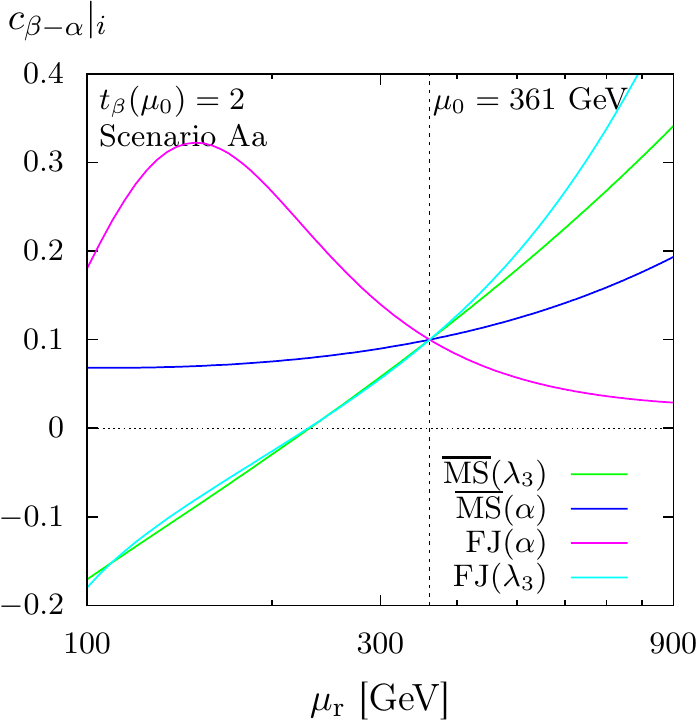}
}
\hspace{15pt}
\subfigure[]{
\label{fig:running_cba-0.1-LM}
\includegraphics{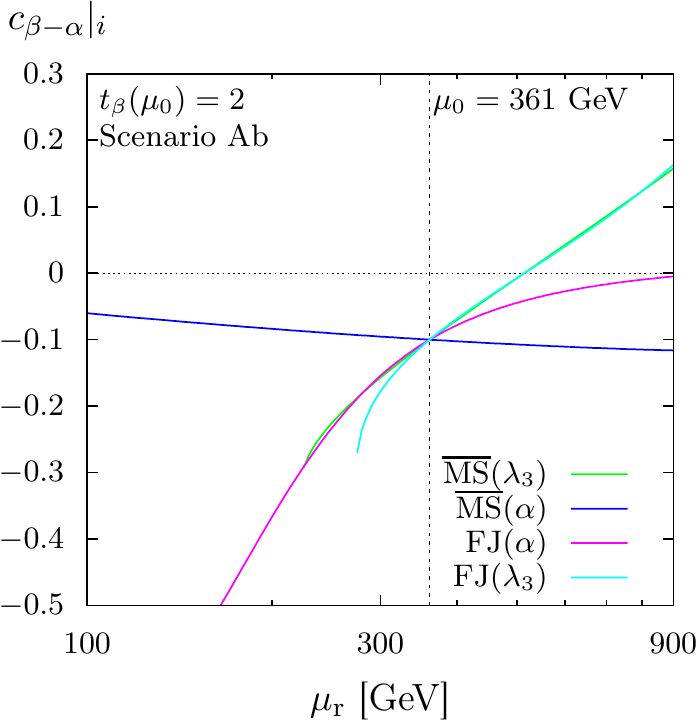}
}
\vspace*{-1em}
  \caption{The running of $c_{\beta-\alpha}$ for scenarios~Aa (a) and Ab (b) in the \MSbar{}$(\alpha)$ (blue), \MSbar{}$(\lambda_3)$ (green), FJ($\alpha$) (pink), and FJ($\lambda_3$) (turquoise) schemes.  }
\label{fig:plotrunningA}
\end{figure}

Parameters renormalized in $\MSbar$ depend on a renormalization scale $\mu_\mr{r}$, where
the dependence is governed by the renormalization group equations (RGEs) of the THDM
(see, e.g., \citeres{Cvetic:1997zd,Branco:2011iw,Bijnens:2011gd,Chakrabarty:2014aya,Ferreira:2015rha}).
For each renormalization scheme we solve the RGEs using a classical Runge--Kutta algorithm. We isolate the effects of the running from the conversion by considering each renormalization scheme separately, but do not convert the input values in this investigation. 
The scale dependence of $c_{\beta-\alpha}$ from $\mu_\mr{r}=100\GeV$ to $900\GeV$ is plotted in Fig.~\ref{fig:plotrunningA}, for the 
scenarios~Aa (l.h.s) and Ab (r.h.s) and input values given at the central scale $\mu_0$. 
It shows that the choice of  the renormalization scheme has a large impact on the scale dependence. While the \MSbar{}$(\alpha)$ scheme introduces only a slow running, 
the other schemes show a much stronger scale dependence so that excluded and unphysical values of input parameters are reached quickly. A similar observation has also been made in supersymmetric models for the parameter $\tan \beta$ (the ratio of the vacuum expectation values of the Higgs doublets in SUSY models) 
in \citere{Freitas:2002um}: 
The gauge-dependent $\MSbar$ schemes with vanishing renormalized tadpoles
have a small scale dependence while replacing the parameters by gauge-independent ones introduces additional terms in the $\beta$-functions, which arrange for a stronger scale dependence of such schemes.
We find similar results comparing the gauge-dependent \MSbar{} schemes with the gauge-independent FJ schemes in \reffi{fig:plotrunningA}.
It is remarkable that the sign of the slope differs for the different renormalization schemes. This is another consequence of the additional terms in the $\beta$-functions and shows that the choice of the renormalization scheme has large effects. 
As some curves hit the $c_{\beta-\alpha}=0$ axis and therefore run into the alignment limit, we explicitly see that
this limit depends both on the renormalization scheme and on the renormalization scale in a given scheme.
In Fig~\ref{fig:running_cba-0.1-LM} one can also see that the curves for the \MSbar{}$(\lambda_3)$ and the FJ($\lambda_3$) scheme terminate around $250\GeV$. At this scale, the running of $\lambda_3$ yields unphysical values for the parameters of the theory. 
This is unique to the $\lambda_3$ running as only there an equation needs to be solved in the relation to $c_{\beta-\alpha}$. For the other cases we prevent the angles from running out of their domain of definition by solving the running for the tangent of the angles.

\subsubsection{Scale variation of the width}
\label{sec:LowMassscalevariation}

\begin{figure}
  \centering
  \subfigure[]{
\label{fig:plot_MUSCAN-Aa-L3MS}
\includegraphics{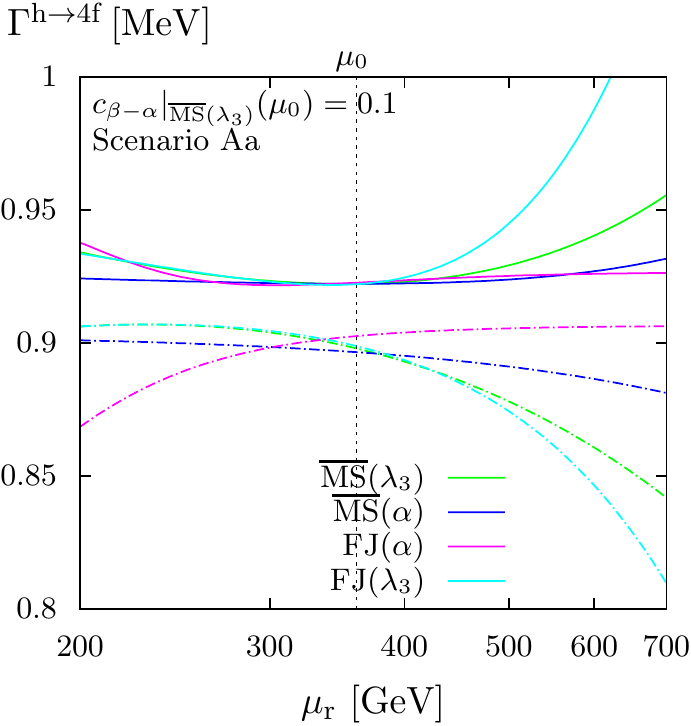}
}
\hspace{15pt}
\subfigure[]{
\label{fig:plot_MUSCAN-Aa-alphaMS}
\includegraphics{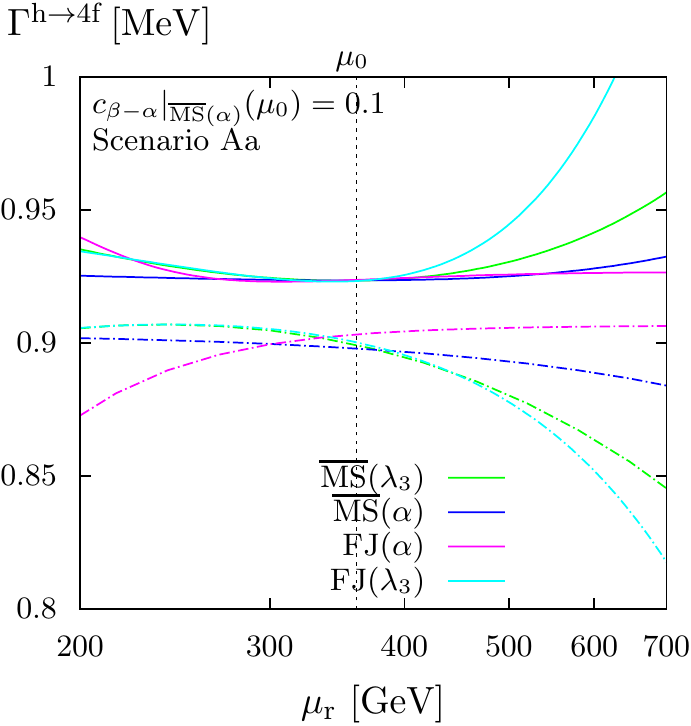}
}
\\[-1em]
  \subfigure[]{
\label{fig:plot_MUSCAN-Aa-FJ}
\includegraphics{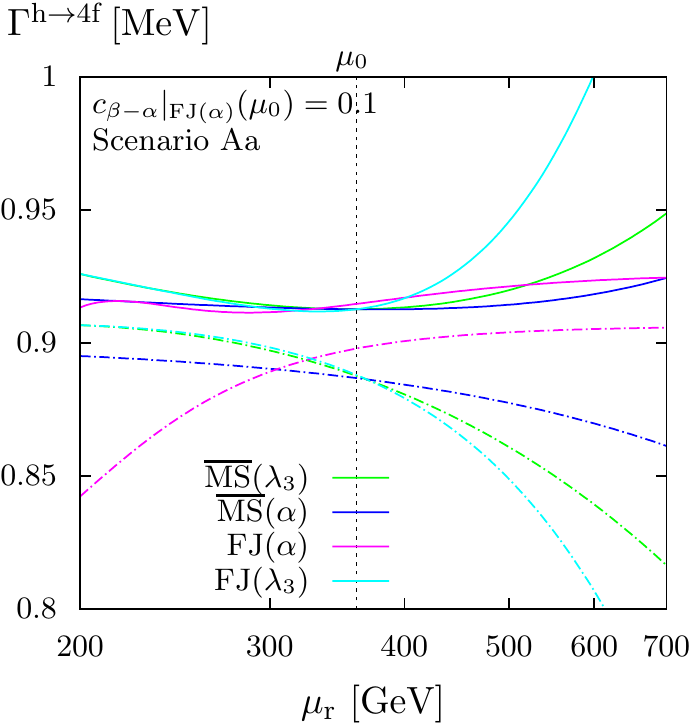}
}
\hspace{15pt}
\subfigure[]{
\label{fig:plot_MUSCAN-Aa-L3MSFJ}
\includegraphics{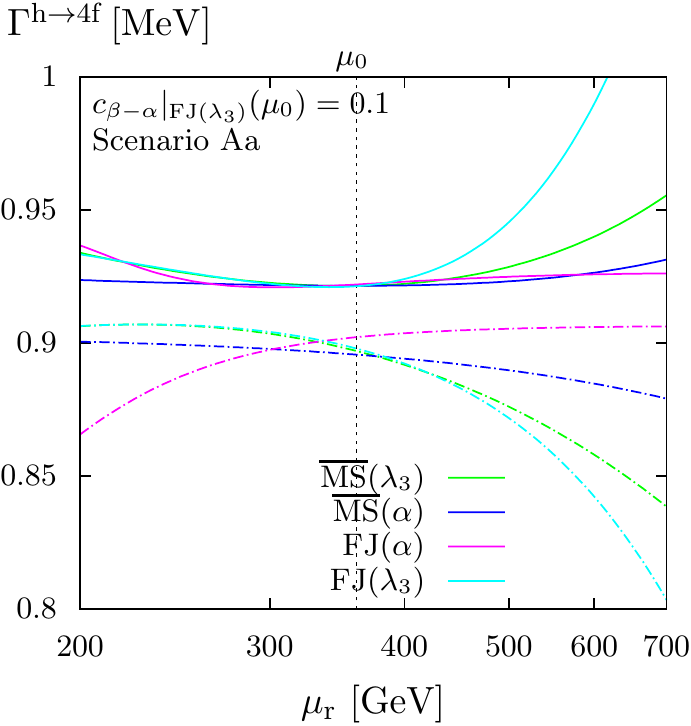}
}
\vspace*{-.5em}
\caption{The $\Ph {\to} 4f$ width at LO (dashed) and NLO EW (solid) for scenario Aa in dependence of the renormalization scale. The panels (a), (b), (c), and (d) correspond to input values defined in the  \MSbar{}$(\lambda_3)$, \MSbar{}$(\alpha)$, FJ($\alpha$), and FJ($\lambda_3$) schemes, respectively. The result is computed in all four different renormalization schemes after converting the input at NLO (also for the LO curves)
and displayed using the colour code of Fig.~\ref{fig:plotrunningA}.  }
\label{fig:plotmuscanAa}
\end{figure}

\begin{figure}
  \centering
  \subfigure[]{
\label{fig:plot_MUSCAN-Ab-alphaMS}
\includegraphics{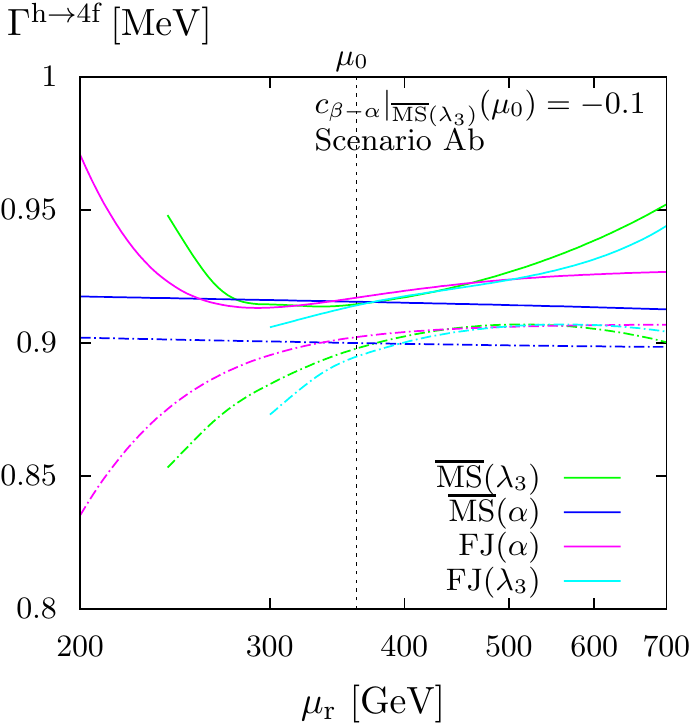}
}
\hspace{15pt}
\subfigure[]{
\label{fig:plot_MUSCAN-Ab-L3MS}
\includegraphics{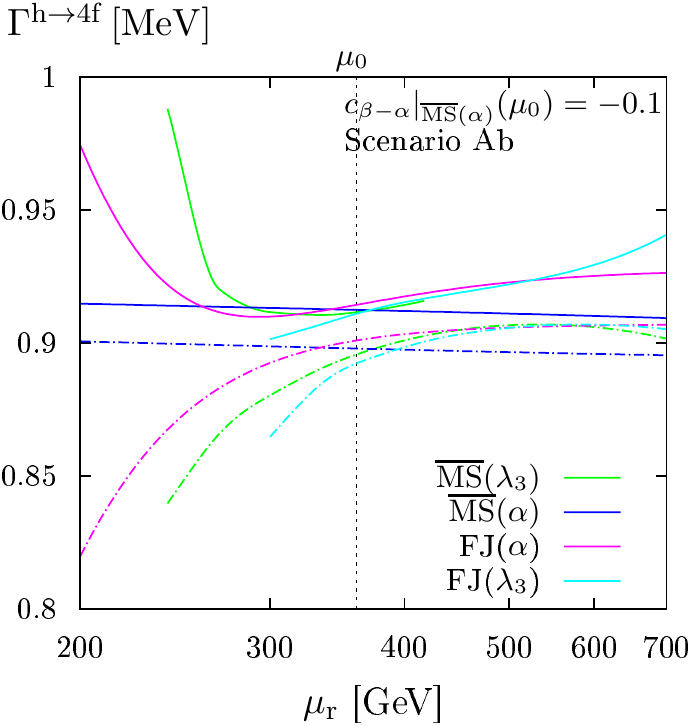}
}
\\[-1em]
  \subfigure[]{
\label{fig:plot_MUSCAN-Ab-FJ}
\includegraphics{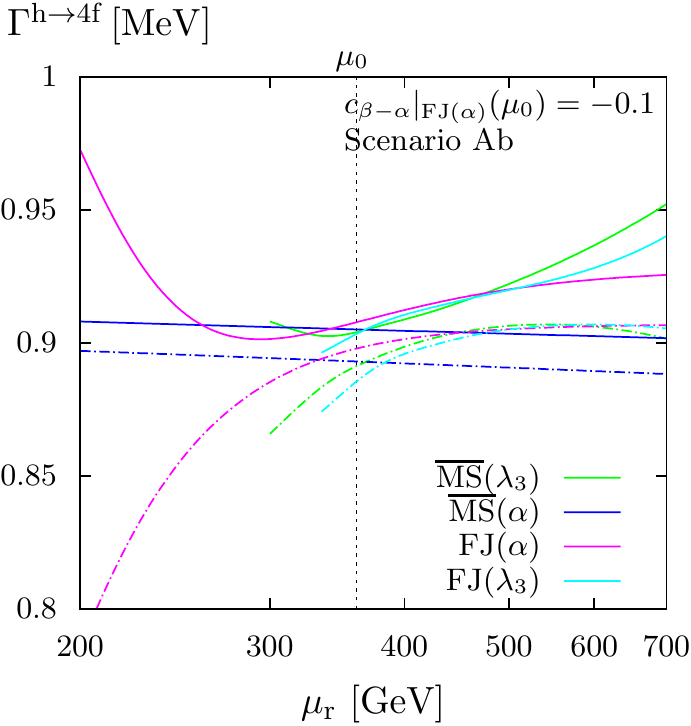}
}
\hspace{15pt}
\subfigure[]{
\label{fig:plot_MUSCAN-Ab-L3MSFJ}
\includegraphics{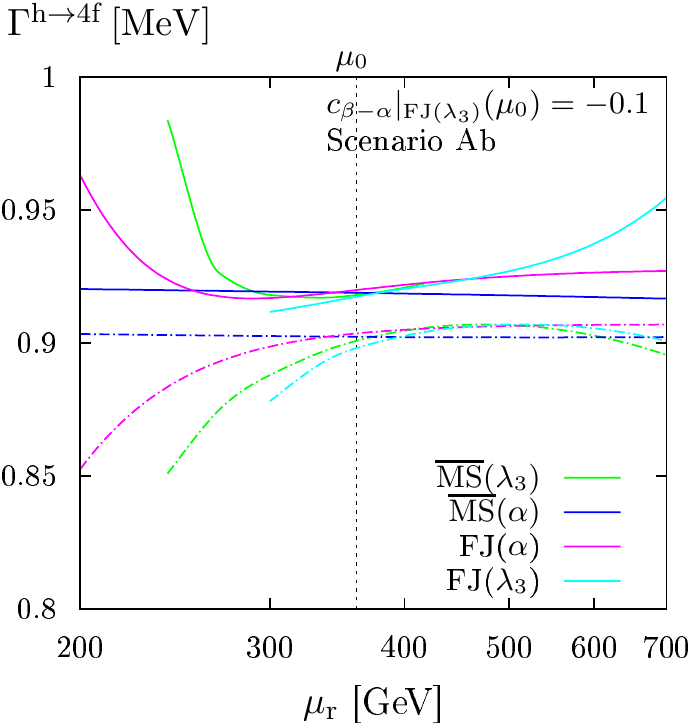}
}
\vspace*{-.5em}
%  \caption{The $\Ph {\to} 4f$ width at LO (dashed) and NLO EW (solid) for scenario~Ab in dependence of the renormalization scale. The panels (a), (b), (c), and (d) correspond to input values  defined in the  \MSbar{}$(\lambda_3)$, \MSbar{}$(\alpha)$, FJ($\alpha$), and FJ($\lambda_3$) schemes, respectively. The result is computed in all four different renormalization schemes after converting the input at NLO (also for the LO curves) and displayed using the colour code of Fig.~\ref{fig:plotrunningA}.}
\caption{As in \reffi{fig:plotmuscanAa}, but for scenario~Ab.}
\label{fig:plotmuscanAb}
\end{figure}

Owing
to the appearance of heavy Higgs bosons in the loop diagrams multiple scales occur in the calculation of the NLO EW corrections in the THDM. Therefore, a naive choice of the central renormalization scale of $\mu_0=\Mh$ might not be appropriate. To choose and to justify our central scale of Eq.~\eqref{eq:centralscale}, and  to estimate the theoretical uncertainties, we compute the total width according to Eq.~\eqref{eq:totwidth} while the scale is varied 
by roughly a factor of two around $\mu_0$.
As a definition of the input parameters in each of the four renormalization schemes represents a physical scenario on its own, we have four input prescriptions (\MSbar{}($\lambda_3$), \MSbar{}$(\alpha)$, FJ($\alpha$), FJ($\lambda_3$)), and for each of them we compute the result in all renormalization schemes. 
After converting the input to the desired renormalization scheme, we evolve the \MSbar{} parameters from the scale $\mu_0$ to
$\mu_\mr{r}$ by solving the RGEs, and finally compute the $\Ph{\to}4f$ width.
The results are shown in Figs.~\ref{fig:plotmuscanAa} and~\ref{fig:plotmuscanAb} at LO (dashed) and NLO EW (solid) for the 
scenarios~Aa and Ab for each of the input prescriptions. 
The QCD corrections are not part of the EW scale variation and therefore omitted in these results. 

The benchmark scenario~Aa 
shows almost textbook-like behaviour, and the results are similar for all input prescriptions so that we discuss all of them simultaneously. First of all, the LO computation shows a strong scale dependence for all renormalization schemes, resulting in sizable differences between the curves. However, each of the NLO curves show a wide extremum with a large plateau, reducing the scale dependence drastically, as it is expected for NLO calculations. 
The central scale $\mu_0=(\Mh+\MH+\MAO+2\MHP)/5$ lies perfectly in the middle of the plateau regions justifying this scale choice. 
In contrast, the naive scale choice $\mu_\mr{r}=\Mh$ is not within the plateau region%
\footnote{For each renormalization scheme (without parameter conversion), we also tested the choice of the running input parameters at the scale $\mu_\mr{r}=\Mh$. 
Some of those results and further explanations can be found in \citere{Altenkamp:2017ldc}.
No plateau region was found for $\mu_\mr{r}=\Mh$. In addition, we found that the conversion of the parameters  became unreliable for this scale choice.},
 leads to 
unnaturally large corrections, and should not be chosen. 
For all renormalization schemes, the plateaus coincide, and the agreement between the renormalization schemes is improved. 
This is expected since results obtained with different renormalization schemes should be equal up to higher-order terms, if the input parameters are properly converted. 
The relative renormalization scheme dependence at the central scale,
\begin{align}
\Delta_\mr{RS}=2 \,\frac{\Gamma^{\Ph {\to} 4 \Pf}_\mr{max}(\mu_0)-\Gamma^{\Ph {\to} 4\Pf}_\mr{min}(\mu_0)}{\Gamma^{\Ph {\to} 4\Pf}_\mr{max}(\mu_0)+\Gamma^{\Ph {\to} 4\Pf}_\mr{min}(\mu_0)},
\label{eq:renschemedep}
\end{align}
 expresses the dependence of the result on the renormalization scheme. It can be computed for a specific input prescription from the difference of the smallest and largest width of the four renormalization schemes, $\Gamma^{\Ph {\to} 4 \Pf}_\mr{min}(\mu_0)$ and $\Gamma^{\Ph {\to} 4\Pf}_\mr{max}(\mu_0)$, 
 normalized to their average. In Tab.~\ref{tab:schemevar}, $\Delta_\mr{RS}$ is given at LO and NLO for each of the input variants and confirms the reduction of the scheme dependence in the NLO calculation. 
  \begin{table}
   \centering
   \renewcommand{\arraystretch}{1.1}
 \begin{tabular}{|cc|cccc|}\hline
   && \MSbar{}$(\lambda_3)$ & \MSbar{}$(\alpha)$ & FJ($\alpha$) & FJ($\lambda_3$)\\\hline
   \multirow{2}{*}{Scenario Aa:}& $\Delta^\mr{LO}_\mr{RS}$[\%]       & 0.67(0)  & 0.59(0)  & 1.25(0)   & 0.73(0) \\
    &$\Delta^\mr{NLO}_\mr{RS}$ [\%]     & 0.08(0)  & 0.06 (0)  & 0.27(0)   & 0.09(0)   \\\hline
   \multirow{2}{*}{Scenario Ab:}&  $\Delta^\mr{LO}_\mr{RS}$[\%]       & 0.84(0)  & 1.00(0)  & 1.31(0)   & 0.63(0)  \\
    &$\Delta^\mr{NLO}_\mr{RS}$ [\%]     & 0.34(0)  & 0.39(0) & 0.49(1)   & 0.28(0) \\\hline
  \end{tabular}  
   \caption{The variation $\Delta_\mr{RS}$ of the $\mr{h} {\to} 4f$ width 
in scenarios~Aa and Ab at the central scale $\mu_0$
using different renormalization schemes (with NLO parameter conversions).
The columns correspond to the schemes in which the input parameters are defined.
The technical uncertainty in brackets is calculated by exploiting the integration errors for the central values corresponding to the maximal and the minimal width.}
 \label{tab:schemevar}
 \end{table}%
In addition, the choice of the \MSbar{}$(\alpha)$ scheme as an input scheme leads to the smallest dependence on the renormalization schemes in scenario~Aa. This fits well to the observation perceived when the running was analyzed that the  \MSbar{}$(\alpha)$ scheme shows the smallest dependence on the renormalization scale, attesting a good absorption of further corrections into the NLO prediction.

The situation for benchmark 
scenario~Ab is more subtle. For negative values of $c_{\beta-\alpha}$ the truncation of the schemes involving $\lambda_3$ at $\mu_\mr{r}=250{-}300\GeV$ 
as well as the breakdown of the running of the FJ($\alpha$) scheme, which both were observed in the running in Fig.~\ref{fig:running_cba-0.1-LM}, are also manifest in the computation of the $\Ph {\to} 4f$ width.
 Therefore the results vary much more, and the extrema with the plateau regions are not as distinct as for the previous 
scenario, and even missing 
for some of
the truncated curves. Nevertheless, the situation improves at NLO, and the relative renormalization scheme dependence reduces, as shown in Tab.~\ref{tab:schemevar}. Also the central scale choice of $\mu_0$ seems to be appropriate in contrast to a naive choice of $\Mh$.
 
For both benchmark scenarios, 
the estimate of the theoretical uncertainties by varying the scale by a factor of two from the central value for an arbitrary renormalization scheme is generally not appropriate. A proper strategy would be to identify the renormalization schemes 
that yield reliable results, and to use only those to quantify the theoretical uncertainties from the scale variation. In addition, the renormalization scheme dependence of those schemes should be investigated. However, this procedure must be performed for each benchmark 
scenario separately, which is beyond the scope of this work for a larger list of benchmark scenarios.

\subsubsection{\boldmath{$c_{\beta-\alpha}$} dependence}
\label{sec:cbascanA}

\begin{figure}
  \centering
  \subfigure[]{
\label{fig:plot_cbascan-A-diffschemes-L3MS}
\includegraphics{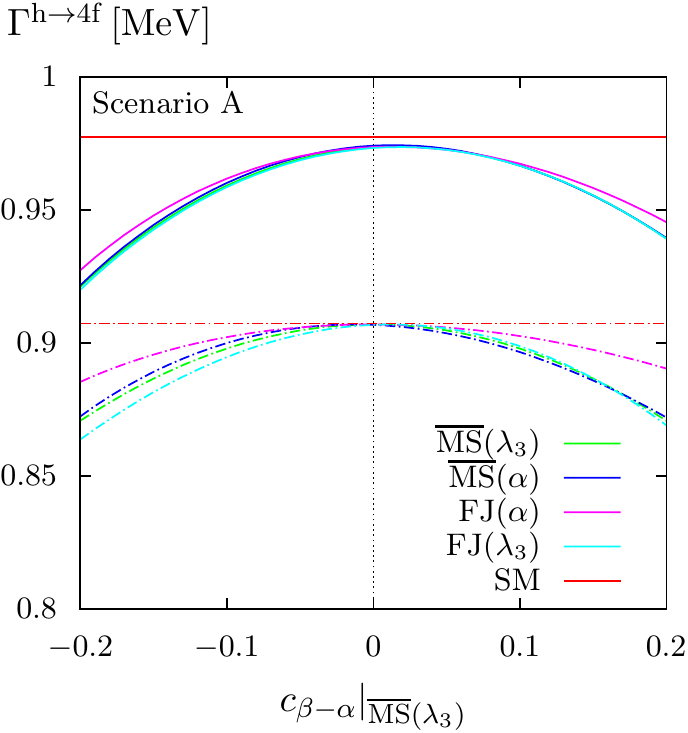}
}
\hspace{15pt}
\subfigure[]{
\label{fig:plot_cbascan-A-diffschemes-alpha}
\includegraphics{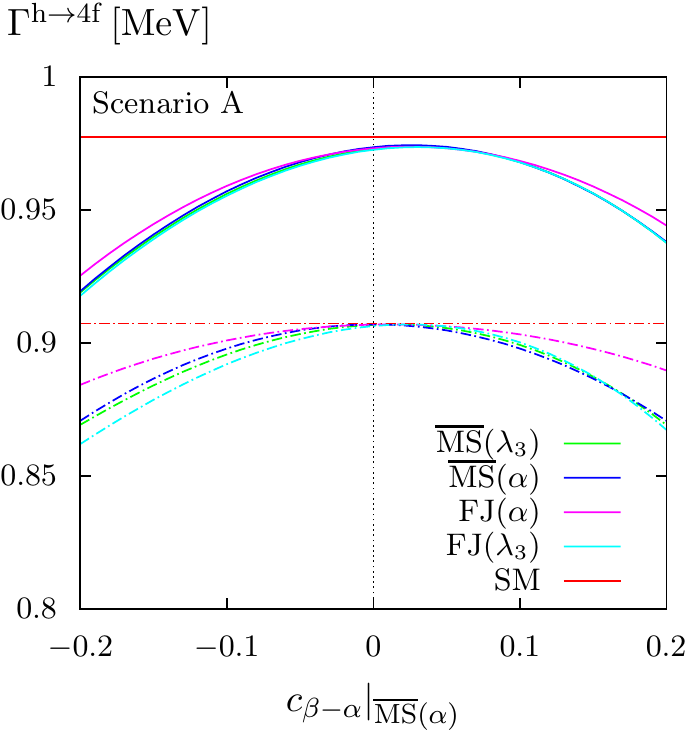}
}
\\[-1em]
  \subfigure[]{
\label{fig:plot_cbascan-A-diffschemes-FJ}
\includegraphics{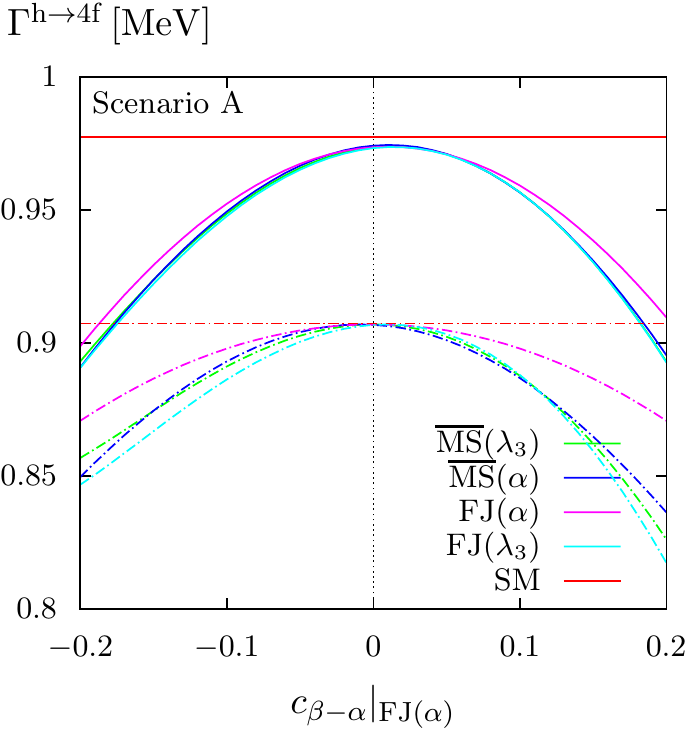}
}
\hspace{15pt}
\subfigure[]{
\label{fig:plot_cbascan-A-diffschemes-L3MSFJ}
\includegraphics{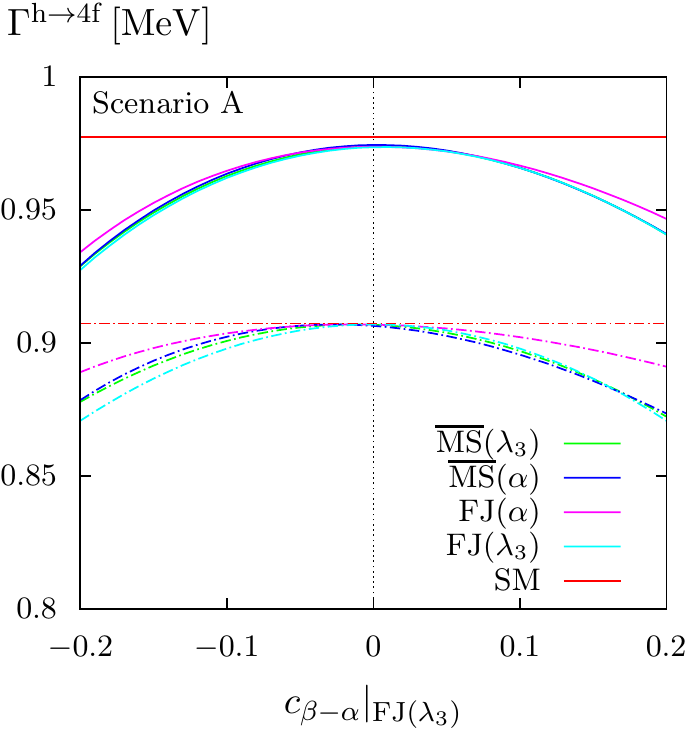}
}
\vspace*{-.5em}
  \caption{The $\mr{h} {\to} 4f$ width at LO and full NLO EW+QCD (solid) for scenario~A in dependence of $c_{\beta-\alpha}$. The panels (a), (b), (c), and (d) correspond to input values  defined in the  \MSbar{}$(\lambda_3)$, \MSbar{}$(\alpha)$, FJ($\alpha$), and FJ($\lambda_3$) schemes, respectively. 
Parameters are consistently converted between the renormalization schemes (both for NLO and LO predictions) by numerically solving the
non-linear matching equations \refeq{eq:pconversion_full}.
Results in the different target schemes are displayed with the colour code of Fig.~\ref{fig:plotrunningA}, and the SM 
(with a SM Higgs-boson mass of $\Mh$) is shown for comparison in red.}
\label{fig:plotscbascanA}
\end{figure}

\begin{figure}
  \centering
\includegraphics[scale=0.98]{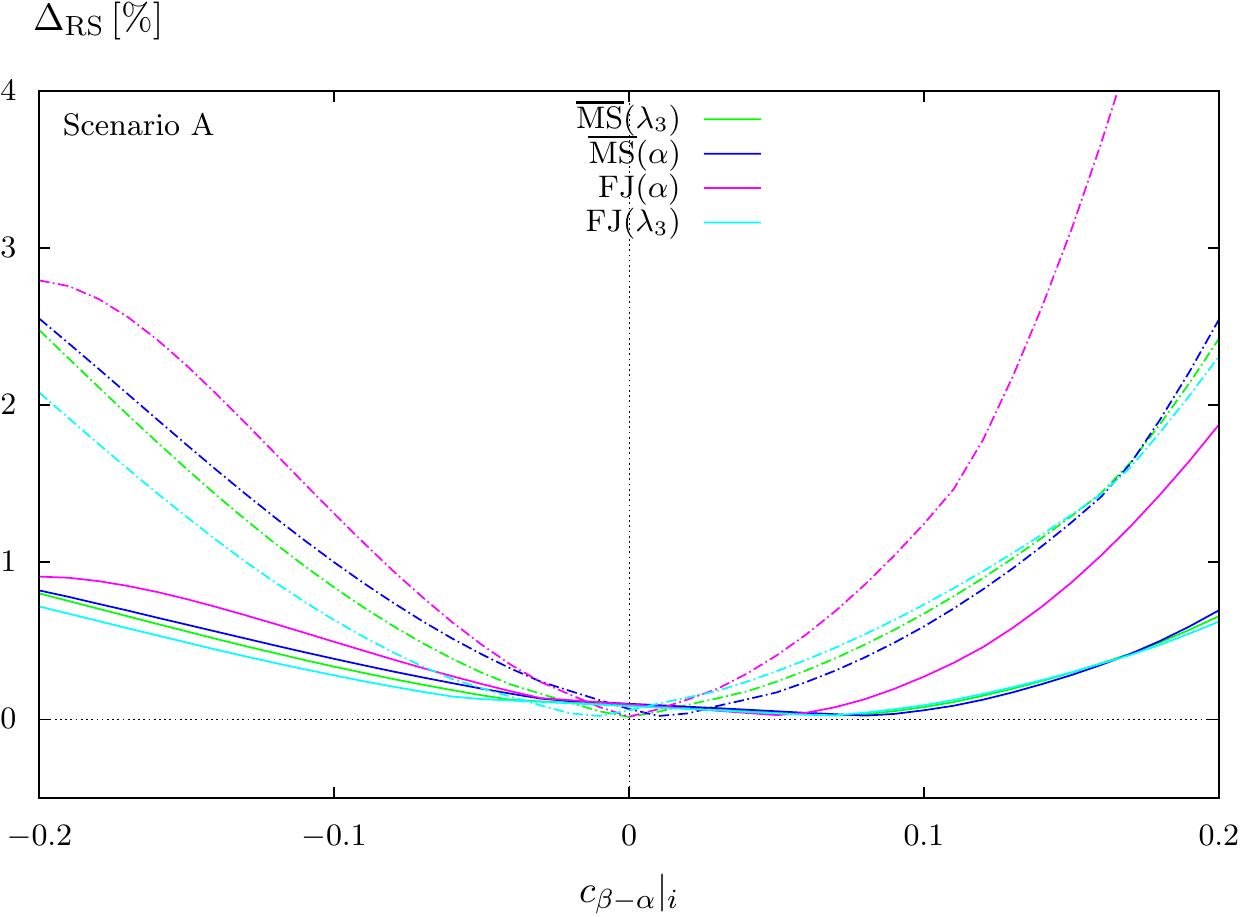}
  \caption{The relative dependence of the $\Ph {\to} 4f$ width on the renormalization schemes as defined in Eq.~\eqref{eq:renschemedep} for the LO (dashed) and NLO EW+QCD (solid) calculation. The different colours correspond to calculations with input values defined in the different renormalization schemes, converted to the other schemes at NLO (converted parameters have also been used for the LO curves).}
\label{fig:plot_schemedependence-A}
\end{figure}

The dependence of the $\mr{h} {\to} 4f$ width on $c_{\beta-\alpha}$ is one of the central results of our analysis, as the decay observables of the Higgs boson into four fermions in the THDM are most sensitive to this THDM parameter.
The $\Ph {\to} 4f$ width in dependence of $c_{\beta-\alpha}$ in scenario~A is shown
in Figs.~\ref{fig:plot_cbascan-A-diffschemes-L3MS}--(d) for the four different input prescriptions. 
The LO width with NLO conversion (dashed) and the full NLO EW+QCD total width (solid) are computed in the different renormalization schemes after the NLO input conversion,  using the constant default scale $\mu_0$ of Eq.~\eqref{eq:centralscale}. The SM values with a SM Higgs-boson mass of $\Mh$ are illustrated in red. 
The results are similar for all input prescriptions so that we discuss them simultaneously. 
At LO they show the suppression w.r.t.\ to the SM with the factor $s_{\beta-\alpha}^2$. The differences at LO
between the renormalization schemes are due to the conversion of the input. A pure LO computation is identical for all renormalization schemes,
since there is no conversion in a pure LO prediction. 
This pure LO prediction is represented in each plot by the LO curve for which no conversion to another scheme is performed. 
The suppression w.r.t.\ to the SM computation does not change at NLO, while the shape becomes slightly asymmetric, and the NLO 
results show a significantly better agreement between the renormalization schemes. This is also confirmed by the relative renormalization scheme dependence shown in Fig.~\ref{fig:plot_schemedependence-A}.

\begin{figure}
  \centering
\includegraphics[scale=0.9]{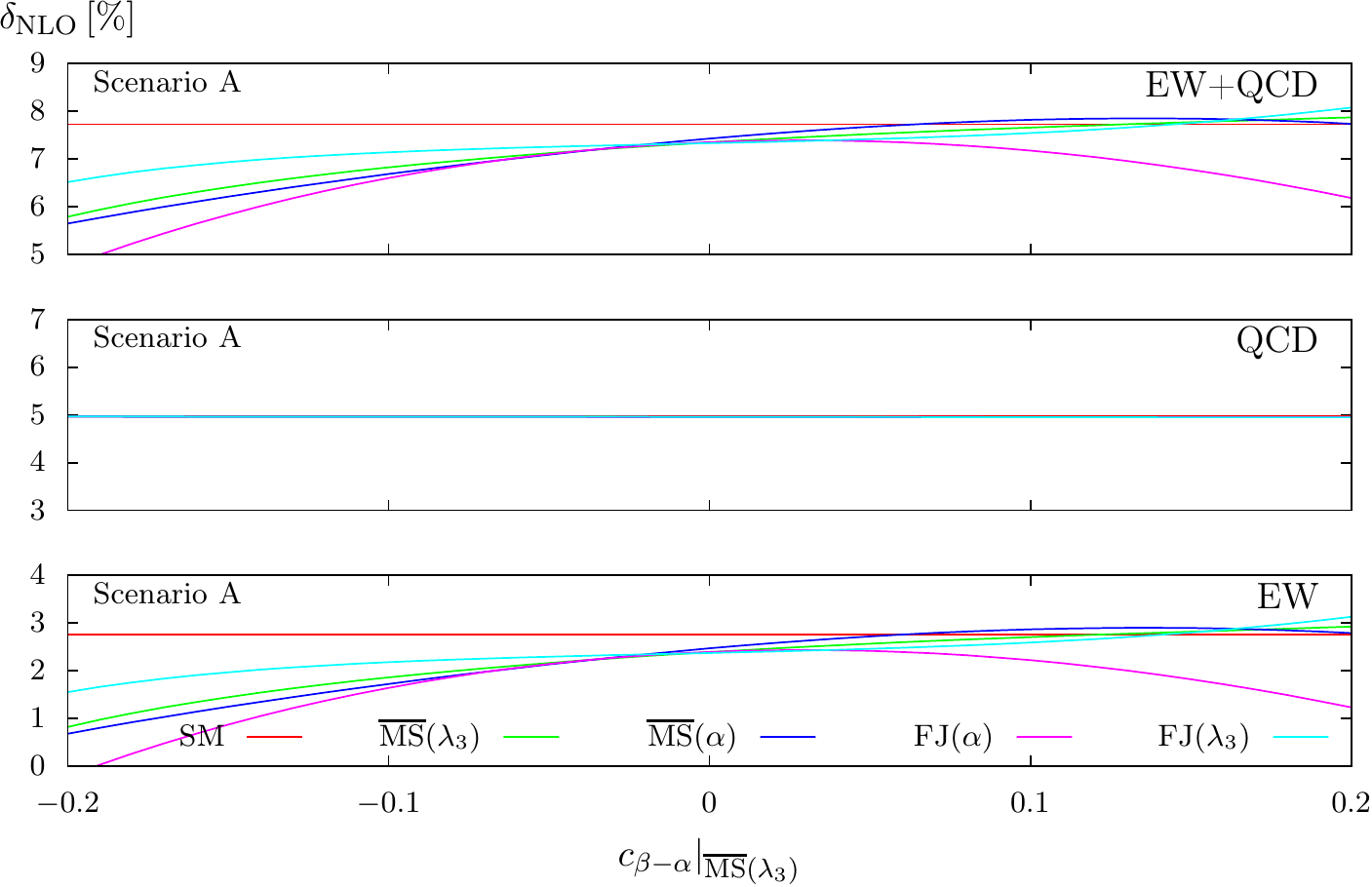}
  \caption{The relative NLO EW+QCD, QCD, and EW corrections to the $\Ph {\to} 4f$ width in scenario~A.
The input values are defined in the  $\MSbar(\lambda_3)$ scheme and 
converted to the other schemes at NLO (also for the LO curves). 
The results computed with different renormalization schemes are displayed with the colour code of Fig.~\ref{fig:plotrunningA}.}
\label{fig:plot_cbascanrel-A-diffschemes-L3MS}
\end{figure}

The relative corrections to the $\Ph{\to}4f$ width, defined by
\begin{align}
\delta_\NLO = \delta_\QCD + \delta_\EW = \frac{\Gamma_\mr{NLO}}{\Gamma_\mr{LO}}-1,
\end{align}
are displayed 
in Fig.~\ref{fig:plot_cbascanrel-A-diffschemes-L3MS} for input parameters defined in the \MSbar{}$(\lambda_3)$ scheme. 
For input parameters defined in the other schemes we obtain similar results, which are not shown. 
The different plots show the full EW+QCD ($\delta_\NLO$), the QCD ($\delta_\QCD$), and the EW corrections ($\delta_\EW$), 
where the first is just the sum of the two individual contributions. The relative QCD corrections lie practically on top of each other, so that only one line is visible even though the calculation was made in all renormalization schemes. 
The QCD corrections are almost identical to the SM case, which is not surprising as the interference of the diagram involving a closed quark
loop (Fig.~\ref{fig:QCDvirt3}) is the only contribution 
in which the THDM amplitudes are not simple rescalings of their SM counterpart
by the factor $s_{\beta-\alpha}$.
Those diagrams contribute only little to the $\Ph {\to} 4f$ width, so that the relative QCD corrections become similar to the SM. The EW corrections with the heavy Higgs bosons in the loop show a small asymmetry w.r.t.\ to the sign of $c_{\beta-\alpha}$ and are between 0 and $3\%$, even exceeding the relative corrections in the SM
in the regions of large $c_{\beta-\alpha}$.

\begin{figure}
  \centering
\includegraphics[scale=0.9]{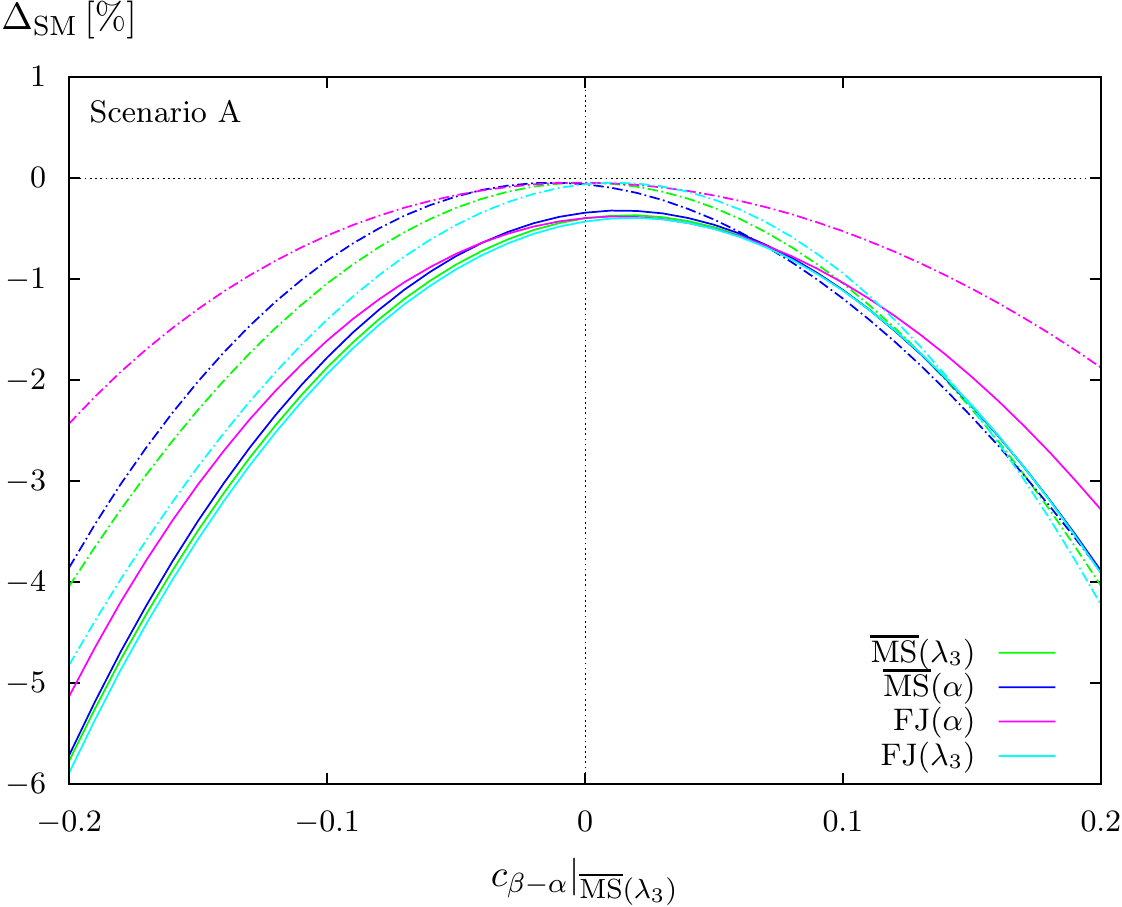}
\caption{The relative difference of the $\Ph {\to} 4f$ decay width in the THDM w.r.t.\ the SM prediction at LO (dashed) and NLO EW+QCD (solid). The input scheme is \MSbar{}$(\lambda_3)$, and the corrections are computed in all four renormalization schemes after converting the input at NLO (also for the LO curves), which are displayed using the colour code of Fig.~\ref{fig:plotrunningA}. }
\label{fig:plot_cbascanrelSM-A-diffschemes-L3MS}
\end{figure}

Deviations of the THDM results from the SM can be investigated when the SM Higgs-boson mass is identified with the mass of the
light Higgs boson~$\Ph$ of the THDM. The relative deviation of the full width from the SM is then
\begin{align}
 \Delta_\mr{SM}= \frac{\Gamma_\mr{THDM}-\Gamma_\mr{SM}}{\Gamma_\mr{SM}},
\end{align}
which is shown in Fig.~\ref{fig:plot_cbascanrelSM-A-diffschemes-L3MS} at LO (dashed) and NLO (solid) 
for parameters defined in the \MSbar{}$(\lambda_3)$ scheme (other input definitions yield similar results). 
The SM exceeds the THDM widths at LO and NLO. At LO the shape of $c_{\beta-\alpha}^2$ can be seen, with modifications due to input conversion. This shape is slightly distorted at NLO by the asymmetry of the EW corrections, and a small offset of $-0.5$\% is visible even in the alignment limit where the diagrams including heavy Higgs bosons still contribute. The NLO computations show larger negative deviations,
which could be used to improve current exclusion bounds or increase their significance. Nevertheless, in the whole scan region the deviation from the SM is within $6\%$ and for the parameter region with $|c_{\beta-\alpha}|<0.1$ even less than $2\%$, which 
will be challenging for experiments to measure.

\begin{figure}
  \centering
\includegraphics[scale=0.9]{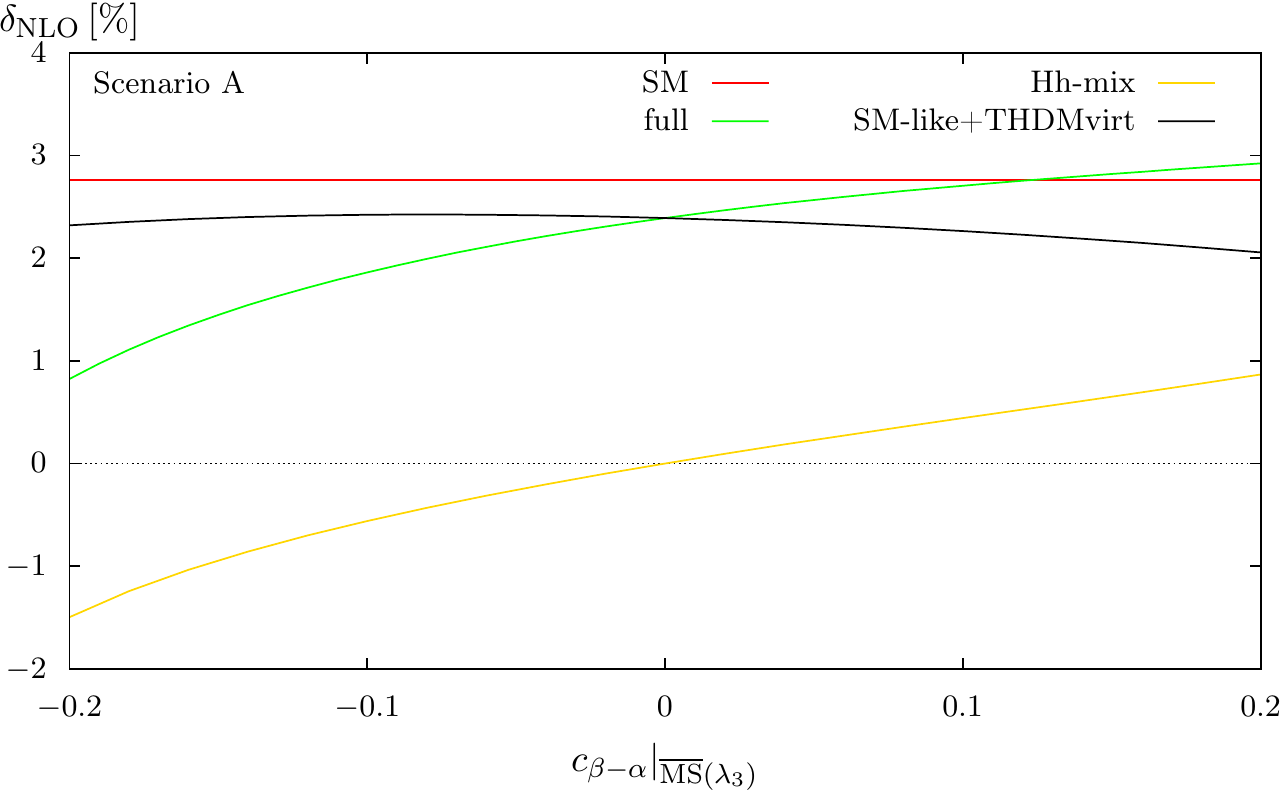}
  \caption{The full relative NLO corrections to $\Ph {\to} 4f$ (green) split into different subcontributions. The SM-like contribution consists of all diagrams that have a SM equivalent, the THDM-virt contribution includes all one-loop and counterterm contributions that involve heavy Higgs bosons 
(sum of the two parts shown in black), and Hh-mixing contribution is displayed in yellow. For comparison, the relative NLO corrections in the SM are shown in red.}
\label{fig:plot_cbascanrel-A-diffschemes-L3MS-ext}
\end{figure}

We also investigate the origin of the relative EW corrections. 
To this end, in Fig.~\ref{fig:plot_cbascanrel-A-diffschemes-L3MS-ext} we plot 
different contributions 
 to the full correction
in the \MSbar{}($\lambda_3$) renormalization scheme 
(the breakup in the other schemes is qualitatively similar).
The contribution called ``SM-like+THDM-virt''
comprises all diagrams that have a SM correspondence as well as the real corrections of Eq.~\eqref{eq:realEW},
and the diagrams of Fig.~\ref{fig:EWvirtHVVbosTHDM} with at least one heavy Higgs boson in the loop and the contributions 
of heavy Higgs bosons to the counterterms. 
The part called ``Hh-mix'' is defined by the contributions of the Hh-mixing field renormalization constant
$\delta Z_{\PH\Ph}$ and the renormalization constant $\delta\alpha$ in Eq.~\refeq{eq:CTampZ}.
Note that this splitting is neither gauge-independent, nor UV-finite.% 
\footnote{The standard UV divergence $\Delta=2/(4-D)-\gamma_{\mathrm{E}}+\ln(4\pi)$
in dimensional regularization is set to zero, and the reference mass scale $\mu$ is identified
with $\mu_0$ here.}
However, the major contribution to
the Hh-mix part is furnished by Higgs-boson loops, which do not depend on the gauge, so that some qualitative
conclusions may be drawn.
The SM-like+THDM-virt diagrams shows a 
small off-set from the SM result and only a small $c_{\beta-\alpha}$ dependence, showing that the modification of the coupling factors in the THDM is small, but grows when the alignment limit is left. In the alignment limit, the off-set originates only from the heavy Higgs boson contribution, 
since only these diagrams introduce differences w.r.t.\ the SM. 
For values of $|c_{\beta-\alpha}|>0.05$ the major deviation from the SM and the shape of the EW corrections are mainly 
due to the Hh~Higgs mixing. These terms factorize from the LO contribution and thus lead to a uniform 
(i.e.\ phase-space independent) and universal (i.e.\ final-state independent) correction factor to the LO prediction.

\subsubsection{Partial widths for individual four-fermion states}
\label{sec:partwidth-LM}

The partial $\Ph {\to} 4f$ widths (as defined in \refse{sec:Prophecy4f}) at NLO
are shown in the \MSbar{}$(\lambda_3)$ scheme
for benchmark 
scenarios~Aa and Ab in Tabs.~\ref{tab:partialwidthsAa} and \ref{tab:partialwidthsAb}, respectively. 
\begin{table}
   \centering
   \renewcommand{\arraystretch}{.94}
 \begin{tabular}{|c|ccccc|}\hline \vphantom{$\Big|$}
  Final state& $\Gamma^{\Ph{\to}4f}_\mr{NLO}$ [MeV] & $\delta_\mr{EW}$ [\%]& $\delta_\mr{QCD}$ [\%] & $\Delta_\mr{SM}^\mr{NLO}$ [\%]&$\Delta_\mr{SM}^\mr{LO}$ [\%]\\\hline
inclusive $\Ph {\to} 4f$ & $ 0.96730 ( 7)$ & $ 2.71 ( 0)$ & $ 4.96 ( 1)$ & $ -1.05 ( 1)$ & $ -1.00 ( 1)$ \\
ZZ & $ 0.106126 ( 6)$ & $ 0.34 ( 0)$ & $ 4.88 ( 0)$ & $ -1.13 ( 1)$ & $ -1.00 ( 0)$ \\
WW & $ 0.86630 ( 8)$ & $ 3.00 ( 0)$ & $ 5.01 ( 1)$ & $ -1.04 ( 1)$ & $ -1.00 ( 1)$ \\
WW/ZZ int. & $ -0.00513( 5)$ & $ 1.3 ( 2)$ & $ 12.0 ( 8)$ & $ -1 ( 1)$ & $ -1 ( 1)$ \\
$ \nu_\Pe \Pe^+ \mu^- \bar{\nu}_\mu$ & $ 0.010201 ( 1)$ & $ 3.03 ( 0)$ & $ 0.00 $ & $ -1.04 ( 1)$ & $ -1.00 ( 1)$ \\
$ \nu_\Pe \Pe^+ \Pu \bar{\Pd}$ & $ 0.031719 ( 4)$ & $ 3.02 ( 0)$ & $ 3.76 ( 1)$ & $ -1.04 ( 2)$ & $ -1.00 ( 1)$ \\
$ \Pu \bar{\Pd} \Ps \bar{\Pc}$ & $ 0.09847 ( 2)$ & $ 2.97 ( 0)$ & $ 7.52 ( 1)$ & $ -1.04 ( 2)$ & $ -1.00 ( 1)$ \\
$\nu_\Pe \Pe^+ \Pe^- \bar{\nu}_\Pe$ & $ 0.010197 ( 1)$ & $ 3.12 ( 0)$ & $ 0.00 $ & $ -1.04 ( 1)$ & $ -1.00 ( 1)$ \\
$ \Pu \bar{\Pd} \Pd \bar{\Pu}$ & $ 0.10048 ( 2)$ & $ 2.85 ( 0)$ & $ 7.35 ( 2)$ & $ -1.06 ( 3)$ & $ -1.00 ( 1)$ \\
$\nu_\Pe \bar{\nu}_\Pe \nu_\mu \bar{\nu}_\mu$ & $ 0.000949 ( 0)$ & $ 3.01 ( 0)$ & $ 0.00 $ & $ -1.14 ( 1)$ & $ -1.00 ( 1)$ \\
$ \Pe^- \Pe^+ \mu^- \mu^+$ & $ 0.000239 ( 0)$ & $ 1.30 ( 1)$ & $ 0.00 $ & $ -1.13 ( 2)$ & $ -1.00 ( 1)$ \\
$\nu_\Pe \bar{\nu}_\Pe \mu^-\mu^+$ & $ 0.000477 ( 0)$ & $ 2.45 ( 1)$ & $ 0.00 $ & $ -1.13 ( 2)$ & $ -1.00 ( 1)$ \\
$\nu_\Pe \bar{\nu}_\Pe \nu_\Pe \bar{\nu}_\Pe$ & $ 0.000569 ( 0)$ & $ 2.90 ( 0)$ & $ 0.00 $ & $ -1.14 ( 2)$ & $ -1.00 ( 1)$ \\
$ \Pe^- \Pe^+ \Pe^-\Pe^+$ & $ 0.000132 ( 0)$ & $ 1.12 ( 1)$ & $ 0.00 $ & $ -1.12 ( 2)$ & $ -1.00 ( 1)$ \\
$\nu_\Pe \bar{\nu}_\Pe \Pu \bar{\Pu}$ & $ 0.001679 ( 0)$ & $ 0.60 ( 1)$ & $ 3.76 ( 1)$ & $ -1.12 ( 2)$ & $ -1.00 ( 1)$ \\
$\nu_\Pe \bar{\nu}_\Pe \Pd \bar{\Pd}$ & $ 0.002177 ( 1%0
 )$ & $ 1.69 ( 0)$ & $ 3.76 ( 1)$ & $ -1.12 ( 2)$ & $ -1.00 ( 1)$ \\
$ \Pe^-\Pe^+ \Pu \bar{\Pu}$ & $ 0.000845 ( 0)$ & $ 0.11 ( 1)$ & $ 3.76 ( 1)$ & $ -1.12 ( 2)$ & $ -1.00 ( 1)$ \\
$ \Pe^- \Pe^+ \Pd \bar{\Pd}$ & $ 0.001088 ( 0)$ & $ 0.47 ( 1)$ & $ 3.76 ( 1)$ & $ -1.12 ( 2)$ & $ -1.00 ( 1)$ \\
$ \Pu \bar{\Pu} \Pc \bar{\Pc}$ & $ 0.002971 ( 0)$ & $ -1.80 ( 1)$ & $ 7.51 ( 1)$ & $ -1.11 ( 2)$ & $ -1.00 ( 1)$ \\
$ \Pd \bar{\Pd} \Pd \bar{\Pd}$ & $ 0.002556 ( 1)$ & $ -0.38 ( 0)$ & $ 4.38 ( 2)$ & $ -1.21 ( 3)$ & $ -1.00 ( 1)$ \\
$ \Pd \bar{\Pd} \Ps \bar{\Ps}$ & $ 0.004956 ( 1)$ & $ -0.36 ( 0)$ & $ 7.51 ( 1)$ & $ -1.12 ( 2)$ & $ -1.00 ( 1)$ \\
$ \Pu \bar{\Pu} \Ps \bar{\Ps}$ & $ 0.003852 ( 1)$ & $ -0.66 ( 1)$ & $ 7.51 ( 1)$ & $ -1.11 ( 2)$ & $ -1.00 ( 1)$ \\
$ \Pu \bar{\Pu} \Pu \bar{\Pu}$ & $ 0.001506 ( 0)$ & $ -1.92 ( 1)$ & $ 4.06 ( 3)$ & $ -1.24 ( 4)$ & $ -1.00 ( 1)$ \\
  \hline
  \end{tabular}  
   \caption{Partial widths for benchmark scenario Aa in the \MSbar{}$(\lambda_3)$ renormalization scheme.}
 \label{tab:partialwidthsAa}
\vspace*{1em}
% \end{table}
%
% \begin{table}
   \centering
   \renewcommand{\arraystretch}{.94}
 \begin{tabular}{|c|ccccc|}\hline\vphantom{$\Big|$}
  Final state& $\Gamma^{\Ph{\to}4f}_\mr{NLO}$ [MeV] & $\delta_\mr{EW}$ [\%]& $\delta_\mr{QCD}$ [\%] & $\Delta_\mr{SM}^\mr{NLO}$ [\%]&$\Delta_\mr{SM}^\mr{LO}$ [\%]\\\hline
inclusive $\Ph {\to} 4f$ & $ 0.95980 ( 7)$ & $ 1.87 ( 0)$ & $ 4.97 ( 1)$ & $ -1.82 ( 1)$ & $ -1.00 ( 1)$ \\
ZZ & $ 0.105464 ( 5)$ & $ -0.34 ( 0)$ & $ 4.90 ( 0)$ & $ -1.75 ( 1)$ & $ -1.00 ( 0)$ \\
WW & $ 0.85938 ( 8)$ & $ 2.14 ( 0)$ & $ 5.01 ( 1)$ & $ -1.83 ( 1)$ & $ -1.00 ( 1)$ \\
WW/ZZ int. & $ -0.00504 ( 5)$ & $ 0.5 ( 1)$ & $ 10.7 ( 8)$ & $ -2 ( 1)$ & $ -1 ( 1)$ \\
$ \nu_\Pe \Pe^+ \mu^- \bar{\nu}_\mu$ & $ 0.010116 ( 1)$ & $ 2.17 ( 1%0
)$ & $ 0.00 $ & $ -1.87 ( 1)$ & $ -1.00 ( 1)$ \\
$ \nu_\Pe \Pe^+ \Pu \bar{\Pd}$ & $ 0.031463 ( 4)$ & $ 2.16 ( 0)$ & $ 3.76 ( 1)$ & $ -1.84 ( 2)$ & $ -1.00 ( 1)$ \\
$ \Pu \bar{\Pd} \Ps \bar{\Pc}$ & $ 0.09770 ( 2)$ & $ 2.11 ( 0)$ & $ 7.52 ( 1)$ & $ -1.81 ( 2)$ & $ -1.00 ( 1)$ \\
$\nu_\Pe \Pe^+ \Pe^- \bar{\nu}_\Pe$ & $ 0.010112 ( 1)$ & $ 2.27 ( 1%0
)$ & $ 0.00 $ & $ -1.87 ( 1)$ & $ -1.00 ( 1)$ \\
$ \Pu \bar{\Pd} \Pd \bar{\Pu}$ & $ 0.09972 ( 2)$ & $ 1.99 ( 0)$ & $ 7.38 ( 2)$ & $ -1.80 ( 2)$ & $ -1.00 ( 1)$ \\
$\nu_\Pe \bar{\nu}_\Pe \nu_\mu \bar{\nu}_\mu$ & $ 0.000943 ( 0)$ & $ 2.34 ( 0)$ & $ 0.00 $ & $ -1.78 ( 1)$ & $ -1.00 ( 1)$ \\
$ \Pe^- \Pe^+ \mu^- \mu^+$ & $ 0.000237 ( 0)$ & $ 0.62 ( 1)$ & $ 0.00 $ & $ -1.79 ( 2)$ & $ -1.00 ( 1)$ \\
$\nu_\Pe \bar{\nu}_\Pe \mu^-\mu^+$ & $ 0.000474 ( 0)$ & $ 1.78 ( 1)$ & $ 0.00 $ & $ -1.78 ( 2)$ & $ -1.00 ( 1)$ \\
$\nu_\Pe \bar{\nu}_\Pe \nu_\Pe \bar{\nu}_\Pe$ & $ 0.000565 ( 0)$ & $ 2.23 ( 0)$ & $ 0.00 $ & $ -1.79 ( 2)$ & $ -1.00 ( 1)$ \\
$ \Pe^- \Pe^+ \Pe^-\Pe^+$ & $ 0.000131 ( 0)$ & $ 0.45 ( 1)$ & $ 0.00 $ & $ -1.78 ( 2)$ & $ -1.00 ( 1)$ \\
$\nu_\Pe \bar{\nu}_\Pe \Pu \bar{\Pu}$ & $ 0.001668 ( 0)$ & $ -0.08 ( 1)$ & $ 3.76 ( 1)$ & $ -1.76 ( 2)$ & $ -1.00 ( 1)$ \\
$\nu_\Pe \bar{\nu}_\Pe \Pd \bar{\Pd}$ & $ 0.002163 ( 0)$ & $ 1.02 ( 0)$ & $ 3.76 ( 1)$ & $ -1.76 ( 2)$ & $ -1.00 ( 1)$ \\
$ \Pe^-\Pe^+ \Pu \bar{\Pu}$ & $ 0.000840 ( 0)$ & $ -0.57 ( 1)$ & $ 3.76 ( 1)$ & $ -1.77 ( 2)$ & $ -1.00 ( 1)$ \\
$ \Pe^- \Pe^+ \Pd \bar{\Pd}$ & $ 0.001081 ( 0)$ & $ -0.21 ( 1)$ & $ 3.76 ( 1)$ & $ -1.76 ( 2)$ & $ -1.00 ( 1)$ \\
$ \Pu \bar{\Pu} \Pc \bar{\Pc}$ & $ 0.002952 ( 0)$ & $ -2.48 ( 1)$ & $ 7.51 ( 1)$ & $ -1.75 ( 2)$ & $ -1.00 ( 1)$ \\
$ \Pd \bar{\Pd} \Pd \bar{\Pd}$ & $ 0.002545 ( 1)$ & $ -1.06 ( 0)$ & $ 4.57 ( 2)$ & $ -1.67 ( 3)$ & $ -1.00 ( 1)$ \\
$ \Pd \bar{\Pd} \Ps \bar{\Ps}$ & $ 0.004925 ( 1)$ & $ -1.04 ( 0)$ & $ 7.51 ( 1)$ & $ -1.74 ( 2)$ & $ -1.00 ( 1)$ \\
$ \Pu \bar{\Pu} \Ps \bar{\Ps}$ & $ 0.003828 ( 1)$ & $ -1.35 ( 1)$ & $ 7.51 ( 1)$ & $ -1.74 ( 2)$ & $ -1.00 ( 1)$ \\
$ \Pu \bar{\Pu} \Pu \bar{\Pu}$ & $ 0.001500 ( 0)$ & $ -2.60 ( 1)$ & $ 4.31 ( 2)$ & $ -1.65 ( 3)$ & $ -1.00 ( 1)$ \\
  \hline
  \end{tabular}  
   \caption{Partial widths for benchmark scenario Ab in the \MSbar{}$(\lambda_3)$ renormalization scheme.}
 \label{tab:partialwidthsAb}
 \end{table}
For other schemes, the numbers  differ slightly, but show the same qualitative pattern, so that we do not show them here. In the tables, we do not only state the full NLO QCD+EW partial widths, but also the relative EW and QCD corrections. 
The qualitative picture is similar for the two benchmark scenarios. 
The WW contribution originating from charged-current final states yields the largest contribution, while the ZZ contribution is minor and the interference term yields a small negative contribution. 
The EW corrections to the WW-mediated final states are uniformly about $2{-}3$\%, which determine the EW corrections to the
partial $\Ph{\to}4f$ width. 
The EW corrections to the neutral-current final states strongly depend on the fermion flavour and range between $\pm3\%$.
The QCD corrections are essentially the strong corrections to $\PW/\PZ \to q\bar{q}$ and therefore amount to 
$\alpha_\mr{s}/\pi$ for each pair of quarks in the final state. 
The $\Pu \bar{\Pu}\Pu \bar{\Pu}$ and $\Pd \bar{\Pd}\Pd \bar{\Pd}$ final states, where interference contributions between
two different ZZ~channels exist, are somewhat exceptional with QCD corrections of only about $4\%$.
The deviations $\Delta_\mr{SM}$ from the SM expectation are shown at NLO and LO in the last two columns.
The LO deviation is due to the suppression factor $s_{\beta-\alpha}$ of the $\Ph VV$ coupling w.r.t.\ the SM
and therefore identical with $s_{\beta-\alpha}^2-1=-c_{\beta-\alpha}^2=-10^{-2}$ for all final states, since $c_{\beta-\alpha}=\pm0.1$. 
It should be noted that the indicated errors are integration errors, and the presented LO results are thus also a consistency check.
At NLO the deviation is slightly larger, though still within only 1.3\% (2\%) for the Aa (Ab) benchmark scenario. 
The deviations from the SM are quite uniform, i.e.\ insensitive to the final state, so that they are described by
the partial $\Ph{\to}4f$ width well within a few per mille.

\subsubsection{Differential distributions}
\label{sec:diffdistr}

The program \textsc{Prophecy4f} provides invariant-mass and angular differential distributions for the $\Ph {\to} 4f$ decays.
The differential decay widths may serve as a
window to observe BSM effects as the shape of distributions might be distorted significantly by new 
coupling structures. This might occur even if the partial widths do not change significantly, 
and therefore, the differential distributions of leptonic and semi-leptonic final states (see \refse{sec:Prophecy4f}) 
are important observables. 
In the following, we study them for both charged- and neutral-current processes, e.g.\ the fully leptonic final states $ \Pe^- \Pe^+ \mu^- \mu^+$ (nc), $ \nu_\Pe \Pe^+ \mu^- \bar{\nu}_\mu$ (cc), and the semi-leptonic final states  $ \Pe^-\Pe^+ q \bar{q}$ (nc), $  \nu_\Pe \Pe^+ \Pd \bar{\Pu}$ (cc). 
Most likely, differential distributions for fully hadronic final states are not experimentally accessible. 
A detailed discussion of the SM distributions at NLO, including issues of final-state radiation
(such as photon recombination),
can be found for the fully leptonic final states in Refs.~\cite{Bredenstein:2006rh,Bredenstein:2006nk,Boselli:2015aha} 
and for semi-leptonic final states in Ref.~\cite{Bredenstein:2006ha}. 
In our study we emphasize the differences between the SM and the THDM results, 
while the features of photonic (and gluonic)
corrections in the THDM and the SM are identical.
The distributions discussed in the following are calculated using the \MSbar{}$(\lambda_3)$ renormalization scheme; 
the other renormalization schemes yield similar results.

\paragraph{Leptonic final states:}
\begin{figure}
  \centering
  \subfigure[]{
\label{fig:plot_mumuee_inv12}
\includegraphics[scale=0.8]{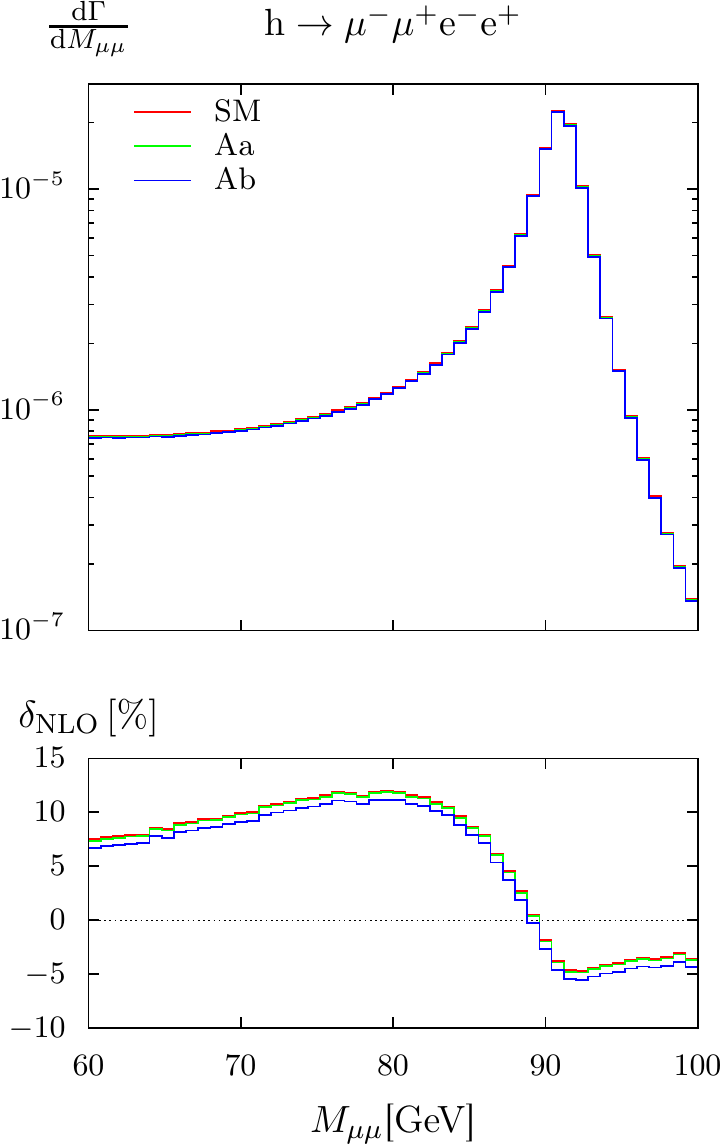}
}
\hspace{12pt}
\subfigure[]{
\label{fig:plot_mumuee_phi}
\includegraphics[scale=0.8]{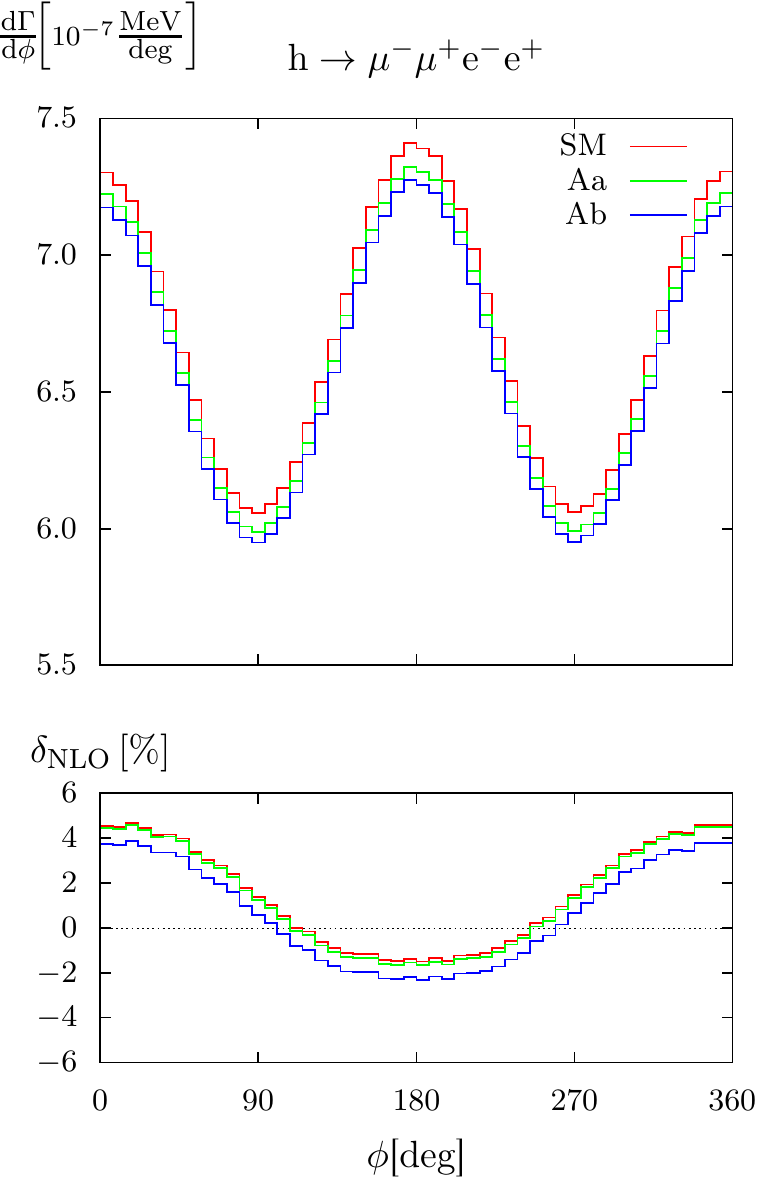}
}
\\[-.7em]
  \caption{Invariant-mass (a) and angular distributions (b) of the leptonic neutral-current decay $\mr{h} \to \mu^- \mu^+ \Pe^- \Pe^+$ for the SM and the THDM benchmark scenarios Aa and Ab. The relative NLO corrections to the distributions are plotted in the lower panels.}
\label{fig:distr_mumuee}
\vspace*{1em}
%\end{figure}
%
%\begin{figure}
  \centering
  \subfigure[]{\hspace{12pt}
\label{fig:plot_vmuev_inv12}
\includegraphics[scale=0.8]{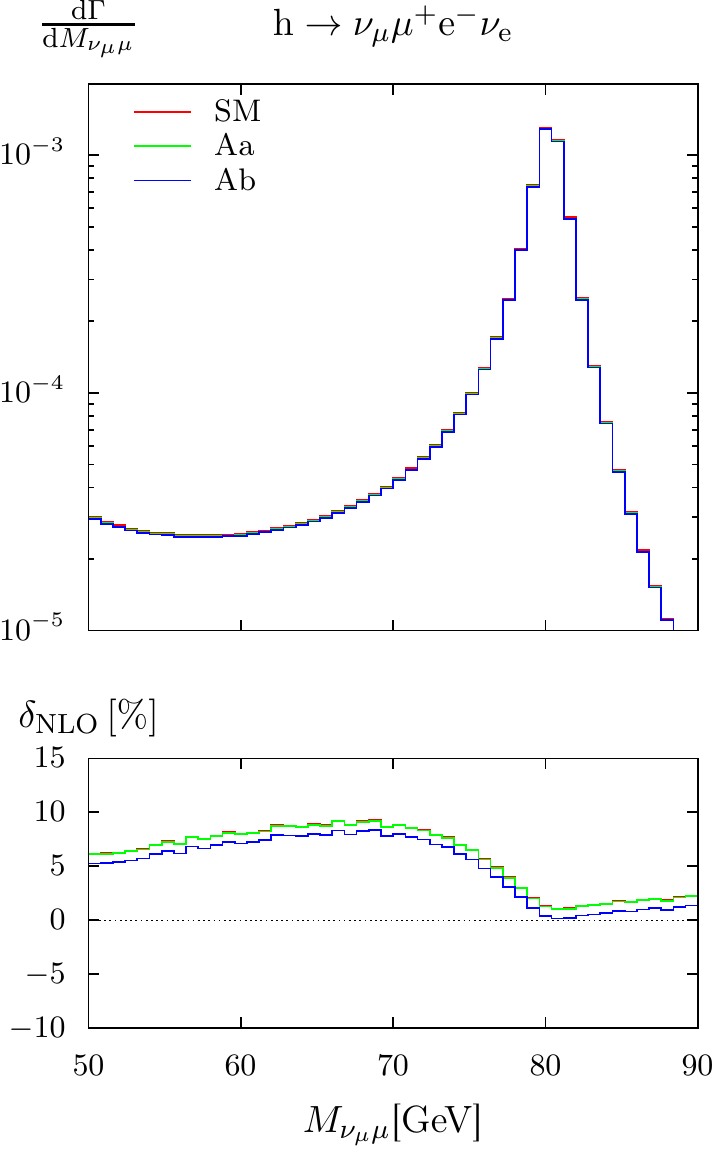}
}
\hspace{12pt}
\subfigure[]{
\label{fig:plot_vmuev_phi}
\includegraphics[scale=0.8]{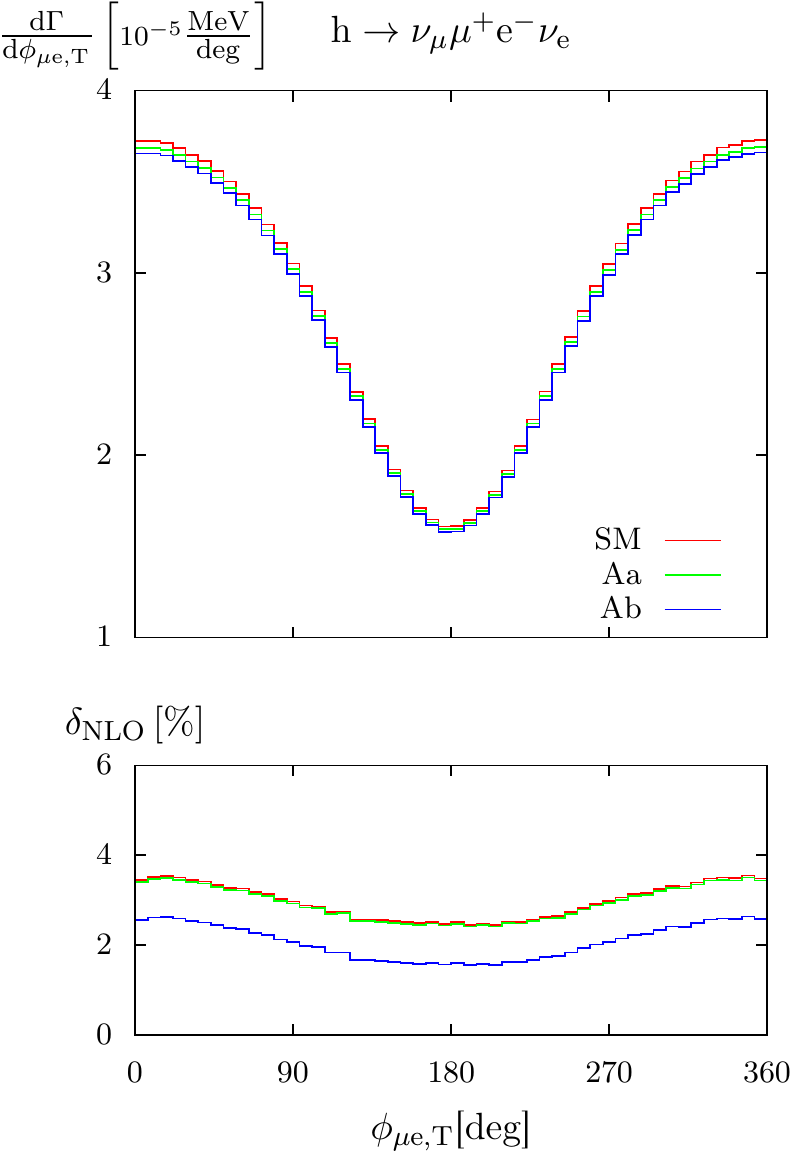}
}
\\[-.7em]
\caption{As in \reffi{fig:distr_mumuee}, but for the leptonic charged-current decay $\mr{h} \to \nu_\mu \mu^+ \Pe^- \bar\nu_\Pe$.}
%  \caption{Invariant-mass (a) and angular distributions (b) of the leptonic charged-current decay $\mr{h} \to \nu_\mu \mu^+ \Pe^- \nu_\Pe$ for the SM and the THDM benchmark scenarios Aa and Ab. The relative NLO corrections to the distributions are plotted in the lower panels.}
\label{fig:distr_vmuev}
\end{figure}
We begin with the leptonic final state $ \Pe^- \Pe^+ \mu^- \mu^+$ which is a decay mediated by Z bosons. 
The invariant mass $M_{f_a \bar{f}_b}$ of a fermion--anti-fermion pair is defined by 
\begin{align}
 M^2_{f_a \bar{f}_b}= (k_a+k_b)^2
\end{align}
with the momentum $k_a$ of the fermion $f_a$ and $k_b$ of the anti-fermion $\bar{f}_b$,
where the photon momentum is already added to the fermion momentum in case of recombination. 
The NLO invariant-mass distributions of the muon pair 
are displayed in the first panel of Fig.~\ref{fig:plot_mumuee_inv12} for the SM and the THDM in scenarios Aa and Ab
and show the Z-boson resonance.
The relative corrections normalized to the LO are illustrated in the second panel and
exhibit the well-known effects of final-state radiation near the Z~resonance:
Photons radiated off a final-state lepton lower the invariant mass of the lepton pair 
and lead to positive corrections---the ``radiative tail''---for invariant masses below the Z-boson peak
and negative corrections above. 
These corrections would contain a logarithm of the form $\alpha\ln(m_\mu/\Mh)$ from collinear photon emission off muons
if no photon recombination was applied. 
However, photon recombination mitigates this large effect by 
shifting events back to larger invariant masses for collinear emission 
and leads to the necessary level of inclusiveness required by the 
Kinoshita--Lee--Nauenberg theorem~\cite{Kinoshita1962,Lee1964} to remove the collinear singularity.
In case of photon recombination, the $\mu^-\mu^+$ and $\Pem\Pep$ 
invariant-mass distributions are, thus, 
identical.
Yet, non-collinear photons, which are not recombined, still lead to a sizable net effect which is observed in the relative corrections.
The shapes of the invariant-mass distributions in SM and THDM are practically identical, i.e.\
the impact of new Lorentz structures in the NLO THDM diagrams is negligible. 
The relative difference between the SM and the THDM distributions is just given by the difference observed already
in the integrated $\Ph{\to}\Pe^- \Pe^+ \mu^- \mu^+$ decay width, i.e.\
$-1.15$\% for scenario~Aa and $-1.81$\% for Ab.
We recall that those differences were traced back to the impact of $\PH\Ph$-mixing effects in \refse{sec:cbascanA},
which are 
independent from the decay kinematics, and thus conclude that those mixing effects are the dominant
higher-order effects visible in the distributions as well.

The differential decay width with respect to the angle $\phi$ between the $\mu^- \mu^+$
 and $\Pe^- \Pe^+$ decay planes 
is defined by
\begin{align}
 \cos \phi=\frac{\left(\left(\mathbf{k}_1+\mathbf{k}_2\right) \times \mathbf{k}_1\right)\cdot((\mathbf{k}_1+\mathbf{k}_2) \times \mathbf{k}_3) }{|(\mathbf{k}_1+\mathbf{k}_2) \times \mathbf{k}_1||(\mathbf{k}_1+\mathbf{k}_2) \times \mathbf{k}_3|}, \label{eq:angdistr1}
\end{align}
with the sign convention
\begin{align}
 \mr{sgn} (\sin \phi)= \mr{sgn}\{\left(\mathbf{k}_1+\mathbf{k}_2\right)
\cdot\left[\left(\left(\mathbf{k}_1+\mathbf{k}_2\right) \times \mathbf{k}_1\right)\times \left(\left(\mathbf{k}_1+\mathbf{k}_2\right) \times \mathbf{k}_3\right)\right]\}
\end{align}
where $\mathbf{k}_1$, $\mathbf{k}_2$, and $\mathbf{k}_3$ are the momenta 
of the muon, the anti-muon and the electron, respectively.
The corresponding distribution is shown in the upper panel of Fig.~\ref{fig:plot_mumuee_phi}. 
One observes a $\cos {(2\phi)}$ pattern in the shape of the distribution, which can be used to 
set bounds on non-standard couplings of the Higgs boson to the EW gauge bosons
(see Refs.~\cite{Nelson:1986ki,Soni:1993jc,Chang:1993jy,Skjold:1993jd,Buszello:2002uu,Arens:1994wd,Choi:2002jk,Boselli:2017pef}). 
Note that the oscillation pattern in the distribution of a pseudo-scalar Higgs boson would have a different sign. 
We again observe that the SM shape is not distorted by THDM effects and that the difference between SM and THDM prediction
just resembles the difference in the integrated widths.

We have also considered the invariant-mass and angular distributions of the $\Pe^- \Pe^+ \Pe^- \Pe^+$ final state (not shown), 
for which interference terms between different ZZ channels appear. 
There, the assignment of the lepton pairs to intermediate Z~bosons is not unique; 
usually the electron and positron with an invariant mass closest to the Z-boson mass is combined to a pair. 
Again we find that the relative difference $\Delta_\mr{SM}$ between THDM and the SM is practically constant
over the phase space and given by its values for the integrated width.

For the W-boson-mediated $ \nu_\Pe \Pe^+ \mu^- \bar{\nu}_\mu$ final state, the respective distributions are shown 
in Fig.~\ref{fig:distr_vmuev}. 
The invariant-mass distribution of $M_{\nu_\mu {\mu}}$ is not experimentally accessible, 
but shown for theoretical interest. 
The plot shows the W~resonance around $M_{\nu_\mu \mu}\approx \MW $ with
the radiative corrections caused by photon radiation as discussed above.
As already observed in the neutral-current final state, 
there is no significant shape distortion in the THDM w.r.t.\ the SM prediction.
As the neutrinos cannot be detected, neither the Higgs nor the W boson can be fully reconstructed. 
However, projecting all lepton momenta into a fixed plane mimics the experimental
situation at the LHC in the centre-of-mass frame of the Higgs boson in the plane transverse to the proton beams, 
where the sum of the neutrino momenta is measurable as missing momentum.
We, thus, analyze the transverse angle $\phi_{\mu\Pe,\rT}$ between the two charged leptons~\cite{Bredenstein:2006rh}, defined by
\begin{align}
 \cos \phi_{\mu\Pe,\rT}&=\frac{\mathbf{k}_{\mu,\rT} \cdot \mathbf{k}_{\Pe,\rT}}{|\mathbf{k}_{\mu,\rT}| | \mathbf{k}_{\Pe,\rT}|},
& \mr{sgn} (\sin \phi_\rT)&=\mr{sgn}\{\mathbf{e}_z \cdot (\mathbf{k}_{\mu,\rT} \times \mathbf{k}_{\Pe,\rT}) \}, 
\label{eq:angdistr3}
\end{align}
where $\mathbf{k}_{i,\rT}$ are the parts of the full lepton momenta $\mathbf{k}_{i}$ orthogonal to the
fixed unit vector $\mathbf{e}_z$ representing the beam direction of the Higgs-boson production process. 
The distribution in $\phi_{\mu\Pe,\rT}$ is shown in the first panel of Fig.~\ref{fig:plot_vmuev_phi}. 
As expected, no shape distortion is seen w.r.t.\ the SM prediction, only 
the constant relative deviation which is 
identical to the deviation in the partial width of $-1.04$\% (Aa) and $-1.87$\% (Ab). 
Other fully leptonic final states show similar patterns, so that their distributions are not separately shown here.

\paragraph{Semi-leptonic final states:}
\begin{figure}
  \centering
  \subfigure[]{
\label{fig:plot_qqee_inv12}
\includegraphics[scale=0.8]{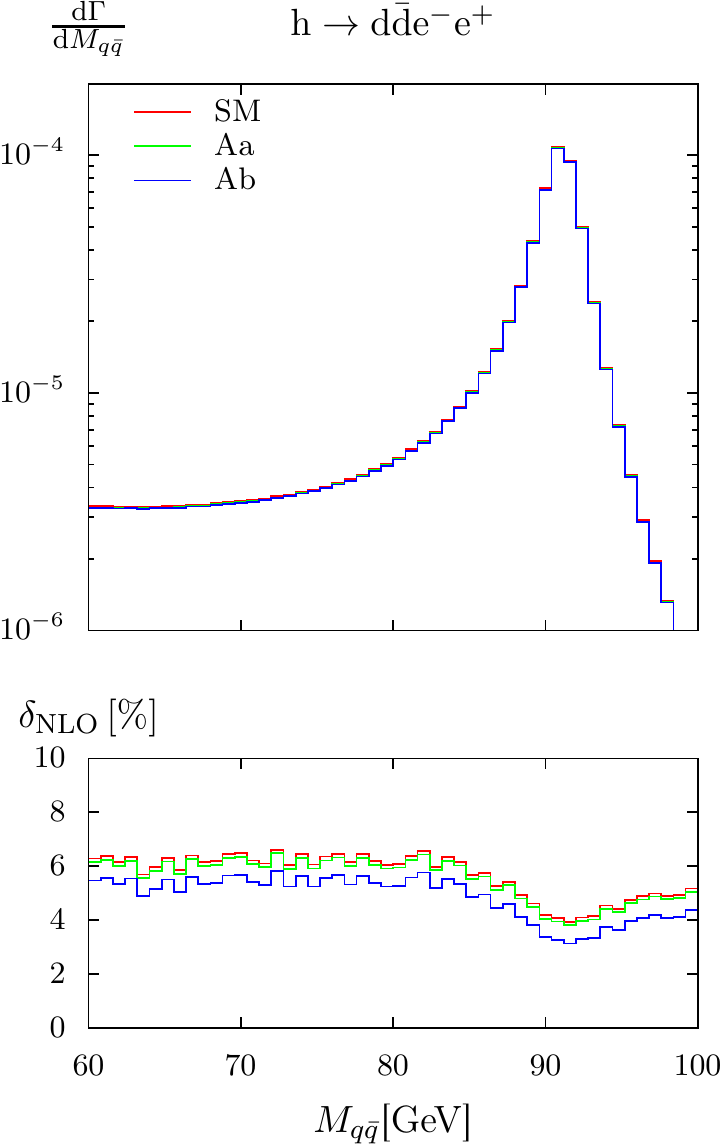}
}
\hspace{12pt}
\subfigure[]{
\label{fig:plot_qqee_phi}
\includegraphics[scale=0.8]{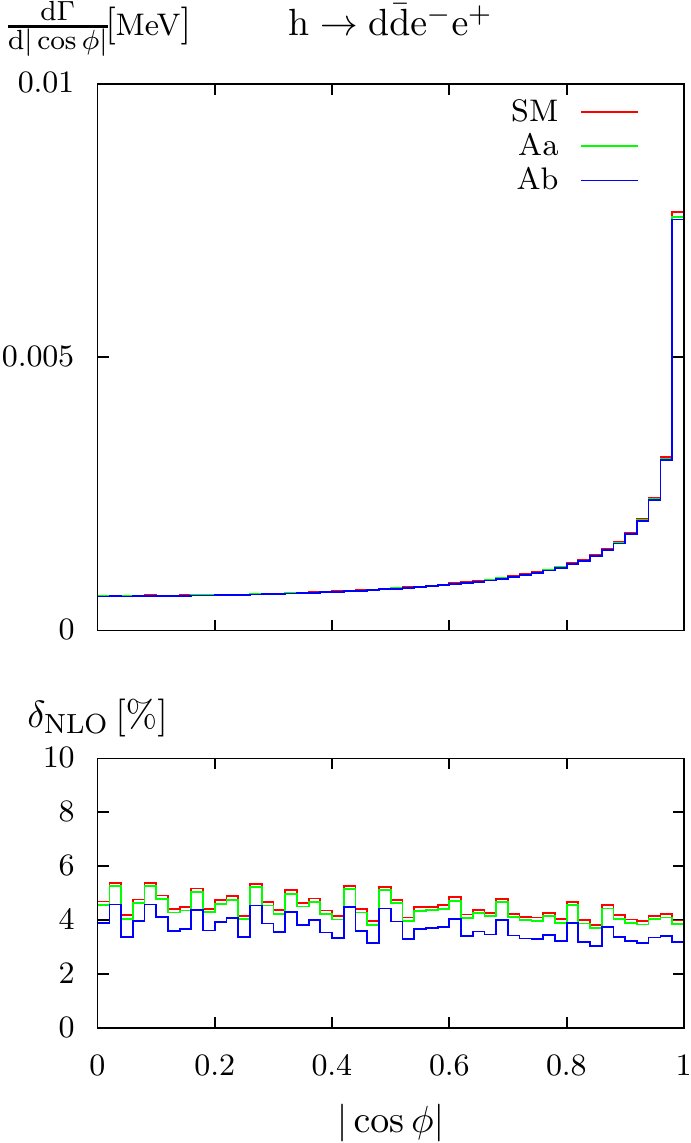}
}
\\[-.5em]
  \caption{Invariant-mass (a) and angular distributions (b) of the charged-current semi-leptonic decay $\mr{h} \to \Pd \bar{\Pd} \Pe^- \Pe^+$ for the SM and the THDM benchmark scenarios Aa and Ab. The relative NLO EW+QCD corrections to the distributions are plotted in the lower panels.}
\label{fig:distr_qqee}
\vspace*{1em}
%\end{figure}
%
%\begin{figure}
  \centering
  \subfigure[]{
\label{fig:plot_veqq_inv12}
\includegraphics[scale=0.8]{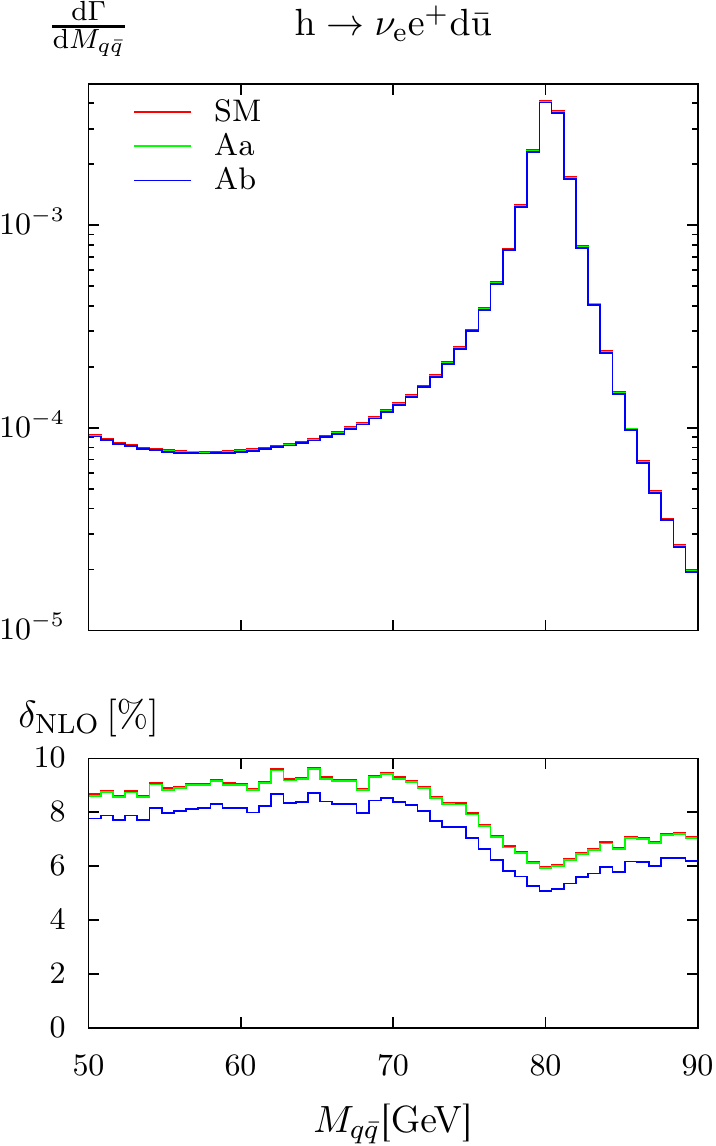}
}
\hspace{12pt}
\subfigure[]{
\label{fig:plot_veqq_phi}
\includegraphics[scale=0.8]{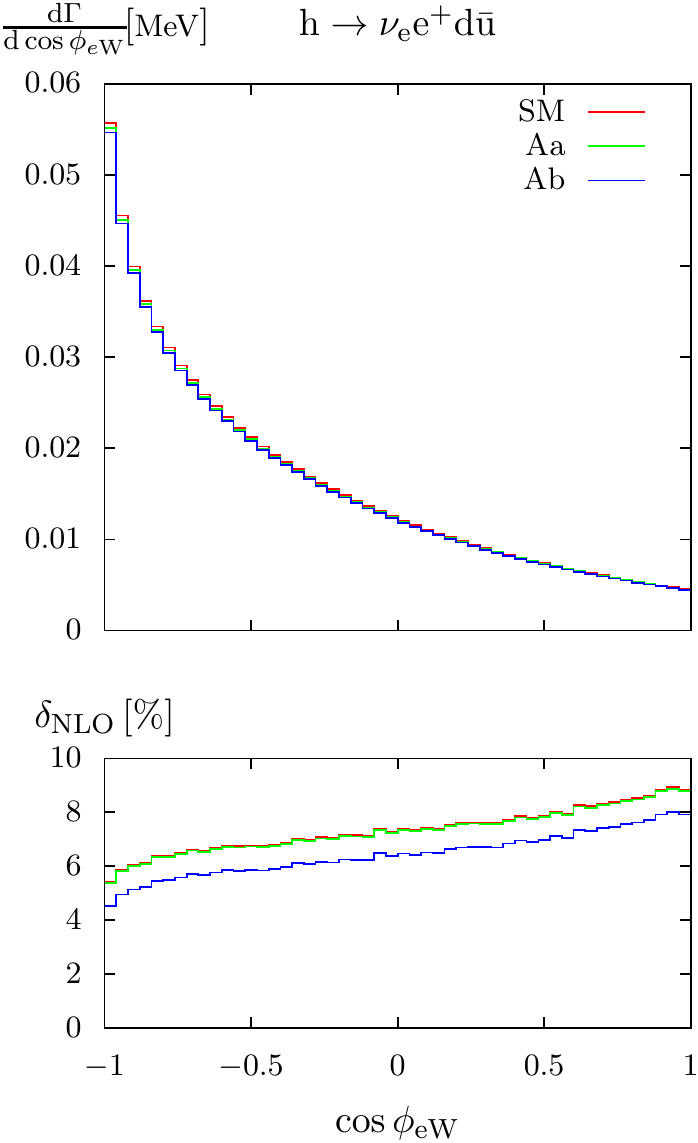}
}
\\[-.5em]
\caption{As for \reffi{fig:distr_qqee}, but for the charged-current semi-leptonic decay $\mr{h} \to   \nu_\Pe \Pe^+ \Pd \bar{\Pu}$.}
%\caption{Invariant-mass (a) and angular distributions (b) of the charged-current semi-leptonic decay $\mr{h} \to   \nu_\Pe \Pe^+ \Pd \bar{\Pu}$ for the SM and the THDM benchmark scenarios Aa and Ab. The relative NLO corrections to the distributions are plotted in the lower panels.}
\label{fig:distr_veqq}
\end{figure}
The invariant-mass distribution 
of the hadronic system of the neutral-current final state $\Pd \bar{\Pd} \Pe^-\Pe^+ $ 
is displayed in Fig.~\ref{fig:distr_qqee} (l.h.s), together with the corresponding NLO EW+QCD corrections.
In case of gluon radiation, the invariant mass is built from the whole hadronic $q\bar q\Pg$ system to obtain an
IR-safe observable.
The distribution and the corrections 
show similar characteristics to the ones of the corresponding leptonic final state: 
Photon radiation leads to a radiative tail, 
but SM and THDM distributions do not show any visible shape difference.
Note that the effect of the photon radiation is less pronounced compared to the leptonic final state, 
as the quark charge factors are smaller than for leptons. 
There is no radiative tail from gluon radiation, because all gluons are recombined with the quark pair, 
so that only a flat QCD correction remains~\cite{Bredenstein:2006ha}. 

In order to analyze angular distributions, we identify the quarks and antiquarks with jets for events without
gluon radiation. In case of gluon radiation, we always combine the two QCD partons with the smallest invariant mass
to a single jet, so that we again obtain an event with two jets.
As the jets cannot be distinguished, 
any observable must be invariant under the permutation of the two jets.
For this reason, the angle $\phi$ between the two Z-boson
decay planes can only be reconstructed up to the sign of $\cos\phi$, so that we define~\cite{Bredenstein:2006ha}
\begin{align}
|\cos \phi| = \left|
\frac{\left(\left(\mathbf{k}_{\mr{jet}_1}+\mathbf{k}_{\mr{jet}_2}\right) 
\times \mathbf{k}_{\Pem}\right)\left(\mathbf{k}_{\mr{jet}_1}
\times\mathbf{k}_{\mr{jet}_2}\right) }
{|\left(\mathbf{k}_{\mr{jet}_1}+\mathbf{k}_{\mr{jet}_2}\right) \times \mathbf{k}_{\Pem}||\mathbf{k}_{\mr{jet}_1}\times\mathbf{k}_{\mr{jet}_2}|} \right|. 
\label{eq:angdistr2}
\end{align}
The corresponding distribution, which is depicted on the r.h.s.\ of Fig.~\ref{fig:distr_qqee}, 
looks rather different from the leptonic case, since $|\cos\phi|$ instead of $\phi$ is used in the binning.
Again, the major finding is the fact that the shape of the distribution does not change in the transition from the
SM to the THDM. Only the flat offsets of $-1.12$\% (Aa) and $-1.76$\% (Ab) already encountered in the
partial width are visible.

The invariant-mass distribution of the hadronic system of the semi-leptonic W-bo\-son-me\-dia\-ted final state 
$\nu_\Pe \Pe^+ \Pd \bar{\Pu}$ is pictured in Fig.~\ref{fig:distr_veqq} (l.h.s) and shows the same characteristics
as the one of the neutral-current final state considered above: 
a moderate radiative tail from photon radiation, flat QCD corrections (not explicitly shown), 
and no shape difference between SM and THDM predictions.
The distribution in the angle between the electron and the hadronically decaying W boson, $\phi_{\Pe\PW}$, 
in the rest frame of the Higgs boson is shown in Fig.~\ref{fig:distr_veqq} (r.h.s). 
The electron is predominantly produced in the direction opposite to the W~boson, 
and the EW corrections slightly distort the shape of the distribution. 
The difference between SM and THDM is described well by the deviation observed for the partial width of 
$-1.05$\% for the Aa and $-1.85$\% for the Ab scenario.

To summarize, the effects of the THDM on the shapes of distributions are negligible, 
only offsets in the normalization are observed.
Thus, distributions for the Higgs decay into four massless fermions 
are not helpful in the search for deviations from the SM induced by effects of the THDM.

\subsection{High-mass scenario B1}
\label{sec:HighMassB1}

The high-mass scenario is divided into two branches which are valid for positive or negative $c_{\beta-\alpha}$ 
and have different $\tan{\beta}$. In this section we cover scenario B1 with positive values of $c_{\beta-\alpha}$,
scenario B2 with negative values is discussed in the subsequent section. 
The perturbativity measure increases with rising $\MH$, as can be seen in Fig.~\ref{fig:perturbativityconstr}, 
restricting the range in $c_{\beta-\alpha}$ and 
potentially affecting the 
stability of the results in the high-mass scenario in a negative way.
Instead of relaxing the situation by moving closer to the alignment limit,
where no problems with too large corrections are expected owing to decoupling,
we delibarately keep this parameter point in order to investigate
the robustness
of the different renormalization schemes by checking the scale uncertainty in
the various schemes and by studying the scheme dependence.
Appendix~\ref{App:scalevar} supplements the discussion of this section by
results with $c_{\beta-\alpha}=0.05$, which are closer to the alignment
limit and show better perturbative stability.
The following 
discussion of the numerical results is structured in the same way as for the previous scenario, beginning with the conversion of the input parameters between
different renormalization schemes.

\subsubsection{Conversion of the input parameters}

\begin{figure}
  \centering
  \subfigure[]{
\label{fig:plot_Umrechnung-von-alpha-B1}
\includegraphics{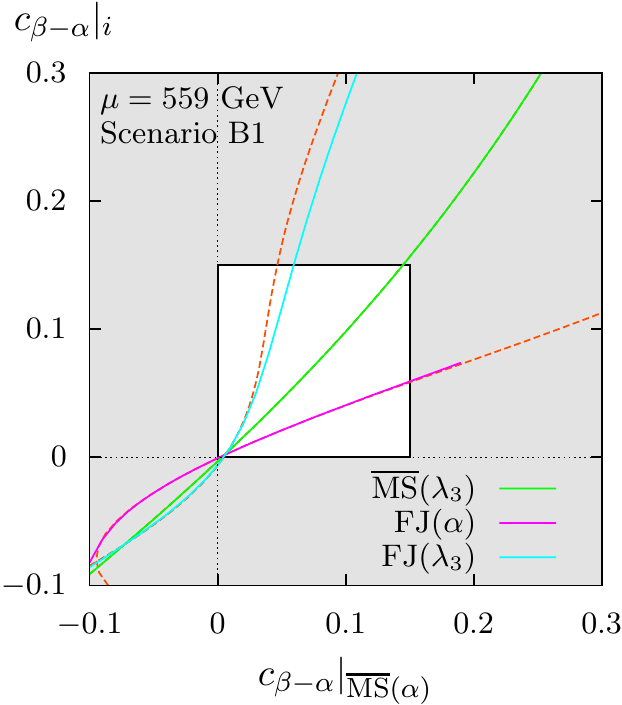}
}
\hspace{15pt}
\subfigure[]{
\label{fig:plot_Umrechnung-nach-alpha-B1}
\includegraphics{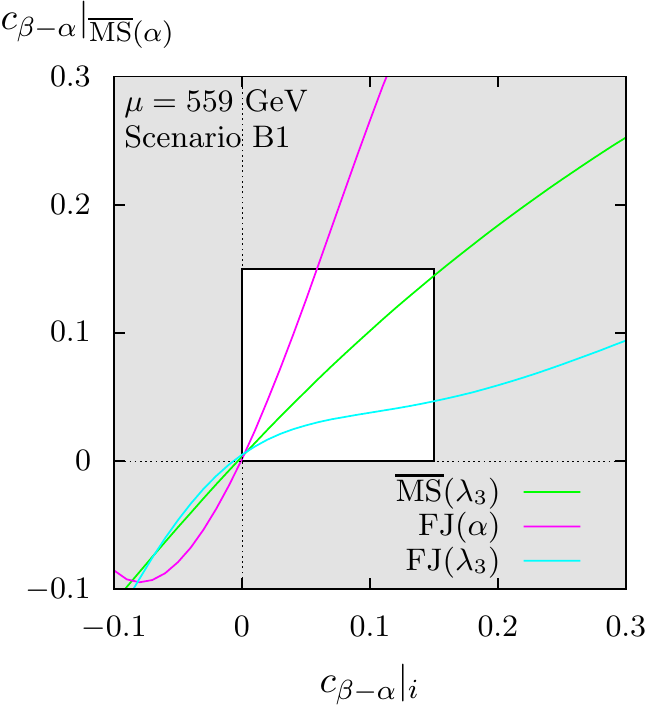}
}
\vspace*{-1em}
  \caption{Panel (a): Conversion of the value of $c_{\beta-\alpha}$ from \MSbar{}$(\alpha)$ to the other schemes for scenario B1 with the colour coding of Fig.~\ref{fig:plotconversionA}. Panel (b) shows the conversion to the \MSbar{}$(\alpha)$ scheme. The solid lines are obtained by solving the implicit equations numerically, the dashed orange lines in (a) correspond to the solution of (b) mirrored at the diagonal. 
The highlighted region shows the phenomenologically most relevant $c_{\beta - \alpha}$ region.}
\label{fig:plotconversionB1}%
\end{figure}

We compute the conversion between the input values in different renormalization schemes for 
$c_{\beta-\alpha}=-0.1$ to $0.3$ 
and use \MSbar{}$(\alpha)$ either as input or as target scheme. 
Using input parameters defined in the FJ$(\alpha)$ scheme leads to particularly large changes in the $c_{\beta-\alpha}$, 
indicating that the NLO terms are large and that the perturbative expansion converges poorly, 
but also for the FJ($\lambda_3$) scheme sizable shifts are observed.
Owing to these large effects, the linearization of the conversion equations suffers from large uncertainties,
and a proper numerical solution is desirable and shown Fig.~\ref{fig:plot_Umrechnung-von-alpha-B1}.
Actually, the two
conversions of Fig.~\ref{fig:plotconversionB1} should be inverse to each other, 
and we perform this consistency check in (a)
by plotting the curves of (b) mirrored at the diagonal with orange dotted lines.
The conversions of the two parameters ($\alpha,\beta$) 
from one scheme to another in fact
is invertible if the implicit equations are solved. In Fig.~\ref{fig:plot_Umrechnung-nach-alpha-B1},
this invertibility is not fully respected, since we consider only a projection 
of the conversion
to the $c_{\beta-\alpha}$ line,
suppressing the changes in $\beta$ 
in the plot, i.e.\ we always take the input values from
Eq.~\refeq{eq:cba_B} in the start scenario of the conversion.

\subsubsection{The running of \boldmath{$c_{\beta-\alpha}$}}

\begin{figure}
  \centering
\includegraphics{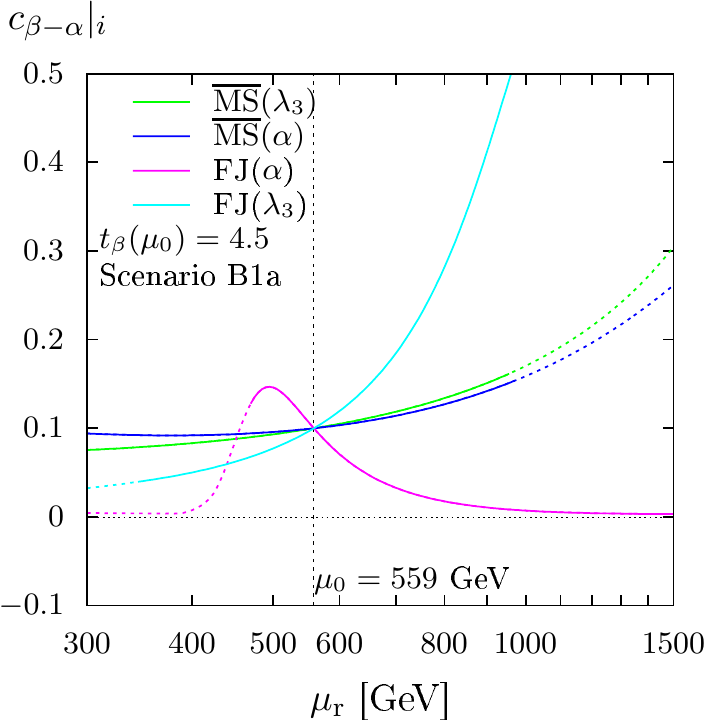}
\vspace*{-.5em}
  \caption{The running of $c_{\beta-\alpha}$ for benchmark scenario~B1a in the \MSbar{}($\alpha$) (blue), \MSbar{}$(\lambda_3)$ (green), FJ($\alpha$) (pink), and FJ($\lambda_3$) (turquoise) schemes. The breakdown of perturbativity ($\lambda_k^{\text{max}}/(4\pi)>1$) is indicated by changing the NLO curve to dotted lines.}
\label{fig:running-B1a}
\end{figure} 
The running of $c_{\beta-\alpha}$ in scenario ~B1a
is investigated in \reffi{fig:running-B1a}
analogously to the low-mass scenario for each renormalization scheme independently, without any conversion. 
The scale dependence of $c_{\beta-\alpha}(\mu_\mr{r})$ with 
$c_{\beta-\alpha}(\mu_0)=0.1$ is computed from $\mu_\mr{r}=300\GeV$ to $1500\GeV$ 
using a Runge--Kutta method. 
We indicated regions where perturbativity is not valid with dotted lines using a slightly different perturbativity measure than in Figs.~\ref{fig:perturbativityconstr} and \ref{fig:perturbativityconstr1000}. Perturbativity is considered to be intact unless the largest of the 
quartic coupling parameters 
$\lambda_k$ of the Higgs potential 
with $k = 1,\dots,5$ becomes too large, $\lambda_k^{\text{max}}/(4\pi) > 1$. Compared to the previously used perturbativity measure, we found that with this measure a slightly larger part of the parameter space fulfills the perturbativity criterion.

In comparison to the low-mass scenario scenario (Fig.~\ref{fig:running_cba0.1-LM}) the scale dependence 
in the FJ($\lambda_3$) scheme increases. For scales above the central scale we obtain large values of 
$c_{\beta-\alpha}$ for which predictions become unreliable. But also the FJ($\alpha$) 
scheme shows a remarkable behaviour as the alignment limit is approached for low as well as for high scales.

\subsubsection{Scale variation of the width}
\label{sec:HighMassB1scalevariation}

We now turn to the calculation of the $\Ph {\to} 4f$ width where we perform a scale variation similarly to the previous section in order to 
investigate the perturbative stability of the results and the validity of
the central scale choice for scenario~B1a. 
The scale is varied from 
$\mu_\mr{r}=300\GeV$ to $1000\GeV$, and the results are shown in Fig.~\ref{fig:plotmuscanB1a} with one plot for each input prescription. 
First, the input values are converted to the target scheme, and afterwards the scale is varied. 
In regions where perturbativity is not valid ($\lambda_k^{\text{max}}/(4\pi)>1$) 
the NLO result is plotted with
dotted lines. The results do not show such a clear and well behaved 
picture as for the low-mass scenario:
\begin{figure}
  \centering
  \subfigure[]{
\label{fig:plot_MUSCAN-B1a-L3MS}
\includegraphics{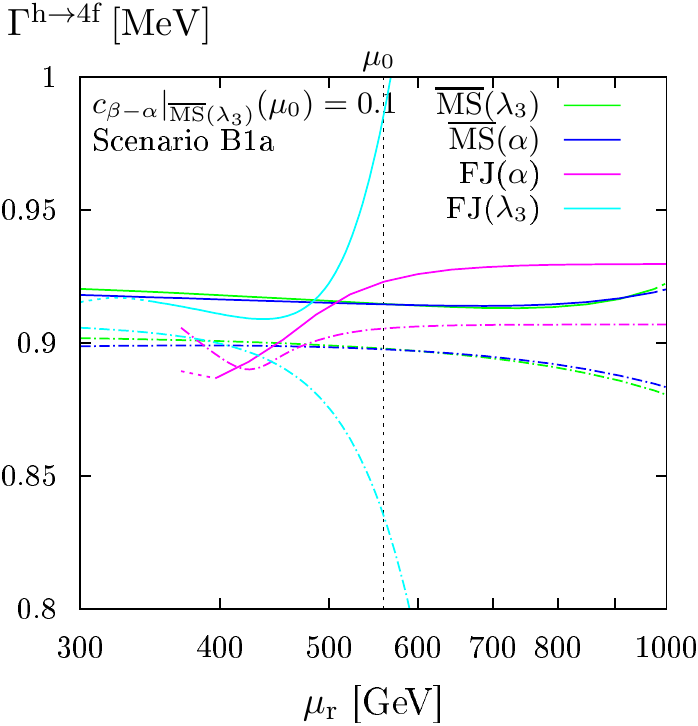}
}
\hspace{15pt}
\subfigure[]{
\label{fig:plot_MUSCAN-B1a-alphaMS}
\includegraphics{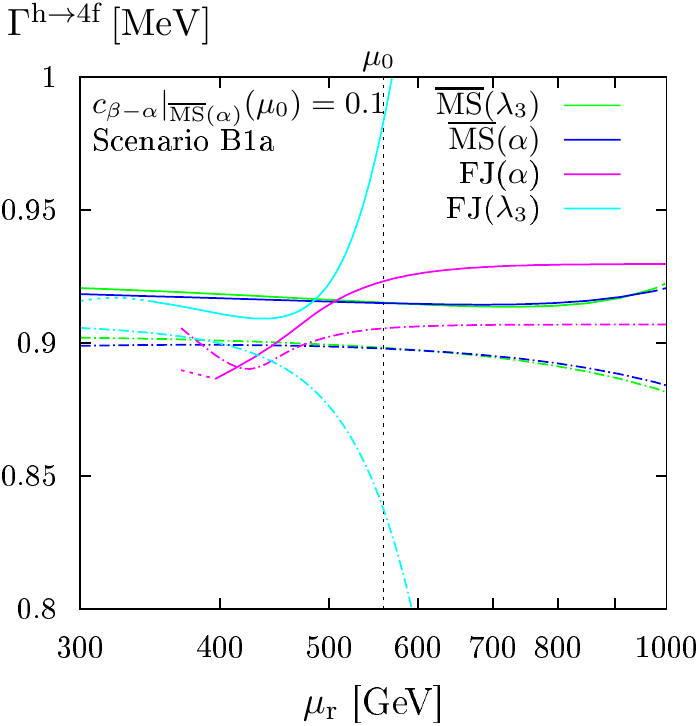}
}
\\[-1em]
  \subfigure[]{
\label{fig:plot_MUSCAN-B1a-FJ}
\includegraphics{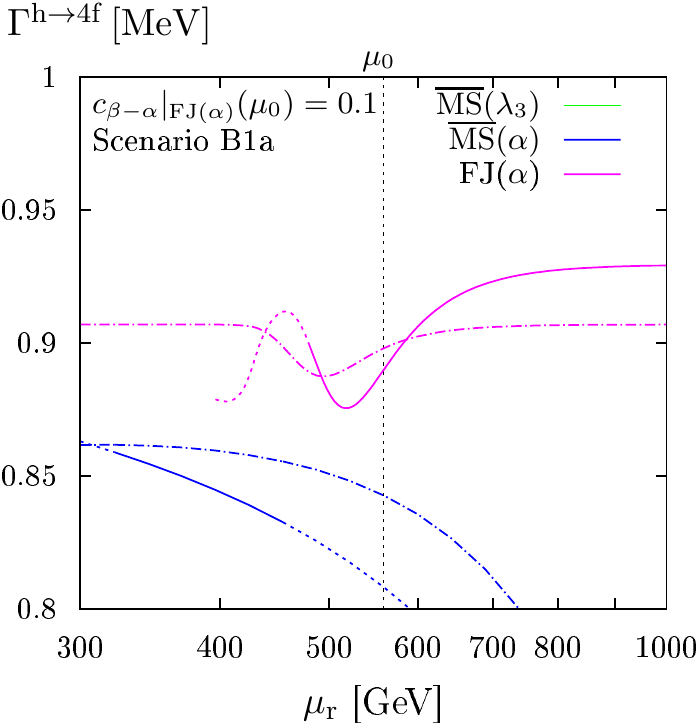}
}
\hspace{15pt}
\subfigure[]{
\label{fig:plot_MUSCAN-B1a-L3MSFJ}
\includegraphics{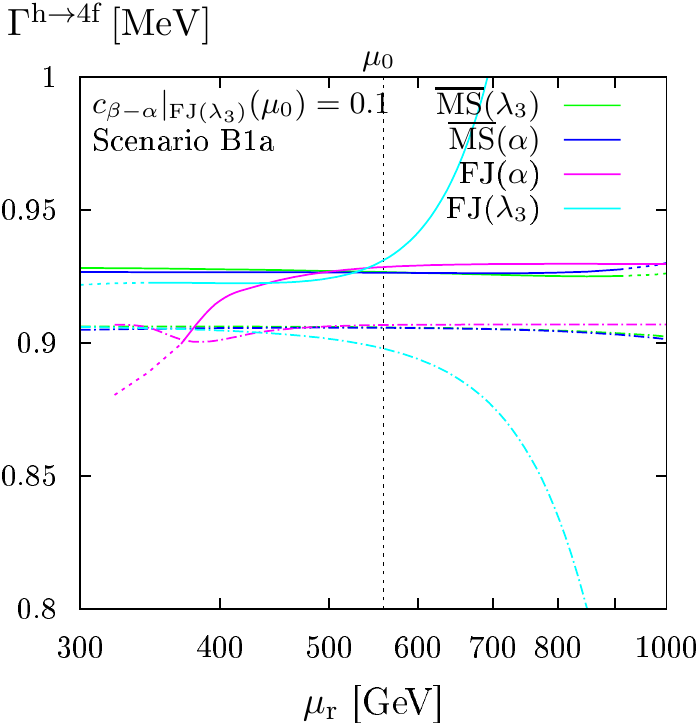}
}
\vspace*{-.5em}
  \caption{The $\Ph {\to} 4f$ width at LO (dashed) and NLO EW (solid) for scenario B1a in dependence of the renormalization scale. The panels (a), (b), (c), and (d) correspond to input values  defined in the  \MSbar{}$(\lambda_3)$, \MSbar{}$(\alpha)$, FJ($\alpha$), and FJ($\lambda_3$) schemes, respectively. The result is computed in all four different renormalization schemes after converting the input at NLO (also for the LO curves)
and displayed using the colour code of Fig.~\ref{fig:plotrunningA}. The breakdown of perturbativity ($\lambda_k^{\text{max}}/(4\pi)>1$) is indicated by changing the NLO curve to dotted lines.}
\label{fig:plotmuscanB1a}
\end{figure}
\begin{itemize}
 \item The \MSbar{}$(\lambda_3)$, Fig.~\ref{fig:plot_MUSCAN-B1a-L3MS}, and the \MSbar{}$(\alpha)$ input prescriptions, 
Fig.~\ref{fig:plot_MUSCAN-B1a-alphaMS}, yield similar results. In both cases, these schemes as target schemes show very good agreement, an extremum and a distinct plateau region in which the central scale fits perfectly. They only
begin to deviate when perturbativity breaks down. The other renormalization schemes do not 
behave as nicely: The 
results of the 
FJ($\alpha$) scheme has a significant offset and drops dramatically for lower scales, until perturbativity breaks down at about $400\GeV$. The FJ($\lambda_3$) scheme suffers from the strong running and diverges 
at high scales
as expected, while it shows relatively good (but not stable) agreement with the other schemes for lower scales.
\item 
For input values defined in the FJ($\alpha$) scheme (Fig.~\ref{fig:plot_MUSCAN-B1a-FJ}), the conversion transports the large NLO corrections to all other schemes, so that perturbativity is not given at all, and all curves disagree. Together with the behaviour of the FJ($\alpha$) scheme in the other plots, we conclude that the perturbative predictions using the FJ($\alpha$) scheme are not trustworthy for this benchmark scenario.
\item The FJ($\lambda_3$) input prescription (Fig.~\ref{fig:plot_MUSCAN-B1a-L3MSFJ}) seems to yield the best agreement between the schemes, however, the conversion to other renormalization schemes results in particularly small values for $c_{\beta-\alpha}$ and therefore corresponds in the other renormalization schemes to a scenario closer to the alignment limit. Such scenarios have smaller couplings and are perturbatively more stable, so that a better agreement is not surprising. 
For the $\Ph {\to} 4f$ width in a high-mass scenario~B1 
with $c_{\beta-\alpha}=0.05$ shown in App.~\ref{App:scalevar} we observe
a reduction of the scale dependence, the development of plateau regions 
for all schemes, and an overlap of the results from the different schemes.
\end{itemize}
In the computation of the relative renormalization scheme dependence at the central scale $\Delta_\mr{RS}$ only reliable renormalization schemes 
should be used. Therefore only the widths computed in the \MSbar{}($\alpha$) and \MSbar{}($\lambda_3$) schemes enter this calculation for which the result is shown in Tab.~\ref{tab:schemevarB1}. 
\begin{table}
   \centering
   \renewcommand{\arraystretch}{1.1}
 \begin{tabular}{|cc|cccc|}\hline
  && \MSbar{}$(\lambda_3)$ & \MSbar{}$(\alpha)$ & FJ($\alpha$) & FJ($\lambda_3$)\\\hline
  \multirow{2}{*}{Scenario B1a} &  $\Delta^\mr{LO}_\mr{RS}$[\%]       & 0.03(0)  & 0.04(0)  & --   & -- \\
   & $\Delta^\mr{NLO}_\mr{RS}$ [\%]     & 0.02(0)  & 0.02(0)  & --   & --   \\\hline
  \end{tabular}  
   \caption{The variation $\Delta_\mr{RS}$ of the $\mr{h} {\to} 4f$ width 
in scenario~B1a at the central scale $\mu_0$
using the reliable renormalization schemes   \MSbar{}($\lambda_3$) and \MSbar{}($\alpha$)
(with NLO parameter conversions).
The columns correspond to the schemes in which the input parameters are defined.
Using parameters defined in the FJ($\alpha$) and FJ($\lambda_3$) schemes, the results are unreliable and a computation of $\Delta_\mr{RS}$ is not meaningful. 
The zeroes in brackets show that the integration errors are negligible.}
 \label{tab:schemevarB1}
 \end{table}%
Omitting the FJ schemes, our estimate of the scheme dependence is, thus, just the difference of the two \MSbar{} schemes,
which is very small both at LO and NLO. Nevertheless a tendency towards a reduction of the scheme dependence is seen
in the transition from LO to NLO.
For the input values defined in one of the FJ schemes, this analysis cannot be performed as the results are unreliable.

\subsubsection{\boldmath{$c_{\beta-\alpha}$} dependence}
\label{sec:cbascanB1}
The $\Ph {\to} 4f$ width of the Higgs decay as a function of positive $c_{\beta-\alpha}$ is shown for all combinations of input prescriptions and renormalization schemes in Fig.~\ref{fig:plotscbascanB1} at the scale $\mu_0$. 
The results from all schemes agree very well in the alignment limits, where
$c_{\beta-\alpha}\to0$.
For $|c_{\beta-\alpha}|>0.05$, differences in the results obtained with different schemes
after conversion from a common parameter input scheme start to become significant.
The patterns 
observed in the investigation of the scale dependence recur. The dashed lines represent the LO result with an NLO input conversion, while at pure LO, i.e.\ without conversion,
all renormalization schemes deliver identical results. The respective curve is the one where no conversion is necessary. The well-known $s_{\beta-\alpha}^2$ pattern can be observed at LO while the conversion into the FJ($\lambda_3$) scheme introduces large corrections leading to a breakdown (see Fig.~\ref{fig:plot_Umrechnung-von-alpha-B1}), so that this scheme is only applicable for very low values of $c_{\beta-\alpha}$. 
The NLO results away from the alignment limit 
are more complicated:
 \begin{figure}
  \centering

  \subfigure[]{
\label{fig:plot_cbascan-B1-diffschemes-L3MS}
\includegraphics{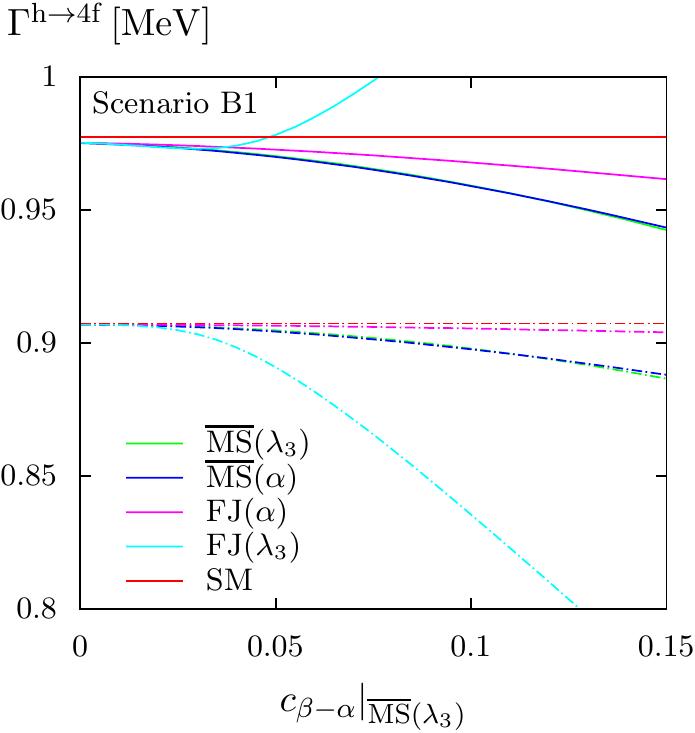}
}
\hspace{15pt}
\subfigure[]{
\label{fig:plot_cbascan-B1-diffschemes-alpha}
\includegraphics{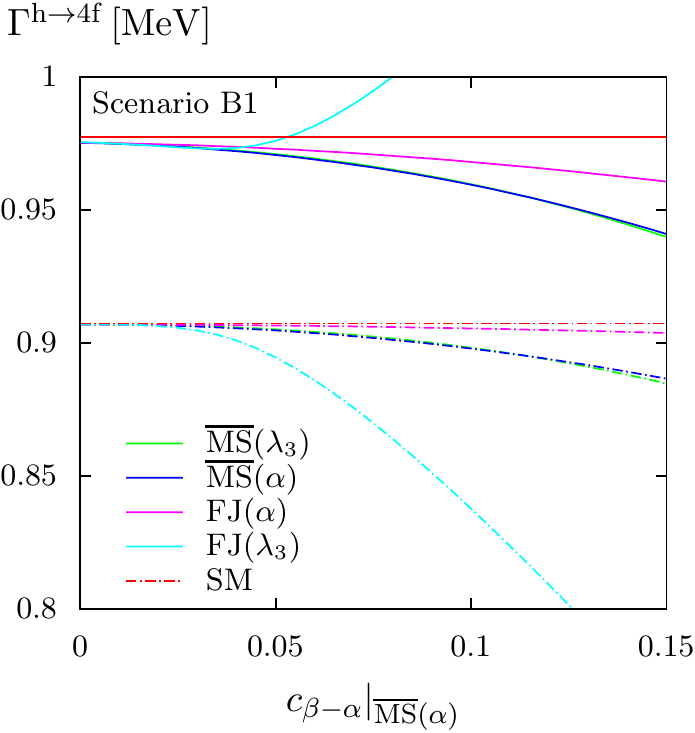}
}
\\[-1em]
  \subfigure[]{
\label{fig:plot_cbascan-B1-diffschemes-FJ}
\includegraphics{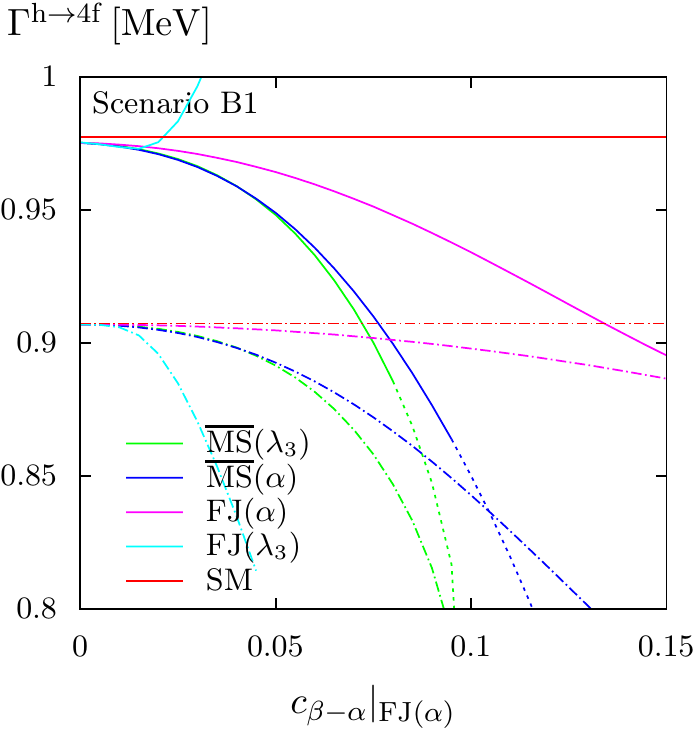}
}
\hspace{15pt}
\subfigure[]{
\label{fig:plot_cbascan-B1-diffschemes-L3MSFJ}
\includegraphics{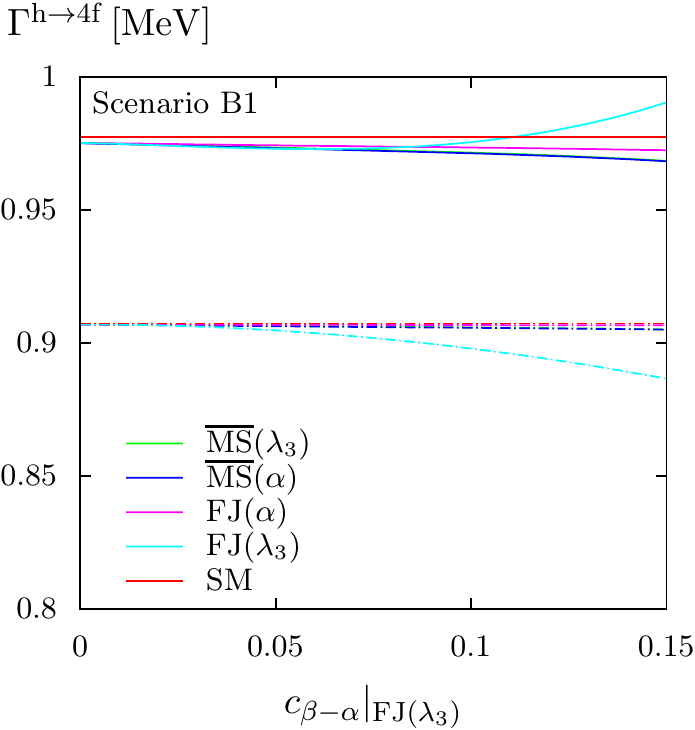}
}
\vspace*{-.5em}
  \caption{The $\Ph {\to} 4f$ width at LO (dashed) and NLO EW+QCD (solid) for scenario~B1 in dependence of $c_{\beta-\alpha}$. The panels (a), (b), (c), and (d) correspond to input values defined in the  \MSbar{}$(\lambda_3)$, \MSbar{}$(\alpha)$, FJ($\alpha$), and FJ($\lambda_3$) schemes, and they are converted to the other schemes at NLO (also for the LO curves). The result is displayed in all four schemes and for the SM using the colour code of Fig.~\ref{fig:plotscbascanA}. The breakdown of perturbativity ($\lambda_k^{\text{max}}/(4\pi)>1$) is visualized by using dotted lines for the NLO curve.}
\label{fig:plotscbascanB1}
\end{figure}

\begin{figure}
  \centering
\includegraphics[scale=0.9]{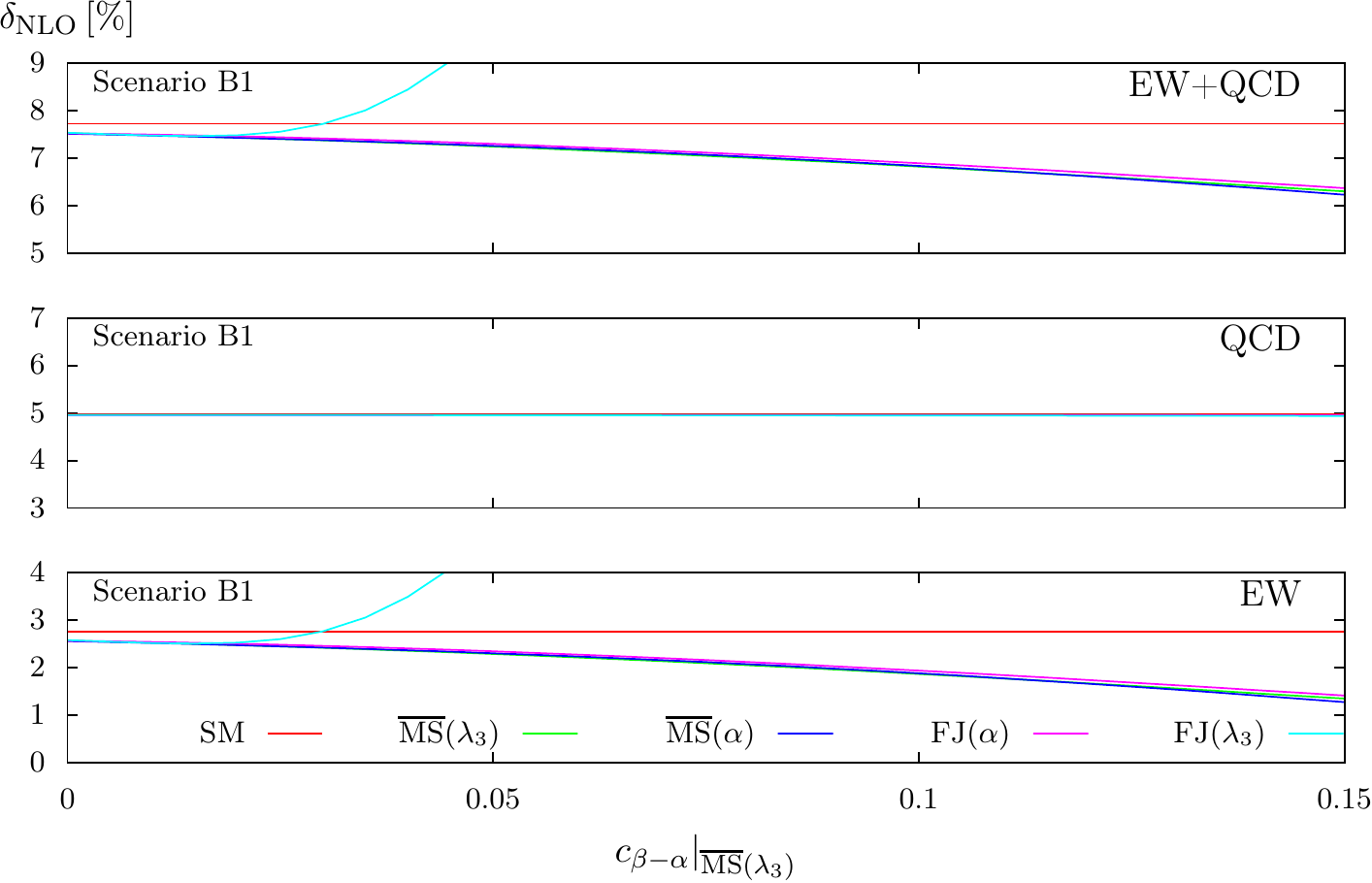}
  \caption{The relative NLO corrections of the full EW+QCD, the QCD, and the EW calculation in scenario B1. The input is defined in the \MSbar{}$(\lambda_3)$ scheme, and the corrections are computed in all four schemes which are displayed using the colour code of Fig.~\ref{fig:plotscbascanA}. For comparison the SM corrections with a SM Higgs-boson mass of $\Mh$ are shown as well.}
\label{fig:plot_cbascanrel-B1-diffschemes-L3MS}
\end{figure}%
\begin{figure}
  \centering
\includegraphics[scale=0.9]{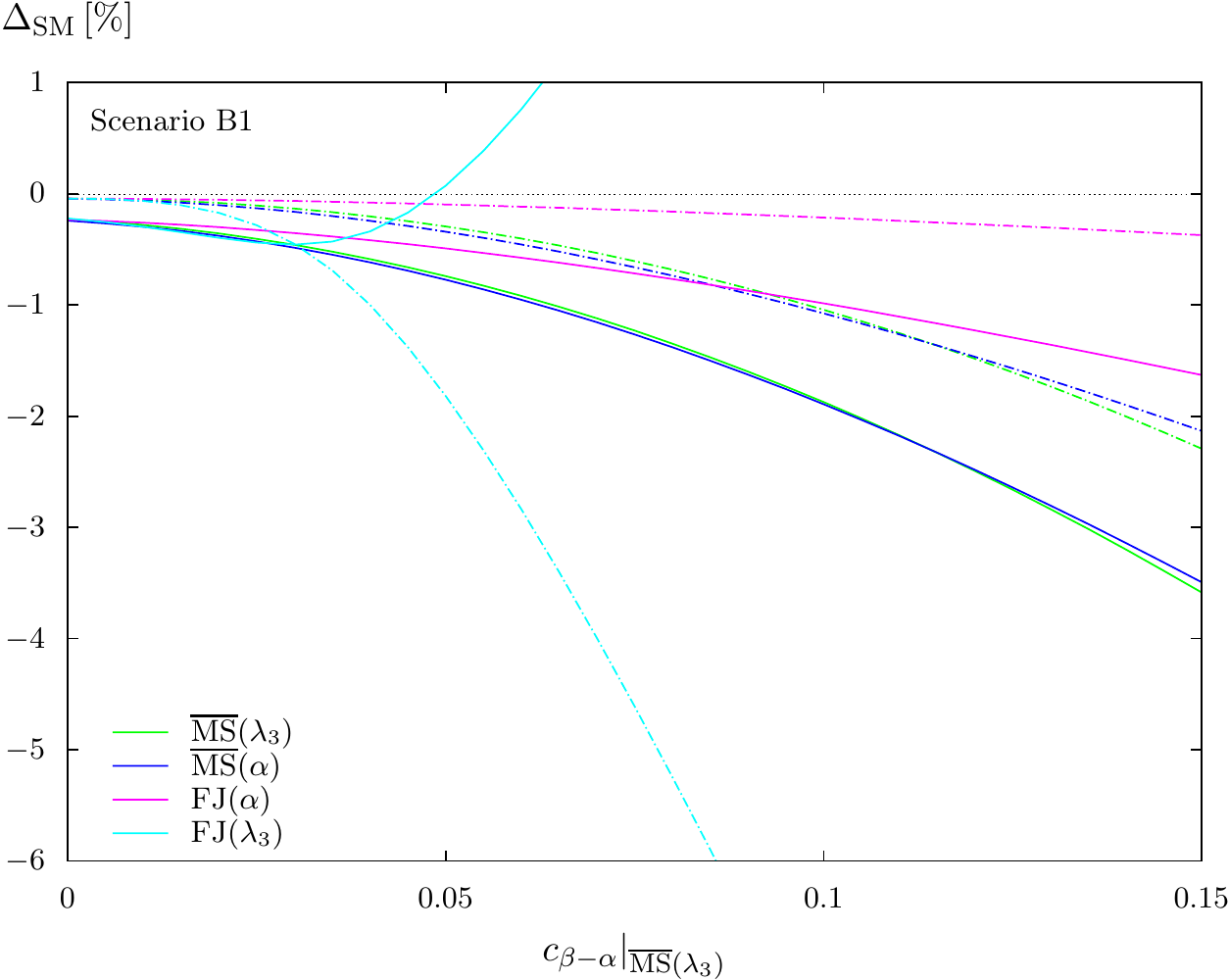}
  \caption{The $h{\to}4f$ width at LO with NLO conversion (dashed) and NLO EW+QCD (solid) in scenario B1, normalized to the respective SM values where the SM Higgs-boson mass is $\Mh$. The input is defined in the \MSbar{}$(\lambda_3)$ scheme, and the results in the four schemes are displayed using the colour code of Fig.~\ref{fig:plotscbascanA}.}
\label{fig:plot_cbascanrelSM-B1-diffschemes-L3MS}\vspace{-6pt}
\end{figure}%

\begin{itemize}
 \item  The \MSbar{}$(\alpha)$ and the \MSbar{}$(\lambda_3)$ input prescriptions (Figs.~\ref{fig:plot_cbascan-B1-diffschemes-L3MS},(b)) have similar characteristics which is due to the small shifts of the parameters in the conversion. The width in the \MSbar{}$(\alpha)$ and the \MSbar{}$(\lambda_3)$ renormalization schemes agree 
very well, and the agreement improves from LO to NLO, as desired. The FJ($\alpha$) scheme 
(as target scheme)
shows differences which can be explained by large higher-order terms shifting the input values towards the alignment limit (see Fig.~\ref{fig:plot_Umrechnung-nach-alpha-B1}). 
Owing to the large corrections at NLO, sizeable corrections beyond NLO are expected in the FJ schemes as well,
i.e.\ for reliable predictions for the B1 scenario in the FJ schemes the inclusion of leading corrections beyond NLO should
be calculated and taken into account.

 \item Using input values defined in the FJ($\alpha$) scheme, Fig.~\ref{fig:plot_cbascan-B1-diffschemes-FJ}, expresses this problem more clearly. The large corrections spread to the other renormalization schemes and affect perturbativity in a negative way. But also within the FJ($\alpha$) scheme the corrections are large and differ from the $s_{\beta-\alpha}^2$ shape seen for other input variants. This confirms the conclusion of the previous section that predictions obtained using the FJ($\alpha$) are not reliable for this scenario.
 \item The good agreement of the renormalization schemes in the FJ($\lambda_3$) input prescription (Fig.~\ref{fig:plot_cbascan-B1-diffschemes-L3MSFJ}) is based on the shift of the input values towards the alignment limit in the conversion. This shrinks the range effectively to 
$0<c_{\beta-\alpha}\lsim0.05$ for the other target schemes after the conversion and pushes the results together.
 \end{itemize}

For the input defined in the \MSbar{}$(\lambda_3)$ scheme, 
the relative corrections separated in EW, QCD, and EW+QCD are shown in Fig.~\ref{fig:plot_cbascanrel-B1-diffschemes-L3MS}. 
Taking the input in the \MSbar{}$(\alpha)$ scheme instead, the results look similar (not shown).
The QCD corrections are similar for all schemes, because only the closed quark-loop diagrams in the $\Ph VV$ vertex corrections
do not factorize from the SM LO amplitude with the coupling factor $\sin(\beta-\alpha)$, but the impact of those diagrams is small.
In contrast to the low-mass scenario, the EW corrections decrease with increasing $c_{\beta-\alpha}$, so that the deviations 
from the SM shown in Fig.~\ref{fig:plot_cbascanrelSM-B1-diffschemes-L3MS} are larger than in the low-mass case, although they remain below 2\%
 for $c_{\beta-\alpha}<0.1$. 
The relative renormalization scheme dependence can only be applied using the \MSbar{}($\alpha$) and $\MSbar(\lambda_3)$ 
schemes, as the results obtained using the two schemes involving FJ prescriptions are only reliable for very small $c_{\beta-\alpha}$. 
From Figs.~\ref{fig:plot_MUSCAN-B1a-L3MS},(b), one can see that the differences between the \MSbar{}($\alpha$) and $\MSbar(\lambda_3)$ 
schemes decrease from LO to NLO, and the scheme dependence is reduced.

\subsubsection{Partial widths for individual four-fermion states}

The partial NLO widths, the relative corrections $\delta_\mr{EW/QCD}$, and the deviations
from the SM, $\Delta_\mr{SM}^\mr{LO/NLO}$, are shown in Tab.~\ref{tab:partialwidthsB1} for benchmark scenario~B1a  
using the \MSbar{}$(\lambda_3)$ renormalization scheme. 
\begin{table}
   \vspace{-3pt}
   \centering
   \renewcommand{\arraystretch}{.94}
 \begin{tabular}{|c|ccccc|}\hline \vphantom{$\Big|$}
  Final state& $\Gamma^{\Ph{\to}4f}_\mr{NLO}$ [MeV] & $\delta_\mr{EW}$ [\%]& $\delta_\mr{QCD}$ [\%] & $\Delta_\mr{SM}^\mr{NLO}$ [\%]&$\Delta_\mr{SM}^\mr{LO}$ [\%]\\\hline
 inclusive $\Ph {\to} 4f$ & $ 0.95976 ( 9)$ & $ 1.88 ( 0)$ & $ 4.96 ( 1)$ & $ -1.82 ( 1)$ & $ -1.00 ( 1)$ \\
ZZ & $ 0.105308 ( 7)$ & $ -0.48 ( 0)$ & $ 4.88 ( 1)$ & $ -1.89 ( 1)$ & $ -1.00 ( 1)$ \\
WW & $ 0.8596 ( 1)$ & $ 2.16 ( 0)$ & $ 5.02 ( 1)$ & $ -1.81 ( 2)$ & $ -1.00 ( 1)$ \\
WW/ZZ int. & $ -0.00514 ( 7)$ & $ 0.3 ( 2)$ & $ 13 ( 1)$ & $ -1 ( 2)$ & $ -1 ( 1)$ \\
$ \nu_\Pe \Pe^+ \mu^- \bar{\nu}_\mu$ & $ 0.010118 ( 1)$ & $ 2.19 ( 0)$ & $ 0.00 $ & $ -1.85 ( 2)$ & $ -1.00 ( 2)$ \\
$ \nu_\Pe \Pe^+ \Pu \bar{\Pd}$ & $ 0.031471 ( 5)$ & $ 2.18 ( 0)$ & $ 3.77 ( 1)$ & $ -1.82 ( 2)$ & $ -1.00 ( 2)$ \\
$ \Pu \bar{\Pd} \Ps \bar{\Pc}$ & $ 0.09772 ( 2)$ & $ 2.14 ( 0)$ & $ 7.52 ( 2)$ & $ -1.79 ( 3)$ & $ -1.00 ( 2)$ \\
$\nu_\Pe \Pe^+ \Pe^- \bar{\nu}_\Pe$ & $ 0.010113 ( 1)$ & $ 2.29 ( 0)$ & $ 0.00 $ & $ -1.85 ( 2)$ & $ -1.00 ( 2)$ \\
$ \Pu \bar{\Pd} \Pd \bar{\Pu}$ & $ 0.09969 ( 2)$ & $ 2.02 ( 0)$ & $ 7.34 ( 2)$ & $ -1.81 ( 4)$ & $ -1.00 ( 2)$ \\
$\nu_\Pe \bar{\nu}_\Pe \nu_\mu \bar{\nu}_\mu$ & $ 0.000941 ( 0)$ & $ 2.19 ( 0)$ & $ 0.00 $ & $ -1.92 ( 2)$ & $ -1.00 ( 2)$ \\
$ \Pe^- \Pe^+ \mu^- \mu^+$ & $ 0.000237 ( 0)$ & $ 0.49 ( 1)$ & $ 0.00 $ & $ -1.94 ( 2)$ & $ -1.00 ( 1)$ \\
$\nu_\Pe \bar{\nu}_\Pe \mu^-\mu^+$ & $ 0.000474 ( 0)$ & $ 1.63 ( 1)$ & $ 0.00 $ & $ -1.91 ( 2)$ & $ -1.00 ( 1)$ \\
$\nu_\Pe \bar{\nu}_\Pe \nu_\Pe \bar{\nu}_\Pe$ & $ 0.000564 ( 0)$ & $ 2.09 ( 0)$ & $ 0.00 $ & $ -1.93 ( 3)$ & $ -1.00 ( 2)$ \\
$ \Pe^- \Pe^+ \Pe^-\Pe^+$ & $ 0.000131 ( 0)$ & $ 0.31 ( 1)$ & $ 0.00 $ & $ -1.92 ( 2)$ & $ -1.00 ( 1)$ \\
$\nu_\Pe \bar{\nu}_\Pe \Pu \bar{\Pu}$ & $ 0.001666 ( 0)$ & $ -0.22 ( 1)$ & $ 3.75 ( 1)$ & $ -1.89 ( 2)$ & $ -1.00 ( 1)$ \\
$\nu_\Pe \bar{\nu}_\Pe \Pd \bar{\Pd}$ & $ 0.002160 ( 0)$ & $ 0.88 ( 1)$ & $ 3.75 ( 1)$ & $ -1.89 ( 2)$ & $ -1.00 ( 2)$ \\
$ \Pe^-\Pe^+ \Pu \bar{\Pu}$ & $ 0.000839 ( 0)$ & $ -0.70 ( 1)$ & $ 3.76 ( 1)$ & $ -1.89 ( 2)$ & $ -1.00 ( 1)$ \\
$ \Pe^- \Pe^+ \Pd \bar{\Pd}$ & $ 0.001080 ( 0)$ & $ -0.35 ( 1)$ & $ 3.76 ( 1)$ & $ -1.89 ( 2)$ & $ -1.00 ( 1)$ \\
$ \Pu \bar{\Pu} \Pc \bar{\Pc}$ & $ 0.002948 ( 1)$ & $ -2.61 ( 1)$ & $ 7.51 ( 2)$ & $ -1.86 ( 3)$ & $ -1.00 ( 1)$ \\
$ \Pd \bar{\Pd} \Pd \bar{\Pd}$ & $ 0.002537 ( 1)$ & $ -1.20 ( 0)$ & $ 4.42 ( 3)$ & $ -1.93 ( 4)$ & $ -1.00 ( 2)$ \\
$ \Pd \bar{\Pd} \Ps \bar{\Ps}$ & $ 0.004918 ( 1)$ & $ -1.17 ( 1)$ & $ 7.50 ( 2)$ & $ -1.86 ( 3)$ & $ -1.00 ( 2)$ \\
$ \Pu \bar{\Pu} \Ps \bar{\Ps}$ & $ 0.003823 ( 1)$ & $ -1.48 ( 1)$ & $ 7.51 ( 2)$ & $ -1.86 ( 3)$ & $ -1.00 ( 1)$ \\
$ \Pu \bar{\Pu} \Pu \bar{\Pu}$ & $ 0.001495 ( 1)$ & $ -2.73 ( 1)$ & $ 4.12 ( 3)$ & $ -1.95 ( 5)$ & $ -1.00 ( 1)$ \\
\hline
  \end{tabular}  
   \caption{Partial widths for benchmark scenario B1a within the \MSbar{}$(\lambda_3)$ scheme.\vspace{-6pt}}
 \label{tab:partialwidthsB1}
 \end{table}
The \MSbar{}$(\alpha)$ scheme yields similar results which differ only at the permille level, whereas the other schemes are not reliable at this benchmark scenario. 
 The widths are slightly smaller than in the low-mass scenario (see \refse{sec:partwidth-LM}) in spite of the identical values of
$c_{\beta-\alpha}$. 
The negative deviation from the SM rises to almost 2\%, however, no final state accounts for distinctively large THDM effects that could be exploited in experiments.
 
In addition, the differential distributions of scenario~B1a, as defined in \refse{sec:Prophecy4f}, 
do not change the shape w.r.t.\ to the SM significantly.
They are shown in App.~\ref{sec:diffdistrHM}, together with the SM ones and the ones from 
scenario~B2b. 
As observed in the low-mass scenario, for each four-fermion final state
the difference between the $\Ph{\to}4f$ widths in the THDM and the SM resembles a
constant shift in all distributions
as well.

\subsection{High-mass scenario B2}
\label{sec:HighMassB2}

To complete the discussion of the high-mass scenario, we turn to negative values of $c_{\beta-\alpha}$ for which the parameter space is strongly reduced by perturbativity, stability, and unitarity constraints, leaving only a small branch around $\tan \beta=1.5$ and leading to scenario B2.
Being in the vicinity of excluded parameter sets potentially
affects the conversion, the scale dependence, and the reliability of the full results. 
Hence, similar to scenario~B1, scenario~B2 is well suited to actually 
address possible problems in that respect.
We discuss the results in the same manner 
as scenario~B1 for positive $c_{\beta-\alpha}$ above and
present results for the less delicate case with 
$c_{\beta-\alpha}=-0.05$ in App.~\ref{App:scalevar}.

\subsubsection{Conversion of the input parameters}

\begin{figure}
  \centering
\subfigure[]{
\label{fig:plot_Umrechnung-von-alpha-B2}
\includegraphics{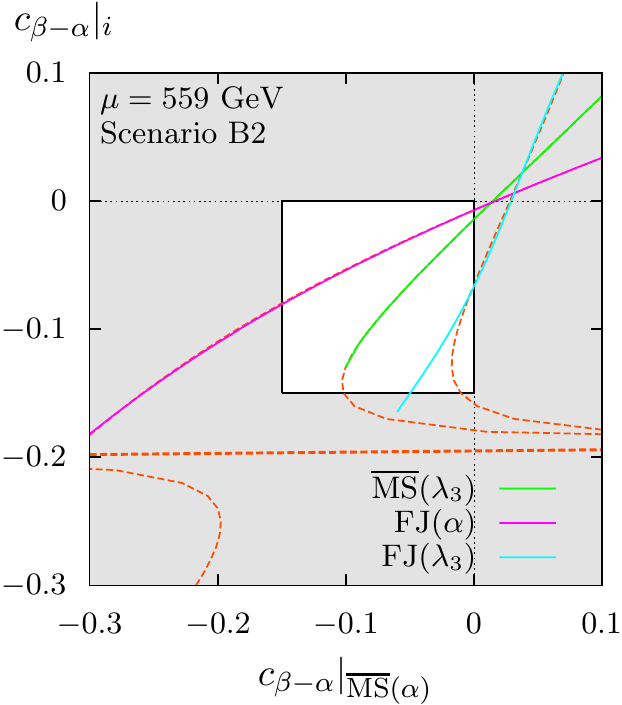}
}
\hspace{15pt}
\subfigure[]{
\label{fig:plot_Umrechnung-nach-alpha-B2}
\includegraphics{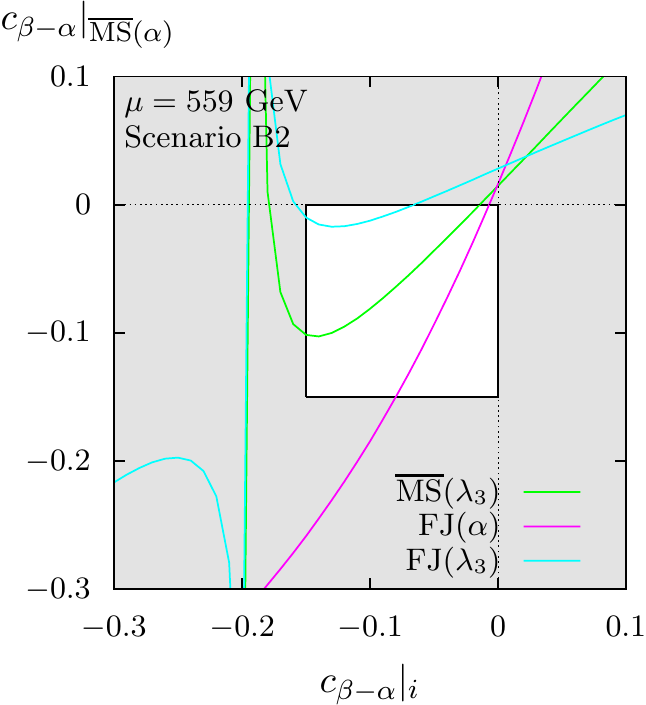}
}
\vspace*{-1em}
  \caption{Panel (a): Conversion of the value of $c_{\beta-\alpha}$ from \MSbar{}$(\alpha)$ to the other renormalization schemes for scenario B2 with the colour coding of Fig.~\ref{fig:plotconversionA}. Panel (b) shows the conversion to the \MSbar{}$(\alpha)$ scheme with the same colour coding. The solid lines are obtained by solving the implicit equations numerically, the dashed orange lines correspond to the solution of (b) mirrored at the diagonal. The linear approximation does not provide reasonable results. The highlighted region shows the phenomenologically most relevant $c_{\beta - \alpha}$ region.}
\label{fig:plotconversionB2}%
\end{figure}

The conversion of the input parameter $c_{\beta-\alpha}$ between different renormalization schemes is shown in Fig.~\ref{fig:plotconversionB2} for an enlarged range with \MSbar{}($\alpha$) 
either as input or target scheme. 
The conversion into the \MSbar{}($\alpha$) scheme shows several ominous features
(\reffi{fig:plotconversionB2}(b)). First of all, divergences for the $\MSbar(\lambda_3)$  and FJ($\lambda_3$) schemes occur at $c_{\beta-\alpha}\approx-0.19$. We have seen such a divergence already in the low-mass scenario outside of the relevant region (c.f.~\refse{sec:conversionA}), which is caused by the singularity at $c_{2\alpha}=0$. Since the ratio of the vacuum expectation values is lower in this scenario, the divergence moves towards the alignment limit and closer to the relevant region. It affects the conversion for $c_{\beta-\alpha}<-0.15$ 
in the $\MSbar(\lambda_3)$ and FJ($\lambda_3$) schemes,
so that such values should be taken with care. If experimental observations favour this region of parameter space, it becomes necessary to redefine the renormalization scheme and choose a different Higgs self-coupling parameter (e.g.~$\lambda_1$) or a combination (e.g.~$\lambda_1+\lambda_2$) 
as independent parameter renormalized in \MSbar{}. The singularity then appears in other parameter regions and allows for predictions 
with $c_{\beta-\alpha}\lsim-0.15$. 
Not only the schemes involving $\lambda_3$ are problematic, but also the conversion from the FJ($\alpha$) scheme, as large shifts indicate problems with the perturbative expansion, 
analogous to scenario B1.

Figure~\ref{fig:plot_Umrechnung-von-alpha-B2} shows 
the results for the ``inverse'' conversion from the \MSbar{}($\alpha$) scheme,
together with the inverse of 
(b) obtained graphically by mirroring the curves at the diagonal (dashed orange). 
Note that the linearized approximation for the conversion would involve large uncertainties here.
Although the comparison of these curves projects the reduction of the conversion to one dimension 
(spanned by $c_{\beta-\alpha}$), which is thus not exact,
it gives a quick overview over the convergence of the numerical solution. 
As expected, the singu\-la\-rity in the relation between $\lambda_3$ and $\alpha$ reduces the domain of definition for the conversion in Fig.~\ref{fig:plot_Umrechnung-von-alpha-B2} to $c_{\beta-\alpha}|_{\MSbar(\alpha)}> -0.1$ for the $\MSbar(\lambda_3)$ 
scheme and $c_{\beta-\alpha}|_{\MSbar(\alpha)}> -0.05$ for the FJ($\lambda_3$) scheme. Values outside this domain cannot be converted into these schemes, and solid predictions cannot be made there.

\subsubsection{The running of \boldmath{$c_{\beta-\alpha}$}}

\begin{figure}
  \centering
\includegraphics{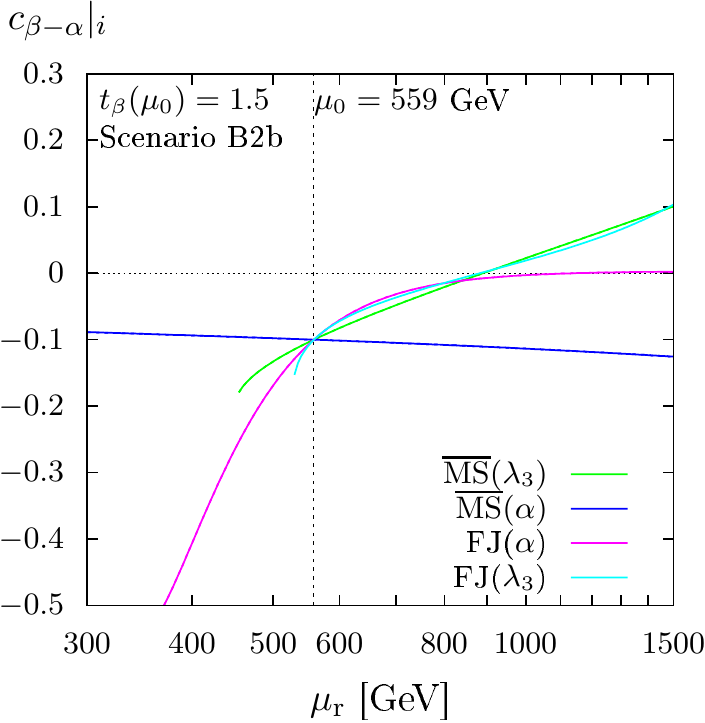}
\vspace*{-.5em}
  \caption{The running of $c_{\beta-\alpha}$ for scenario~B2b for the different schemes in the colour code of Fig.~\ref{fig:plotrunningA}.}
\label{fig:running-B2b}
\end{figure}

The running of $c_{\beta-\alpha}(\mu_\mr{r})$ with $c_{\beta-\alpha}(\mu_0)=-0.1$ is computed from 
$\mu_\mr{r}=300\GeV$ to $1500\GeV$ with a Runge--Kutta method for scenario~B2b. 
The result is shown in Fig.~\ref{fig:running-B2b} for each renormalization scheme without any conversion. 
The curves look similar to the ones of the low-mass scenario pictured in Fig.~\ref{fig:running_cba-0.1-LM} 
(although the range of $\mu_\mr{r}$ is different) and we observe 
the same effects: the truncation of the $\MSbar(\lambda_3)$ and FJ($\lambda_3$) schemes, but also the strong scale dependence of the FJ($\alpha$) scheme and the good stability of the \MSbar{}($\alpha$) scheme.

\subsubsection{Scale variation of the width}
\label{sec:HighMassb1scalevariation}

\begin{figure}
  \centering
  \subfigure[]{
\label{fig:plot_MUSCAN-B2b-L3MS}
\includegraphics{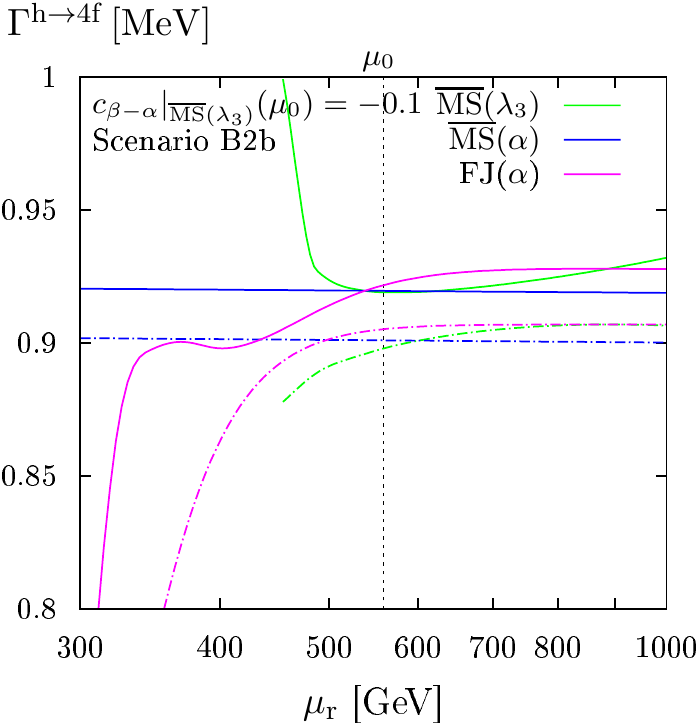}
}
\hspace{15pt}
\subfigure[]{
\label{fig:plot_MUSCAN-B2b-alphaMS}
\includegraphics{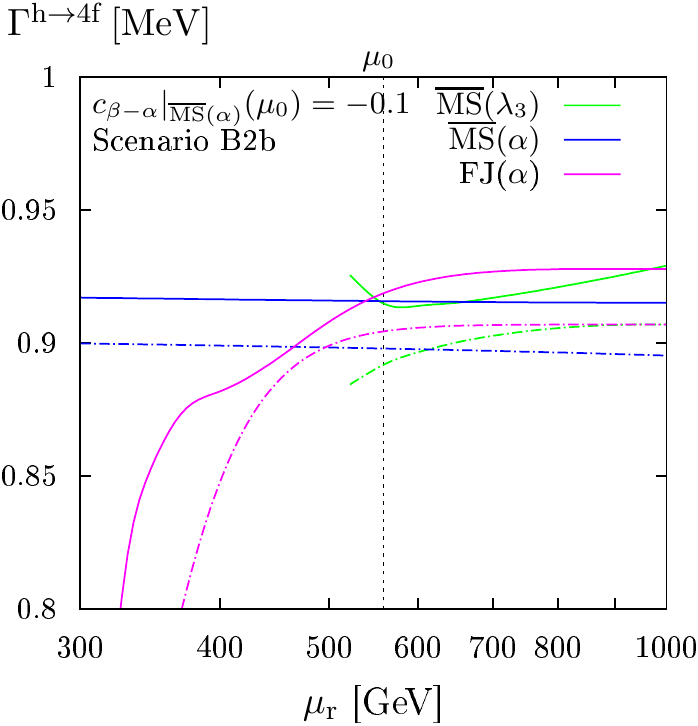}
}
\\[-1em]
  \subfigure[]{
\label{fig:plot_MUSCAN-B2b-FJ}
\includegraphics{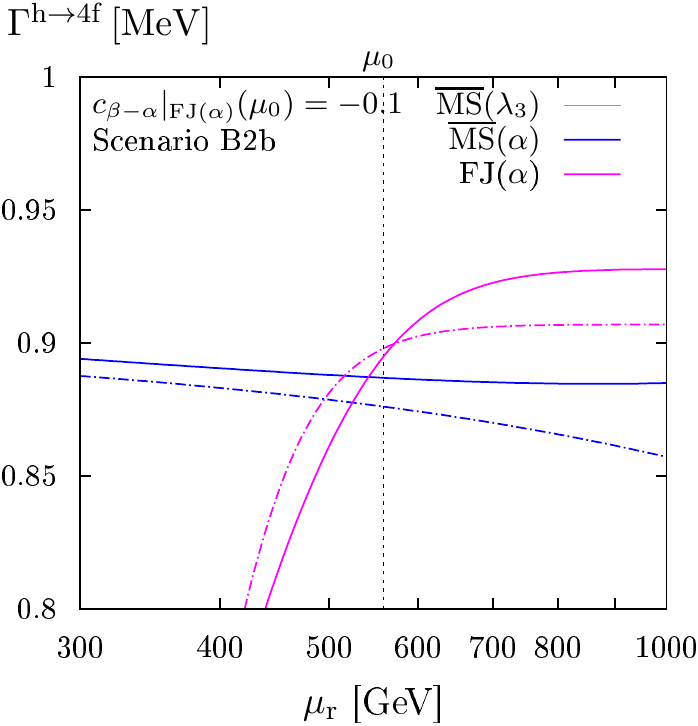}
}
\hspace{15pt}
\subfigure[]{
\label{fig:plot_MUSCAN-B2b-L3MSFJ}
\includegraphics{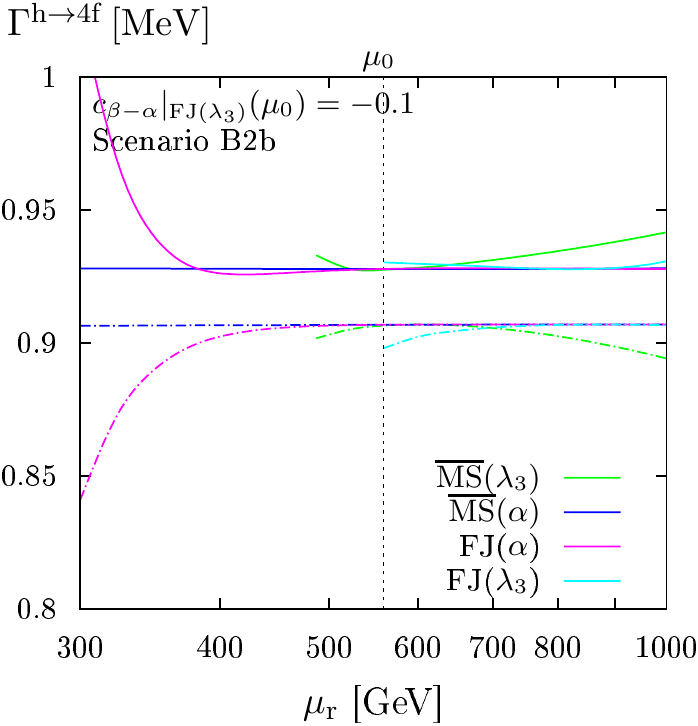}
}
\vspace*{-.5em}
  \caption{The $\Ph {\to} 4f$ width at LO (dashed) and NLO EW (solid) for scenario B2b in dependence of the renormalization scale. The panels (a), (b), (c), and (d) correspond to input values  defined in the  \MSbar{}$(\lambda_3)$, \MSbar{}$(\alpha)$, FJ($\alpha$), and FJ($\lambda_3$) schemes, respectively. For each of them, the result is computed in all four different renormalization schemes after converting the input at NLO (also for the LO curves) and displayed using the colour code of Fig.~\ref{fig:plotrunningA}. The FJ($\lambda_3$) scheme is not defined as target schemes due to the singular relation between $\alpha$ and $\lambda_3$.}
\label{fig:plotmuscanB2b}
\end{figure}

For the $\Ph {\to} 4f$ width we again perform a scale variation in order to estimate theoretical uncertainties and to motivate the central scale choice. The method is as described in \refse{sec:LowMassscalevariation}, and the results are shown in Fig.~\ref{fig:plotmuscanB2b}. 
The FJ($\lambda_3$) renormalization scheme is not included 
as target scheme
here, since it is not possible to convert input values to it for $c_{\beta-\alpha}=-0.1$  (see Fig.~\ref{fig:plot_Umrechnung-von-alpha-B2}), however, it can be used when the input values are defined in it. The observations correspond in the most cases to the ones of scenario~B1: 
\begin{itemize}
 \item 
The first two plots using parameters defined in the $\MSbar(\lambda_3)$ 
(Fig.~\ref{fig:plot_MUSCAN-B2b-L3MS}) and \MSbar{}($\alpha$) (Fig.~\ref{fig:plot_MUSCAN-B2b-alphaMS}) 
schemes show, as in the previous scenarios, similar characteristics. The result obtained with the \MSbar{}($\alpha$) renormalization scheme shows almost no scale dependence, and its value agrees with the extremum of the the $\MSbar(\lambda_3)$ renormalization scheme which lies at the central scale. However, through the truncation of the running a 
broad plateau region cannot be observed for the latter scheme
with input in \MSbar{}$(\alpha)$. 
The width in the FJ($\alpha$) scheme is consistent with the results of the
$\MSbar(\lambda_3)$ and \MSbar{}($\alpha$) schemes at the central scale, but
shows an offset at the plateau and decreases 
for scales below $\mu_0$, as expected from the running of $c_{\beta-\alpha}$. 
 \item 
The results using the FJ($\alpha$) input prescription (Fig.~\ref{fig:plot_MUSCAN-B2b-FJ}) 
are not conclusive,
since large corrections from the conversion spread to all other schemes.
 \item The scale variation of the FJ($\lambda_3$) input prescription (Fig.~\ref{fig:plot_MUSCAN-B2b-L3MSFJ}) corresponds again to an aligned scenario 
in the other renormalization schemes. Closer to the alignment, the renormalization scheme dependence decreases,
which can also be seen from the separate scale variation of a more aligned scenario with $c_{\beta-\alpha}=-0.05$ given in App.~\ref{App:scalevar}.
\end{itemize}
Generically,
we obtain a somewhat better improvement compared to 
benchmark scenario~B1a, which probably originates from smaller perturbativity measures (see Fig.~\ref{fig:perturbativityconstr}). The central scale of Eq.~\eqref{eq:centralscale} is a justifiable choice and suggests that this scale is 
a good candidate for 
the THDM Higgs decay into four fermions in general, although the scale choice should be 
better checked for consistency in any new scenario.
The renormalization schemes \MSbar{}($\alpha$) and $\MSbar{}(\lambda_3)$ 
yield trustworthy and comparable results, even though one should respect 
the domain of definition of the latter. 
Results based on an input in the FJ($\alpha$) scheme do not seem reliable;
the FJ($\lambda_3$) scheme cannot even be applied for this input procedure.
The renormalization scheme dependence at the central scale
reduces from LO to NLO as shown in Tab.~\ref{tab:schemevarB2}. 
For the input renormalization schemes  \MSbar{}$(\lambda_3)$ and \MSbar{}$(\alpha)$, we did not take the  FJ($\lambda_3$) scheme into account when evaluating the renormalization scheme dependence while for the input scheme FJ($\lambda_3$), all four renormalization schemes have been considered. This corresponds to the results shown 
in Fig.~\ref{fig:plotmuscanB2b}.
\begin{table}
   \centering
   \renewcommand{\arraystretch}{1.1}
\begin{tabular}{|cc|cccc|}\hline
   & & \MSbar{}$(\lambda_3)$ & \MSbar{}$(\alpha)$ & FJ($\alpha$) & FJ($\lambda_3$)\\\hline
  \multirow{2}{*}{Scenario B2b} &   $\Delta^\mr{LO}_\mr{RS}$[\%]       & 0.81(0)  & 1.40(0)  & --   & 0.99(0) \\
   &$\Delta^\mr{NLO}_\mr{RS}$ [\%]     & 0.31(0)  & 0.46(0)  & --   & 0.25(0)   \\\hline
 \end{tabular}  
\caption{The variation $\Delta_\mr{RS}$ of the $\mr{h} {\to} 4f$ width 
in scenario~B2b at the central scale $\mu_0$
using the renormalization schemes \MSbar{}($\alpha$), \MSbar{}($\lambda_3$),
and FJ($\alpha$)
(with NLO parameter conversions).
The columns correspond to the schemes in which the input parameters are defined.
Using parameters defined in the FJ($\alpha$) scheme, the results are unreliable, and a computation of $\Delta_\mr{RS}$ is not meaningful. 
The zeroes in brackets show that the integration errors are negligible.}
\label{tab:schemevarB2}
 \end{table}%

\subsubsection{\boldmath{$c_{\beta-\alpha}$} dependence}

\begin{figure}
  \centering
  \subfigure[]{
\label{fig:plot_cbascan-B2-diffschemes-L3MS}
\includegraphics{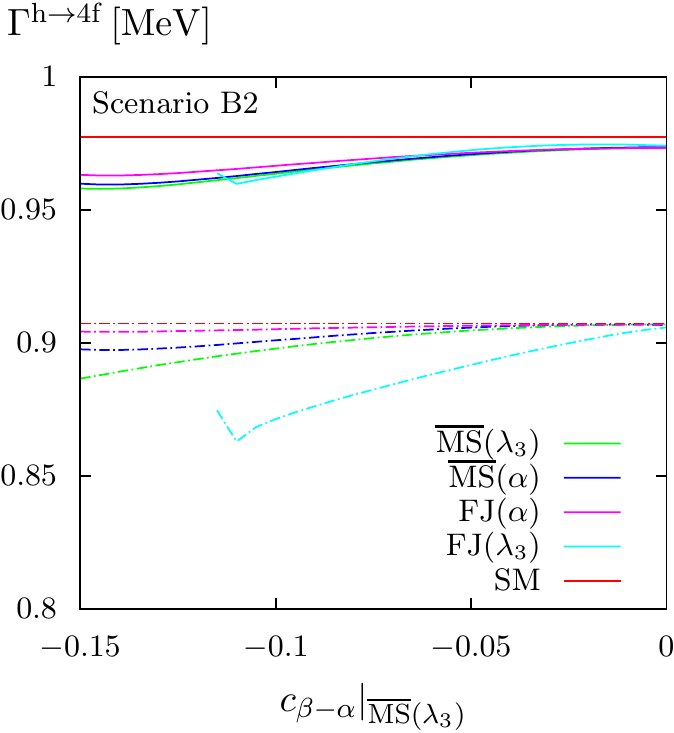}
}
\hspace{15pt}
\subfigure[]{
\label{fig:plot_cbascan-B2-diffschemes-alpha}
\includegraphics{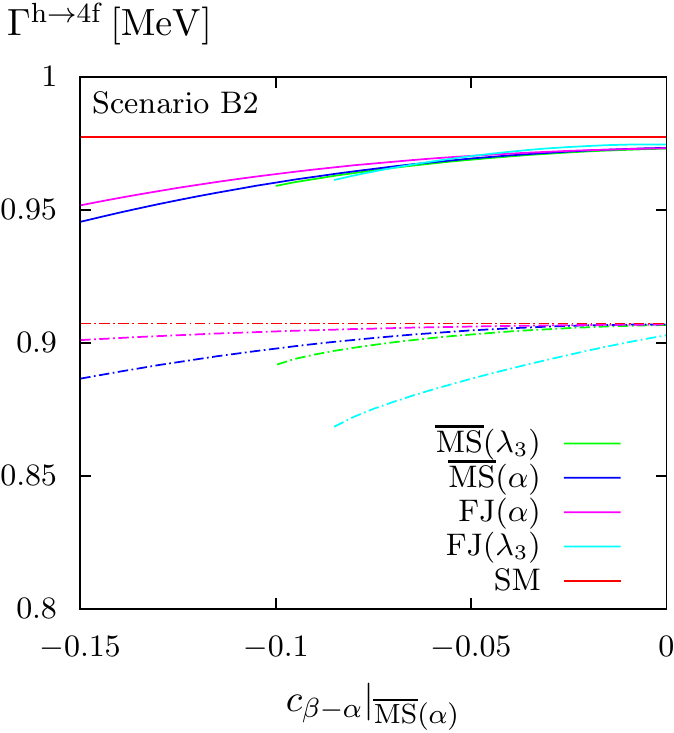}
}
\\[-1em]
  \subfigure[]{
\label{fig:plot_cbascan-B2-diffschemes-FJ}
\includegraphics{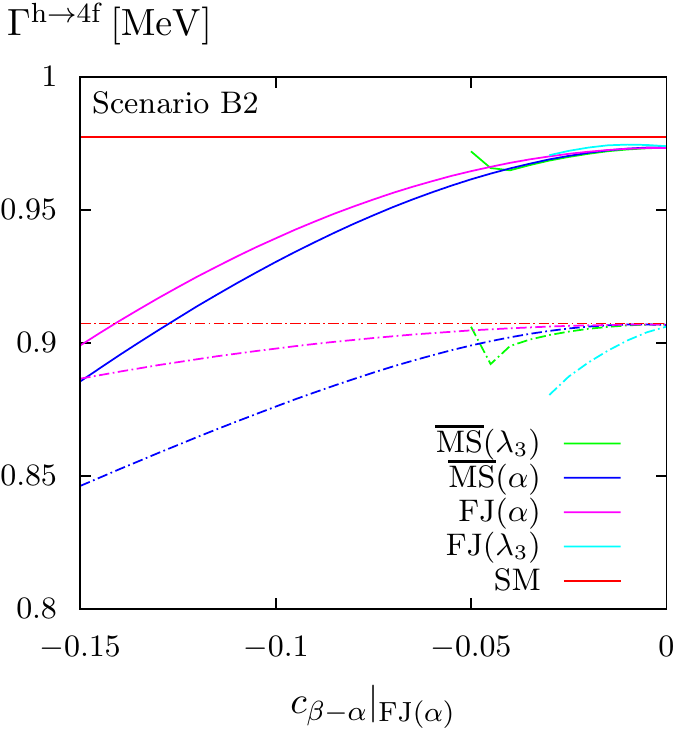}
}
\hspace{15pt}
\subfigure[]{
\label{fig:plot_cbascan-B2-diffschemes-L3MSFJ}
\includegraphics{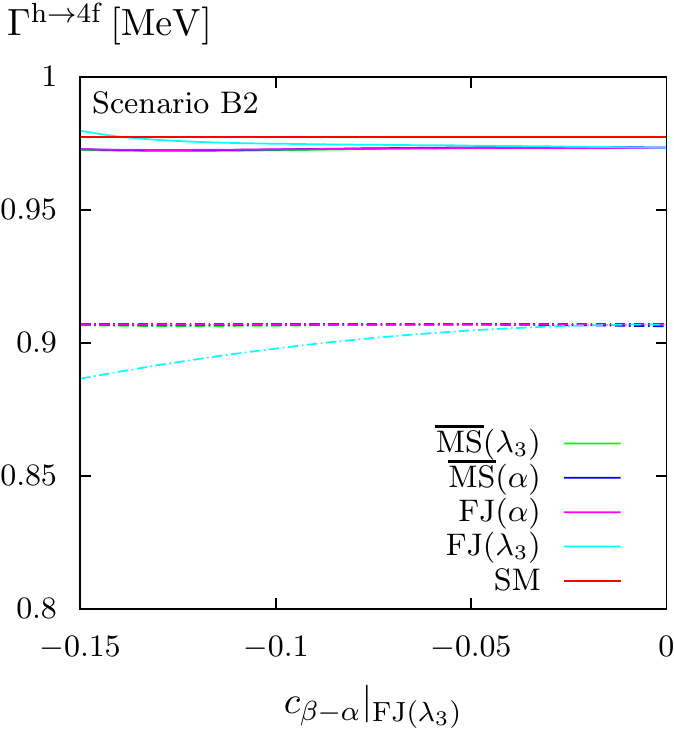}
}
\vspace*{-.5em}
  \caption{The $\Ph {\to} 4f$ width at LO (dashed) and NLO EW+QCD (solid) for scenario~B2 in dependence of $c_{\beta-\alpha}$. The panels (a), (b), (c), and (d) correspond to input values  defined in the  \MSbar{}$(\lambda_3)$, \MSbar{}$(\alpha)$, FJ($\alpha$), and FJ($\lambda_3$) schemes, respectively. The input values are converted to the desired target scheme (colour code of Fig.~\ref{fig:plotrunningA}) in which the calculation is performed. The SM prediction is shown for comparison in red.}
\label{fig:plotscbascanB2}
\end{figure}

The dependence of the $\Ph {\to} 4f$ width on $c_{\beta-\alpha}$ 
is shown for the different input prescriptions in the four panels of Fig.~\ref{fig:plotscbascanB2}, for all renormalization schemes. 
Close to the alignment limit, $-0.05\lsim c_{\beta-\alpha}<0$,
the results from different renormalization schemes agree  nicely.
Away from this limit, however,
the results deviate significantly, 
demanding some discussion:
\begin{itemize}
\item 
The curves obtained using the $\MSbar(\lambda_3)$ and the FJ($\lambda_3$) input prescriptions, 
Fig.~\ref{fig:plot_cbascan-B2-diffschemes-L3MS} and Fig.~\ref{fig:plot_cbascan-B2-diffschemes-L3MSFJ}, 
show the largest deviation from the $s_{\beta-\alpha}^2$ dependence of the LO width because of the
large corrections inherited 
from the parameter conversions to the other schemes, which were
observed in \reffi{fig:plot_Umrechnung-nach-alpha-B2}.
Defining the input in the FJ($\lambda_3$) scheme, the NLO width even slightly increases with smaller
$c_{\beta-\alpha}$ values, since the corresponding $c_{\beta-\alpha}$ value in the $\MSbar(\alpha)$ scheme
stays close to the alignment limit. Owing to the large conversion effects, especially the predictions in the 
FJ($\lambda_3$) scheme involve large uncertainties, which could only be reduced by systematically including 
the leading effects beyond NLO.
\item 
Using input values defined in the \MSbar{}($\alpha$) scheme yields the smooth curves of Fig.~\ref{fig:plot_cbascan-B2-diffschemes-alpha} 
which have the expected $s_{\beta-\alpha}^2$ shape. The relative renormalization scheme dependence reduces from LO to NLO, 
while the breakdown of the $\MSbar(\lambda_3)$ and FJ($\lambda_3$) schemes is manifest, since values
of $c_{\beta-\alpha}$ smaller than $\sim -0.1$ or $\sim-0.05$ in the $\MSbar(\alpha)$ scheme cannot be 
converted into the $\MSbar(\lambda_3)$ or FJ($\lambda_3$) schemes, respectively (cf.\ \reffi{fig:plot_Umrechnung-von-alpha-B2}).
\item 
The FJ$(\alpha)$ input 
prescription shows largest deviations from the SM as large NLO contributions spread to the other schemes through the conversion, shifting the values away from the alignment limit and increasing the deviations from the SM prediction.
\end{itemize}
\begin{figure}
  \centering
\includegraphics[scale=0.9]{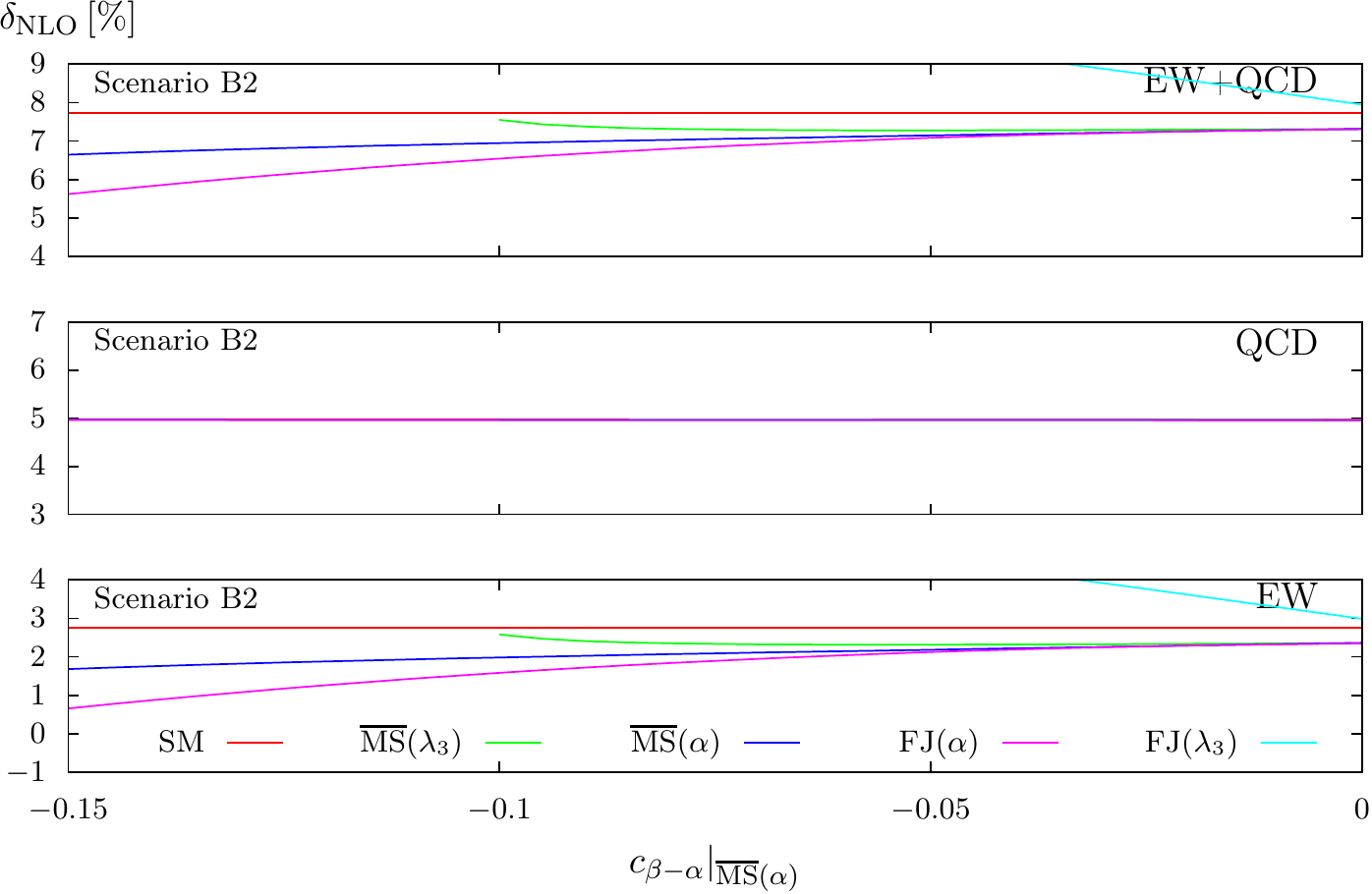}
  \caption{The relative NLO EW+QCD, QCD, and EW corrections to the $\Ph {\to} 4f$ width in scenario~B2.
The input is defined in the \MSbar{}$(\alpha)$ scheme and the corrections are computed in all four schemes which are displayed together with the SM corrections using the colour code of Fig.~\ref{fig:plotscbascanA}.}
\label{fig:plot_cbascanrel-B2-diffschemes-alpha}
\end{figure}%
\begin{figure}
  \centering
\includegraphics[scale=0.9]{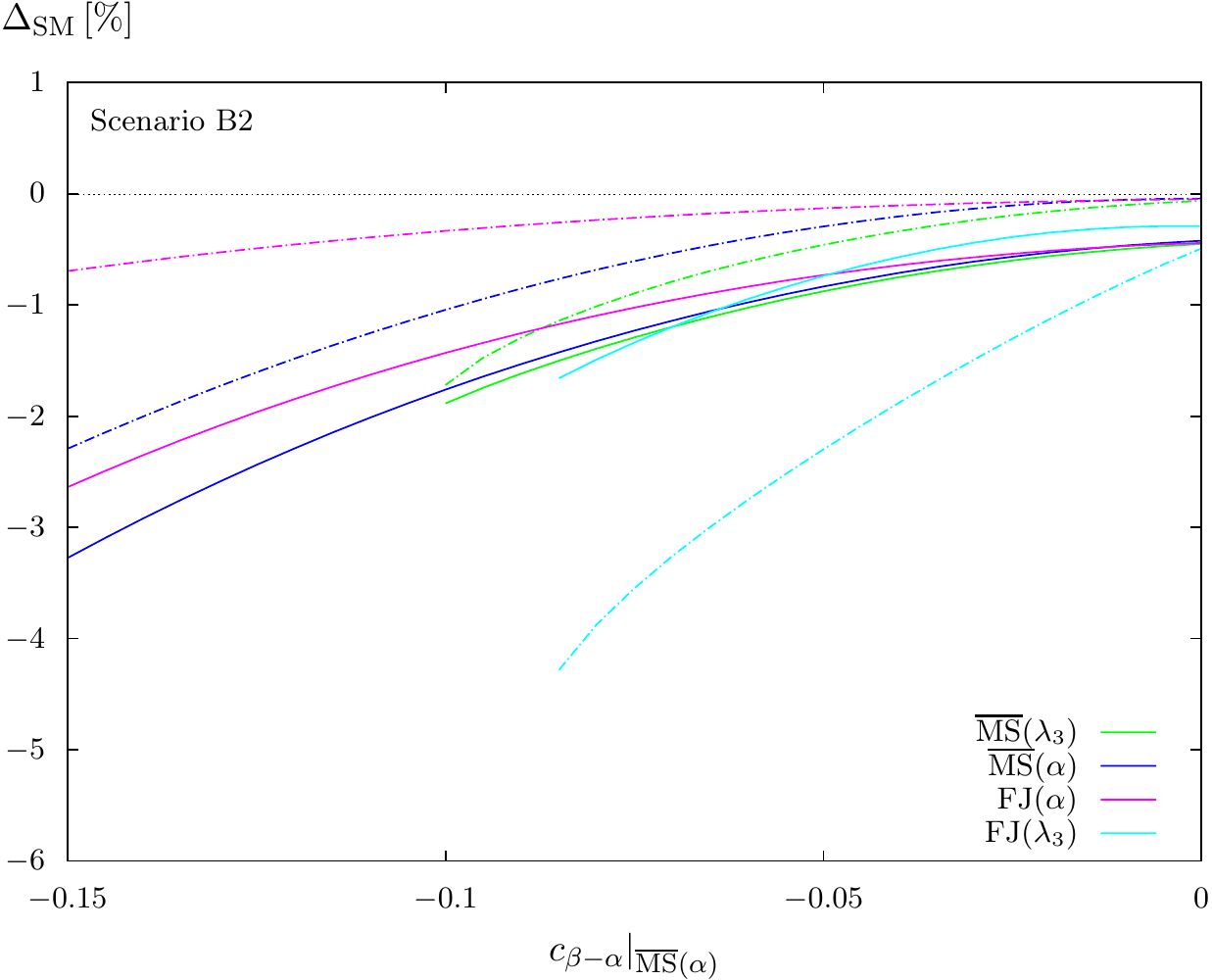}
  \caption{The $\Ph {\to} 4f$ width at LO (dashed) and NLO EW+QCD (solid) in the THDM scenario~B2, normalized to the respective SM values. The input is defined in the \MSbar{}$(\alpha)$ scheme, and the corrections are computed in all four schemes which are displayed using the colour code of Fig.~\ref{fig:plotscbascanA}.}
\label{fig:plot_cbascanrelSM-B2-diffschemes-alpha}
\end{figure}%

However, all results show a significantly better agreement between the renormalization schemes at NLO than for LO
for all regions including the problematic ones, suggesting that the perturbative expansion works for this scenario 
in the vicinity of our central scale $\mu_0$ in spite of partially large NLO terms. 
As the $\MSbar(\lambda_3)$ scheme has a limited region of applicability, we show in Fig.~\ref{fig:plot_cbascanrel-B2-diffschemes-alpha} the relative corrections using the \MSbar{}($\alpha$) scheme, which is reliable over the whole scan range. The QCD corrections are similar to the SM and renormalization scheme independent, while the EW corrections show the breakdown of the $\MSbar(\lambda_3)$ and FJ($\lambda_3$) schemes. The difference between the FJ($\alpha$) and the \MSbar{}($\alpha$) schemes is slightly larger than in the low-mass case, however, the sizes of the corrections are almost equal. This results in similar deviations from the SM as can be seen by 
comparing Fig.~\ref{fig:plot_cbascanrelSM-B2-diffschemes-alpha} with Fig.~\ref{fig:plot_cbascanrelSM-A-diffschemes-L3MS} (it should, however, be noted that a different input scheme has been used),
 so that it is difficult to distinguish these scenarios using the Higgs decay into four fermions.

\subsubsection{Partial widths for individual four-fermion states}

We give the partial widths in Tab.~\ref{tab:partialwidthsB2} for scenario~B2b in the \MSbar{}($\alpha$) scheme, as this scheme provides reliable results for $c_{\beta-\alpha}=-0.1$. 
\begin{table}
   \centering
   \renewcommand{\arraystretch}{.94}
 \begin{tabular}{|c|ccccc|}\hline \vphantom{$\Big|$}
  Final state& $\Gamma^{\Ph{\to}4f}_\mr{NLO}$ [MeV] & $\delta_\mr{EW}$ [\%]& $\delta_\mr{QCD}$ [\%] & $\Delta_\mr{SM}^\mr{NLO}$ [\%]&$\Delta_\mr{SM}^\mr{LO}$ [\%]\\\hline
inclusive $\Ph {\to} 4f$ & $ 0.96086 ( 9)$ & $ 1.99 ( 0)$ & $ 4.97 ( 1)$ & $ -1.71 ( 1)$ & $ -1.00 ( 1)$ \\
ZZ & $ 0.105584 ( 7)$ & $ -0.22 ( 0)$ & $ 4.90 ( 1)$ & $ -1.63 ( 1)$ & $ -1.00 ( 1)$ \\
WW & $ 0.8604 ( 1)$ & $ 2.26 ( 0)$ & $ 5.02 ( 1)$ & $ -1.72 ( 2)$ & $ -1.00 ( 1)$ \\
WW/ZZ int. & $ -0.00509 ( 7)$ & $ 0.5 ( 2)$ & $ 11 ( 1)$ & $ -2 ( 2)$ & $ -1 ( 1)$ \\
$ \nu_\Pe \Pe^+ \mu^- \bar{\nu}_\mu$ & $ 0.010128 ( 1)$ & $ 2.29 ( 0)$ & $ 0.00 $ & $ -1.76 ( 2)$ & $ -1.00 ( 2)$ \\
$ \nu_\Pe \Pe^+ \Pu \bar{\Pd}$ & $ 0.031499 ( 5)$ & $ 2.28 ( 0)$ & $ 3.77 ( 1)$ & $ -1.73 ( 2)$ & $ -1.00 ( 2)$ \\
$ \Pu \bar{\Pd} \Ps \bar{\Pc}$ & $ 0.09781 ( 2)$ & $ 2.23 ( 0)$ & $ 7.52 ( 2)$ & $ -1.70 ( 3)$ & $ -1.00 ( 2)$ \\
$\nu_\Pe \Pe^+ \Pe^- \bar{\nu}_\Pe$ & $ 0.010123 ( 1)$ & $ 2.39 ( 0)$ & $ 0.00 $ & $ -1.75 ( 2)$ & $ -1.00 ( 2)$ \\
$ \Pu \bar{\Pd} \Pd \bar{\Pu}$ & $ 0.09981 ( 2)$ & $ 2.12 ( 0)$ & $ 7.37 ( 2)$ & $ -1.69 ( 4)$ & $ -1.00 ( 2)$ \\
$\nu_\Pe \bar{\nu}_\Pe \nu_\mu \bar{\nu}_\mu$ & $ 0.000944 ( 0)$ & $ 2.46 ( 0)$ & $ 0.00 $ & $ -1.67 ( 2)$ & $ -1.00 ( 2)$ \\
$ \Pe^- \Pe^+ \mu^- \mu^+$ & $ 0.000237 ( 0)$ & $ 0.74 ( 1)$ & $ 0.00 $ & $ -1.69 ( 2)$ & $ -1.00 ( 1)$ \\
$\nu_\Pe \bar{\nu}_\Pe \mu^-\mu^+$ & $ 0.000475 ( 0)$ & $ 1.89 ( 1)$ & $ 0.00 $ & $ -1.66 ( 2)$ & $ -1.00 ( 1)$ \\
$\nu_\Pe \bar{\nu}_\Pe \nu_\Pe \bar{\nu}_\Pe$ & $ 0.000566 ( 0)$ & $ 2.35 ( 0)$ & $ 0.00 $ & $ -1.68 ( 3)$ & $ -1.00 ( 2)$ \\
$ \Pe^- \Pe^+ \Pe^-\Pe^+$ & $ 0.000131 ( 0)$ & $ 0.57 ( 1)$ & $ 0.00 $ & $ -1.66 ( 2)$ & $ -1.00 ( 1)$ \\
$\nu_\Pe \bar{\nu}_\Pe \Pu \bar{\Pu}$ & $ 0.001670 ( 0)$ & $ 0.04 ( 1)$ & $ 3.75 ( 1)$ & $ -1.65 ( 2)$ & $ -1.00 ( 1)$ \\
$\nu_\Pe \bar{\nu}_\Pe \Pd \bar{\Pd}$ & $ 0.002165 ( 0)$ & $ 1.13 ( 0)$ & $ 3.75 ( 1)$ & $ -1.65 ( 2)$ & $ -1.00 ( 2)$ \\
$ \Pe^-\Pe^+ \Pu \bar{\Pu}$ & $ 0.000841 ( 0)$ & $ -0.45 ( 1)$ & $ 3.76 ( 1)$ & $ -1.65 ( 2)$ & $ -1.00 ( 1)$ \\
$ \Pe^- \Pe^+ \Pd \bar{\Pd}$ & $ 0.001082 ( 0)$ & $ -0.09 ( 1)$ & $ 3.76 ( 1)$ & $ -1.65 ( 2)$ & $ -1.00 ( 1)$ \\
$ \Pu \bar{\Pu} \Pc \bar{\Pc}$ & $ 0.002955 ( 1)$ & $ -2.36 ( 1)$ & $ 7.51 ( 2)$ & $ -1.63 ( 3)$ & $ -1.00 ( 1)$ \\
$ \Pd \bar{\Pd} \Pd \bar{\Pd}$ & $ 0.002548 ( 1)$ & $ -0.94 ( 0)$ & $ 4.59 ( 3)$ & $ -1.52 ( 4)$ & $ -1.00 ( 2)$ \\
$ \Pd \bar{\Pd} \Ps \bar{\Ps}$ & $ 0.004930 ( 1)$ & $ -0.92 ( 0)$ & $ 7.50 ( 2)$ & $ -1.63 ( 3)$ & $ -1.00 ( 2)$ \\
$ \Pu \bar{\Pu} \Ps \bar{\Ps}$ & $ 0.003832 ( 1)$ & $ -1.23 ( 1)$ & $ 7.51 ( 2)$ & $ -1.62 ( 3)$ & $ -1.00 ( 1)$ \\
$ \Pu \bar{\Pu} \Pu \bar{\Pu}$ & $ 0.001502 ( 0)$ & $ -2.48 ( 1)$ & $ 4.35 ( 3)$ & $ -1.49 ( 5)$ & $ -1.00 ( 1)$ \\
\hline
  \end{tabular}  
   \caption{Partial widths for benchmark scenario B2b in the \MSbar{}$(\alpha)$ renormalization scheme.}
 \label{tab:partialwidthsB2}
 \end{table}%
All the partial widths are similar to the ones of the low-mass scenario~Ab (Tab.~\ref{tab:partialwidthsAb}) in size (note, however, different input schemes have been used). 
This observation applies to the 
EW and QCD corrections and to the differences to the SM predictions as well. 
Again, there is no final state particularly sensitive to the THDM contributions. The differential distributions analogous to \refse{sec:diffdistr} are shown together with the distributions of the high-mass benchmark 
scenario~B1 in App.~\ref{sec:diffdistrHM} and yield no significant shape distortion w.r.t.\ the SM, but only constant shifts that match the deviation of the respective partial widths.

\subsection{Different THDM types}

In this section, we compare the $\Ph {\to} 4f$ decay widths of the Type~I, Type~II, lepton specific, and flipped THDMs for the two 
scenarios~Aa and B1a using the \MSbar{}$(\lambda_3)$ renormalization scheme. 
We do not expect large differences in the results, because the considered THDM versions differ only in the Yukawa couplings
of Higgs bosons to the leptons and to down-type quarks, which are not enhanced by large fermion masses. 
The by far largest contributions involving Yukawa couplings, however, result from diagrams with top-quark--Higgs
couplings, which are identical in all four THDM versions for the $\Ph {\to} 4f$ processes with massless external fermions.
The results are shown in Tab.~\ref{tab:DiffTypes} with the numerical errors in parentheses, confirming our expectation: 
  \begin{table}
   \centering
   \renewcommand{\arraystretch}{1.1}
 \begin{tabular}{|c|ccc|}\hline
  \multicolumn{4}{|l|}{\textbf{Scenario Aa}}\\\hline
  Model& $\Gamma^{\Ph{\to}4f}_\mr{NLO}$ [MeV] &$\delta_\mr{EW}[\%]$ & $\delta_\mr{QCD}[\%]$ \\\hline
  Type I         & $   0.96730(7)$& $    2.711(1) $ & $ 4.962(5) $ \\
    Type II        & $   0.96729(7)$& $  2.711(1) $ & $ 4.962(5) $ \\
  Lepton-specific& $   0.96730(7)$& $   2.711(1) $ & $ 4.962(5) $ \\
  Flipped        & $   0.96729(7)$& $   2.711(1) $ & $ 4.962(5)$ \\\hline
  \multicolumn{4}{|l|}{\textbf{Scenario B1a}}\\\hline
  Model& $\Gamma^{\Ph{\to}4f}_\mr{NLO}$ [MeV] &$\delta_\mr{EW}[\%]$ & $\delta_\mr{QCD}[\%]$ \\\hline
  Type I          & $   0.95981(7)$& $   1.878(3) $ & $ 4.961(5)$  \\
  Type II         & $   0.95980(7)$& $  1.879(3) $ & $ 4.959(5)$    \\
  Lepton-specific & $   0.95981(7)$& $  1.878(3) $ & $ 4.961(5)$ \\
  Flipped         & $  0.95980(7)$& $   1.879(3) $ & $ 4.959(5)$  \\\hline
 \end{tabular}  
   \caption{The $\Ph {\to} 4f$ widths for the different types of THDM for scenarios Aa and B1a using the \MSbar{}$(\lambda_3)$ renormalization scheme. The numerical errors are given in parentheses.}
 \label{tab:DiffTypes}
 \end{table}
The differences originating from the different THDM types in the NLO corrections are below a permille and
not even significant over the integration error (although we employ large statistics with 190 million phase-space points). 
The $\Ph {\to} 4f$ decay observables are, thus, rather insensitive to the different types of THDMs, 
so that our predictions are universally valid for all types.

\subsection{Benchmark plane}

\begin{figure}
\centering
\includegraphics[width=0.8\columnwidth]{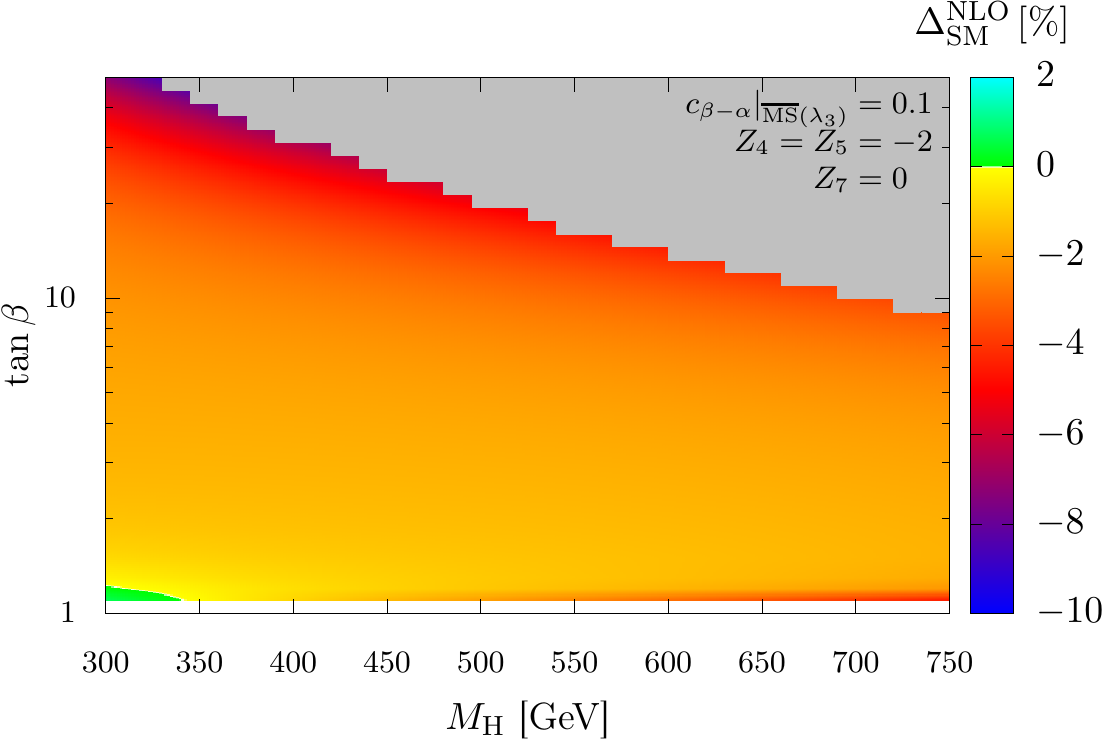}
\caption{The relative deviation $\Delta_\SM$ of the $\Ph {\to} 4f$ width w.r.t.\ the SM at NLO for the benchmark plane scenario in the 
$\MSbar(\lambda_3)$ scheme. Gray areas are excluded by non-perturbativity while the size of the deviations is indicated by the colour. We interpolate linearly between computed points to obtain a smooth picture.}
\label{fig:plane}
\end{figure}

For the benchmark plane scenario defined in \refse{sec:setup} we analyze only the relative deviation 
$\Delta_\SM$ 
of the $\Ph {\to} 4f$ width
with respect to the SM in the $\MSbar(\lambda_3)$ scheme. At LO, this is $-0.01$ as $c_{\beta-\alpha}=0.1$ is kept constant. 
The NLO corrections differentiate this picture as they are dependent on both the heavy-Higgs-boson masses and $\tan{\beta}$. 
We show the result for a wide range in the $(\MH,\tan{\beta})$
plane in Fig.~\ref{fig:plane} where the colour of the parameter point indicates the 
deviation $\Delta_\mr{SM}^\mr{NLO}$
and gray areas are excluded by perturbativity constraints ($\lambda_k^{\text{max}}/(4 \pi)>1$). 
We interpolate between the computed parameter points to obtain a smooth result, however, the original grid can be seen at the border 
between the computed area and the area excluded by non-perturbativity. 
The major deviation is between 
$0$ and $-5\%$ and grows 
in magnitude with increasing $\tan{\beta}$. 
For very large values of this parameter and close to the perturbative exclusion, 
values up to $-8\%$ occur. 
Very interesting is also the region with a small $\tan{\beta}$, as very small enhancements with respect to the SM can be found around $\MH=300\GeV$ (displayed in green). However, this region has a strong mass dependence because for large masses the negative corrections become $-5$\%. We note that this effect is also visible using the \MSbar{}($\alpha$) scheme  and therefore not an artifact of the singularity of the $\MSbar(\lambda_3)$ scheme.

\subsection{Baryogenesis}

In this section we discuss the results for the benchmark sets BP3~\cite{Dorsch:2014qja,deFlorian:2016spz}, 
as defined in \refse{sec:setup},
which were proposed as a possible solution to the problem of baryogenesis. 
The results shown in Tab.~\ref{tab:ResBP3} are computed in the $\MSbar(\lambda_3)$ scheme 
without considering the other schemes. In spite of the large distance to the alignment limit, 
the small heavy-Higgs-boson masses render both scenarios perturbatively stable 
with perturbativity measures of about $0.4$. 
Already at tree level we observe a large negative deviation from the SM caused by the large values for $c_{\beta-\alpha}$ suppressing the $\Ph VV$ coupling. These effects are enhanced at NLO for which we observe an increase of the negative deviation by 3 percentage points.
 This should be used in experiments measuring the Higgs decay into four fermions to put stronger bounds on these scenarios.
   \begin{table}[h]
   \centering
   \renewcommand{\arraystretch}{1.1}
 \begin{tabular}{|c|ccccc|}\hline
                             & $\Gamma^{\Ph{\to}4f}_\mr{NLO}$ [MeV] & $\delta_\mr{EW}[\%]$ & $\delta_\mr{QCD}[\%]$ &$\Delta_\mr{SM}^\mr{LO}[\%]$ &$\Delta_\mr{SM}^\mr{NLO}[\%]$    \\\hline
  $\mr{BP3}_{B1}$	     & $   0.86042(8)$& $  -0.76(0) $ & $ 4.96(1) $& $      -9.00(1)$& $     -11.98(1)$\\
   $\mr{BP3}_{B2}$           & $   0.70240(7)$& $  -1.73(0) $ & $ 4.94(1)$& $     -25.00(1)$& $     -28.15(1)$\\
\hline
  \end{tabular}  
   \caption{The $\Ph {\to} 4f$ widths in the $\MSbar(\lambda_3)$ scheme, 
including the EW and QCD corrections in the benchmark scenarios $\mr{BP3}_{B1,B2}$ (with the numerical errors in parentheses). 
The last two columns show the deviation from the SM prediction at LO and NLO.}
 \label{tab:ResBP3}
 \end{table}

\subsection{Fermiophobic heavy Higgs}

The results for the fermiophobic heavy Higgs scenario \cite{Hespel:2014sla,deFlorian:2016spz}, as defined in \refse{sec:setup}, 
are shown in Tab.~\ref{tab:ResBP6} for the $\MSbar(\lambda_3)$ scheme. 
The three scenarios have a perturbativity measure of $\lambda_k^{\text{max}} = 0.6$ 
and differ by their value of $\tan{\beta}$.
Note that all those scenarios are close to the alignment limit, 
so that the SM width is almost reached at LO. 
The NLO corrections increase the deviation from the SM by about $1\%$ for all scenarios. 
   \begin{table}[h]
   \centering
   \renewcommand{\arraystretch}{1.1}
 \begin{tabular}{|c|ccccc|}\hline
                  & $\Gamma^{\Ph{\to}4f}_\mr{NLO}$ [MeV] & $\delta_\mr{EW}[\%]$ & $\delta_\mr{QCD}[\%]$ &$\Delta_\mr{SM}^\mr{LO}[\%]$ &$\Delta_\mr{SM}^\mr{NLO}[\%]$    \\\hline
   $\mr{BP6}_{a}$ & $   0.96456(9)$& $  1.40(0) $ & $ 4.97(1) $ & $ -0.06(1) $& $      -1.33(1)$\\
   $\mr{BP6}_{b}$ & $   0.96304(9)$& $   1.43(0) $ & $ 4.97(1)$& $      -0.25(1)$& $      -1.49(1)$\\
   $\mr{BP6}_{c}$ & $   0.95701(9)$& $   1.56(0) $ & $ 4.97(1) $& $      -0.99(1)$& $      -2.10(1)$\\
\hline
  \end{tabular}  
   \caption{The $\Ph {\to} 4f$ widths the $\MSbar(\lambda_3)$ scheme including the EW and QCD corrections of the benchmark scenarios $\mr{BP6}_{a{-}c}$ (with the numerical errors in parentheses). The last two columns show the deviation from the SM prediction at LO and NLO.}
 \label{tab:ResBP6}
 \end{table}

\section{Conclusions}
\label{se:conclusion}

We have investigated the decay processes  $\Ph \to \PW\PW/\PZ\PZ \to 4f$ in the THDM, 
where we identify the light neutral CP-even Higgs boson $\Ph$ with the discovered Higgs boson of mass $\Mh=125\GeV$. 
This signature contributed to the discovery of the Higgs boson and is important in the experimental investigation of the 
properties of the Higgs boson, such as the measurement of its couplings to other particles. 
The corresponding decay observables allow for precision tests of the SM and, thus, contribute to the search for
any deviations from SM predictions.
The calculation of strong and electroweak corrections in specific SM extensions, such as the one presented 
in this paper in the THDM, is an important theory input to successful data analyses.

In our phenomenological discussion of numerical results, we have considered
several THDM benchmark scenarios proposed by the LHC Higgs Cross Section Working Group. 
For the investigated scenarios, we generally 
observe that the THDM predictions for the $\Ph{\to}4f$ width are bounded from above
by the SM 
prediction and that the deviations from the SM typically increase at NLO, which might be used to improve exclusion limits in the THDM
parameter space. The individual partial widths show similar deviations from the SM for all final states, 
but the shapes of differential distributions are not distorted by THDM contributions, 
so that the latter are not helpful to identify traces of the THDM.
Moreover, we find that the $\Ph{\to}4f$ widths do not discriminate between different types of THDMs 
(Types~I and II, lepton-specific and flipped).

We employ different renormalization schemes to define the THDM (i.e.\ the precise physical meaning of its
input parameters) at NLO.
Specifically, we apply four different schemes which have in common that we use
as many as possible input quantities that are directly accessible by experiment, such as the (on-shell) masses of all
five Higgs bosons of the THDM. For the remaining three free parameters, which are Higgs mixing angles and
Higgs self-couplings, we adopt \MSbar{} prescriptions in four different variants.
In detail, the $\MSbar(\alpha)$ scheme defines the two mixing angles $\alpha$ and $\beta$
of the CP-even and CP-odd Higgs bosons, respectively, as well as the quartic Higgs self-coupling parameter $\lambda_5$
in the \MSbar{} scheme, and FJ$(\alpha)$ is a modified variant of this scheme with a different treatment of
tadpole contributions in such a way that no gauge dependence between input parameters and predicted observables
is introduced. Similarly, we define the two schemes $\MSbar(\lambda_3)$ and FJ$(\lambda_3)$ in which we replace
the angle $\alpha$ by another self-coupling parameter $\lambda_3$ as input.
For a consistent comparison of results obtained in
the different renormalization schemes, the \MSbar{}-renormalized parameters
have to be properly converted between the schemes. 
Depending on the scenario,
we observe sizeable conversion effects on those parameters which can grow very large
in scenarios close to the experimental exclusion limits or in parameter regions where perturbative stability deteriorates.
These corrections, in particular, imply that the so-called alignment limit, in which one of the CP-even Higgs
bosons of the THDM is SM-like, corresponds to different Higgs mixing angles in different
renormalization schemes (even to different angles of a given renormalization scheme if the 
renormalization scale is changed).
This shows that a proper definition of parameters at NLO is mandatory in future predictions and
parameter fits in the THDM when precision is at stake.

While we observe a reduction of both the renormalization scheme and renormalization scale
dependence of the $\Ph{\to}4f$ width in the
transition from LO to NLO as long as all Higgs-boson masses are moderate and the distance to the alignment limit is not too large,
some renormalization schemes prove unreliable, i.e.\ prone to large corrections beyond NLO,
for scenarios with heavy Higgs bosons or away from the alignment limit. 
Generically,
the comparison of the different schemes reveals that the gauge-depen\-dent \MSbar{}($\alpha$) 
scheme shows a minimal scale dependence which reflects good perturbative stability. 
The $\MSbar(\lambda_3)$ scheme deviates only slightly from the former, yields reliable results and 
in addition is gauge independent at one loop in $R_\xi$ gauges. 
However, a singular region in the THDM parameter space exists in which the scheme is not defined.
If this region is experimentally favoured, it is necessary to redefine the scheme 
by replacing $\lambda_3$ by another scalar self-interaction $\lambda_{k\ne3}$, so
that the singularity is avoided. 
The gauge-independent FJ schemes partially suffer from large corrections and can only be applied for parameter points with 
sufficiently small coupling factors. Since the different schemes do not yield reliable results for all 
scenarios, self-consistency checks should be performed for every scenario when higher-order corrections are computed.
In cases where NLO fails to be predictive, NLO calculations 
should be stabilized upon including 
the leading (renormalization-scheme-specific) corrections beyond NLO, a task that is, however, beyond the scope of this paper.

In more detail,
in the low-mass scenario with a heavy CP-even Higgs boson of $300\GeV$ we obtain textbook-like results for 
the scale dependence, i.e.\ an improvement of the scale uncertainty and a reduction of differences between 
all four renormalization schemes at NLO, which indicates that perturbation theory works well. 
The deviation of the $\Ph{\to}4f$ width from the SM prediction
are, depending on the parameter set, between 0\% and $-6$\%, to which the NLO corrections contribute about
$1{-}2$\%. 
In high-mass scenarios with heavy Higgs bosons with masses of about $600\GeV$, 
the coupling factors are larger, resulting in less predictive results and larger 
differences between the renormalization schemes. 
Close to the alignment limit, the results of all four schemes are self-consistent and nicely agree,
but away from it differences occur.
While the \MSbar{}($\alpha$) and the $\MSbar(\lambda_3)$ 
(in its domain of definition) schemes still yield trustworthy results, the FJ($\alpha$) and the FJ($\lambda_3$) schemes 
suffer from large corrections, and their results should be taken with care. 
The deviations from the SM are similar to the low-mass case, and the NLO corrections similarly contribute
$1{-}2$\% to the deviations. The other investigated scenarios support the described picture as they yield similar results.

The calculated NLO corrections to all $\Ph \to \PW\PW/\PZ\PZ \to 4f$ decays are integrated in a new
version of the Monte Carlo program \textsc{Prophecy4f}, extending its applicability to the THDM.
The new code can be obtained from the authors upon request and will be available from the public
webpage%
\footnote{https://prophecy4f.hepforge.org/} 
soon.

\subsection*{Acknowledgements}

We would like to thank Ansgar Denner, Howard Haber and Jean-Nicolas Lang for helpful discussions
and especially Jean-Nicolas for an independent check of some one-loop matrix elements
against the crossing-related amplitudes used in \citeres{Denner:2016etu,Denner:2017vms}.
HR's work is partially funded by the Danish National Research Foundation, grant number DNRF90. HR acknowledges also support by the National Science Foundation under Grant No. NSFPHY11-25915. 
We thank the German Research Foundation~(DFG) and 
Research Training Group GRK~2044 for the funding and the support and
acknowledge support by the state of Baden--W\"urttemberg through bwHPC and the 
DFG through grant no INST 39/963-1 FUGG.
This research was also supported by the Munich Institute for Astro- and Particle 
Physics (MIAPP) of the DFG cluster of excellence ``Origin and Structure of the Universe''.

\vspace*{2em}
\appendix

\section*{Appendix}
\section{Further results for the high-mass scenario}

In this appendix, we show additional results on the scale variation and differential distributions in the high-mass scenario
for further illustration.
All the figures are similar to ones discussed already in \refse{sec:results} and further support our major conclusions. 

\subsection{Scale variation}
\label{App:scalevar}

\begin{figure}
  \centering
  \subfigure[]{
\label{fig:plot_MUSCAN-B1c-L3MS}
\includegraphics{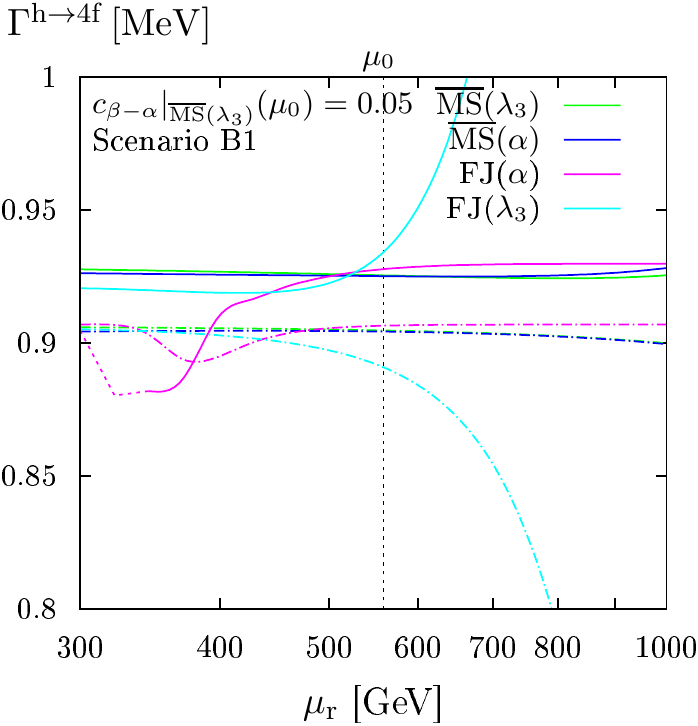}
}
\hspace{15pt}
\subfigure[]{
\label{fig:plot_MUSCAN-B1c-alphaMS}
\includegraphics{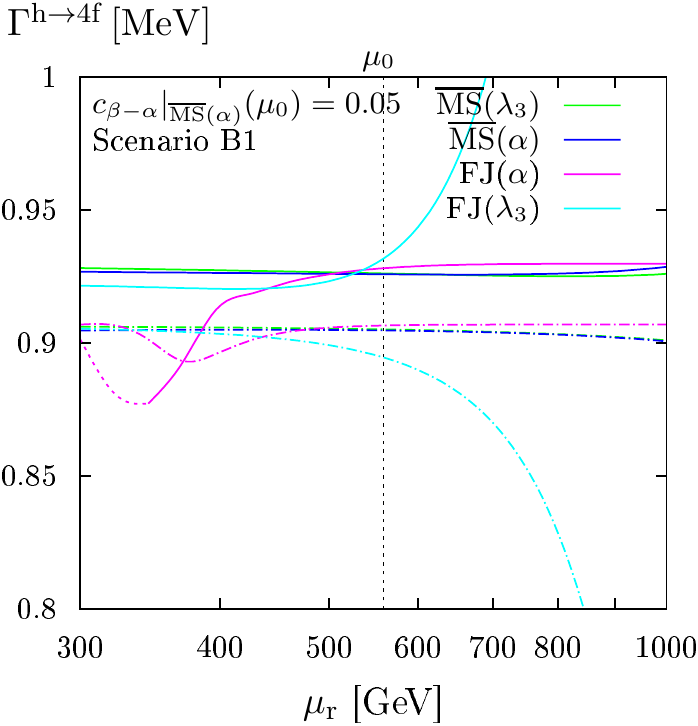}
}
\\[-1em]
  \subfigure[]{
\label{fig:plot_MUSCAN-B1c-FJ}
\includegraphics{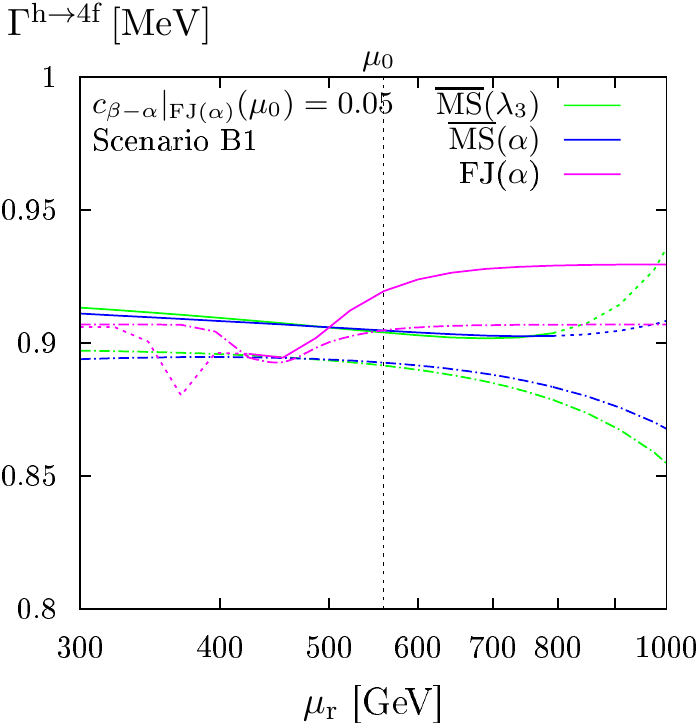}
}
\hspace{15pt}
\subfigure[]{
\label{fig:plot_MUSCAN-B1c-L3MSFJ}
\includegraphics{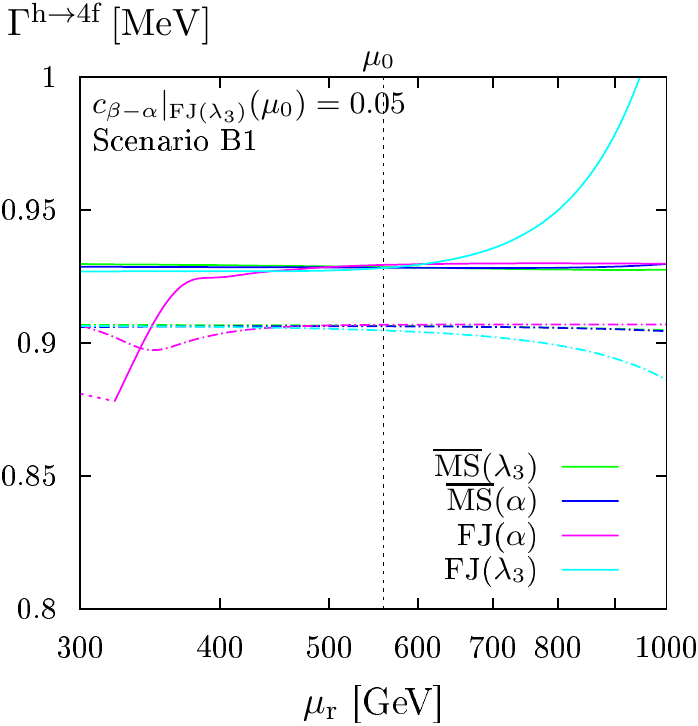}
}
\vspace*{-.5em}
\caption{As in \reffi{fig:plotmuscanB1a}, but for scenario~B1 with $c_{\beta-\alpha}=0.05$.}
%  \caption{The $\Ph{\to} 4f$ cross section at LO (dashed) and NLO EW (solid) for scenario~B1 with $c_{\beta-\alpha}=0.05$ in dependence of the renormalization scale. The panels (a), (b), (c), and (d) correspond to input values  defined in the  \MSbar{}$(\lambda_3)$, \MSbar{}$(\alpha)$, FJ($\alpha$), and FJ($\lambda_3$) schemes, respectively. The result is computed in all four different renormalization schemes after converting the input at NLO (also for the LO curves) and displayed using the colour code of Fig.~\ref{fig:plotrunningA}. The breakdown of perturbativity ($\lambda/(4\pi)>1$) is indicated by changing the NLO curve to dotted lines.}
\label{fig:plotmuscanB1c}
\end{figure}

\begin{figure}
  \centering
  \subfigure[]{
\label{fig:plot_MUSCAN-B2d-L3MS}
\includegraphics{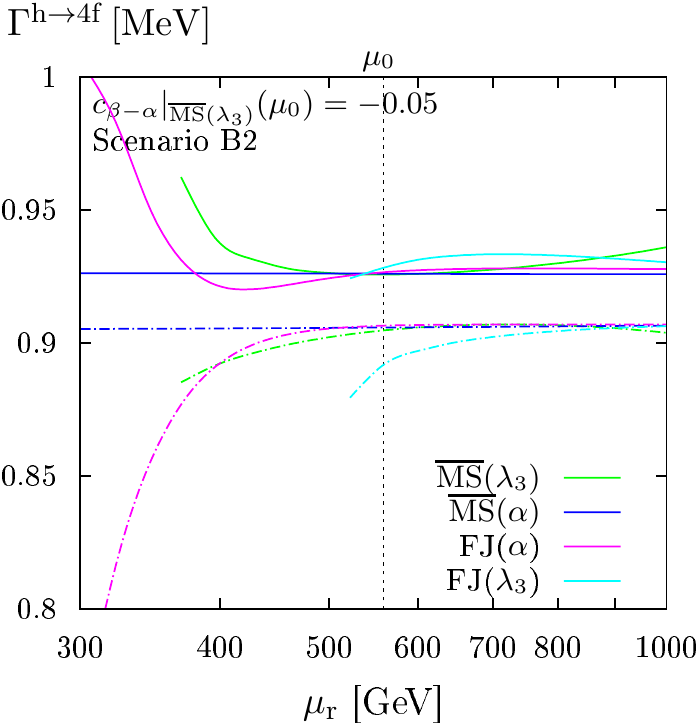}
}
\hspace{15pt}
\subfigure[]{
\label{fig:plot_MUSCAN-B2d-alphaMS}
\includegraphics{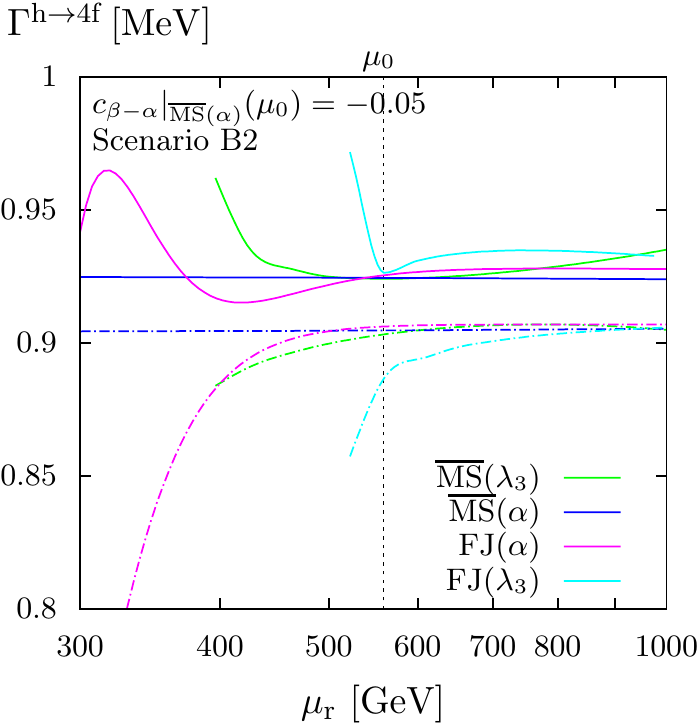}
}
\\[-1em]
  \subfigure[]{
\label{fig:plot_MUSCAN-B2d-FJ}
\includegraphics{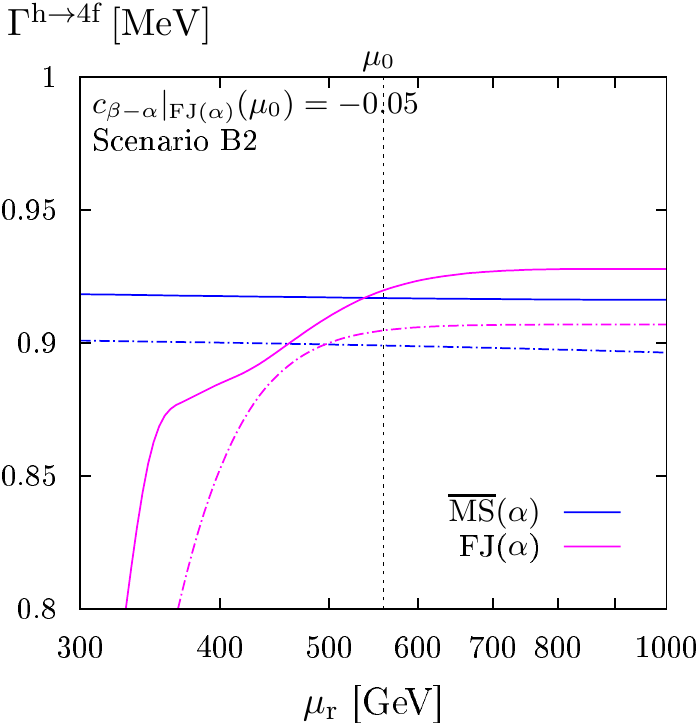}
}
\hspace{15pt}
\subfigure[]{
\label{fig:plot_MUSCAN-B2d-L3MSFJ}
\includegraphics{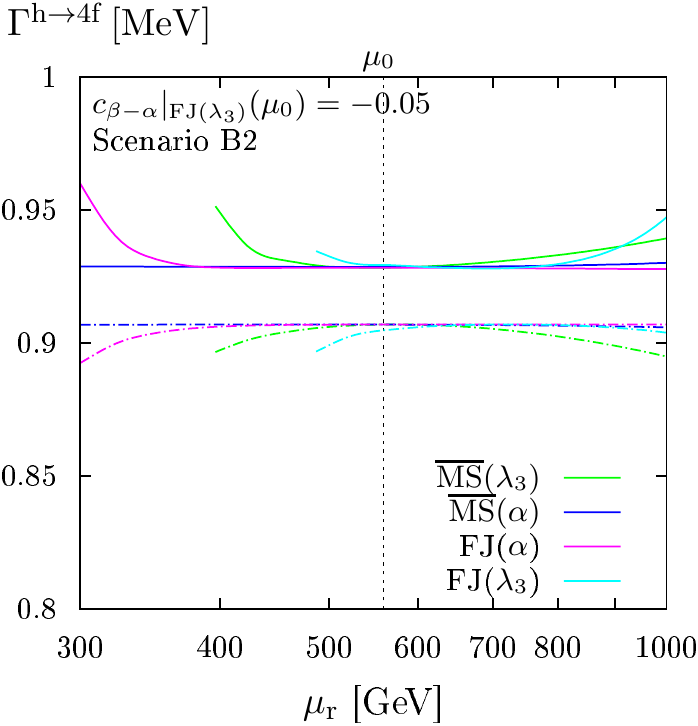}
}
\vspace*{-.5em}
\caption{As in \reffi{fig:plotmuscanB2b}, but for scenario~B2 with $c_{\beta-\alpha}=-0.05$.}
%  \caption{The $\Ph{\to} 4f$ cross section at LO (dashed) and NLO EW (solid) for scenario~B2 with $c_{\beta-\alpha}=-0.05$ in dependence of the renormalization scale. The panels(a), (b), (c), and (d) correspond to input values  defined in the  \MSbar{}$(\lambda_3)$, \MSbar{}$(\alpha)$, FJ($\alpha$), and FJ($\lambda_3$) schemes, respectively. The result is computed in all four different renormalization schemes after converting the input at NLO (also for the LO curves) and displayed using the colour code of Fig.~\ref{fig:plotrunningA} and for such small $c_{\beta-\alpha}$ the FJ($\lambda_3$) scheme can be  defined as target scheme. }
\label{fig:plotmuscanB2d}
\end{figure}

As pointed out in \refse{sec:HighMassb1scalevariation}, the reduction of the scale 
and
 renormalization scheme dependence in the transition from LO to NLO works better 
for scenarios closer to the alignment limit. 
To show this, we perform a scale variation using the benchmark scenarios B1 and B2 with $c_{\beta-\alpha}=\pm 0.05$
in Figs.~\ref{fig:plotmuscanB1c} and \ref{fig:plotmuscanB2d}.
These results should be compared to the ones shown in Figs.~\ref{fig:plotmuscanB1a} and \ref{fig:plotmuscanB2b}
for scenarios B1a and B2b; the reduced scale and scheme dependence is clearly visible.
Moreover, the conversion into the FJ($\lambda_3$) is possible when the alignment limit is approached, 
so that this scheme is included in the comparison of Fig.~\ref{fig:plotmuscanB1c}.

\subsection{Differential distributions}
\label{sec:diffdistrHM}

\begin{figure}
  \centering
  \subfigure[]{
\label{fig:plot_HM_mumuee_inv12}
\includegraphics[scale=0.8]{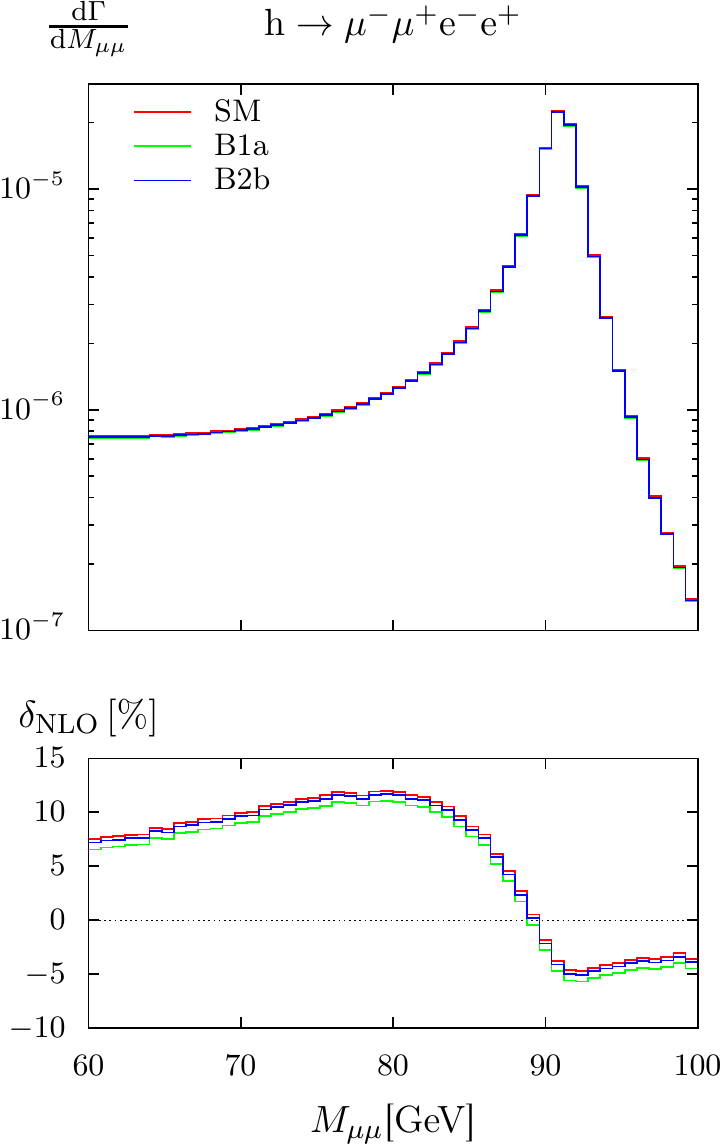}
}
\hspace{12pt}
\subfigure[]{
\label{fig:plot_HM_mumuee_phi}
\includegraphics[scale=0.8]{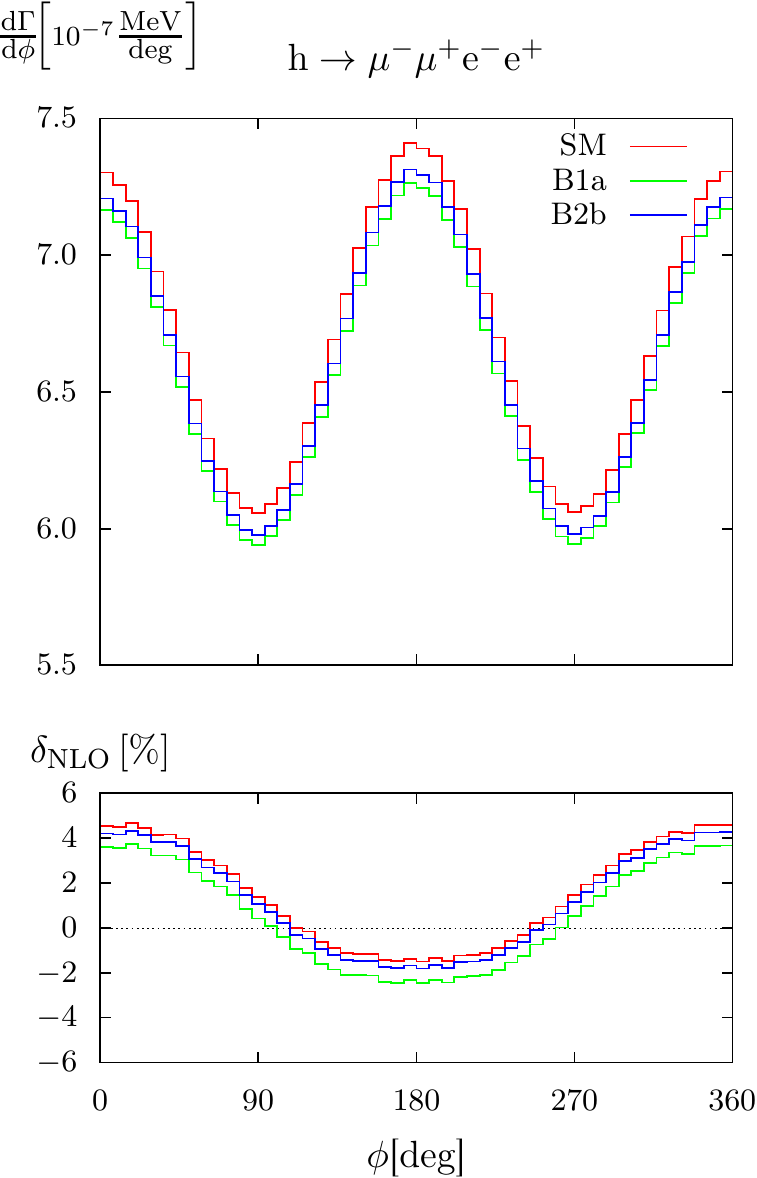}
}
\\[-.7em]
  \caption{Invariant-mass (a) and angular distributions (b) of the leptonic neutral-current decay $\mr{h} \to \mu^- \mu^+ \Pe^- \Pe^+$ for the SM and the THDM scenarios~B1b and B2b. The relative NLO corrections to the distributions are plotted in the lower panels.}
\label{fig:distr_HM_mumuee}
\vspace*{1em}
%\end{figure}
%
%\begin{figure}
  \centering
  \subfigure[]{\hspace{12pt}
\label{fig:plot_HM_vmuev_inv12}
\includegraphics[scale=0.8]{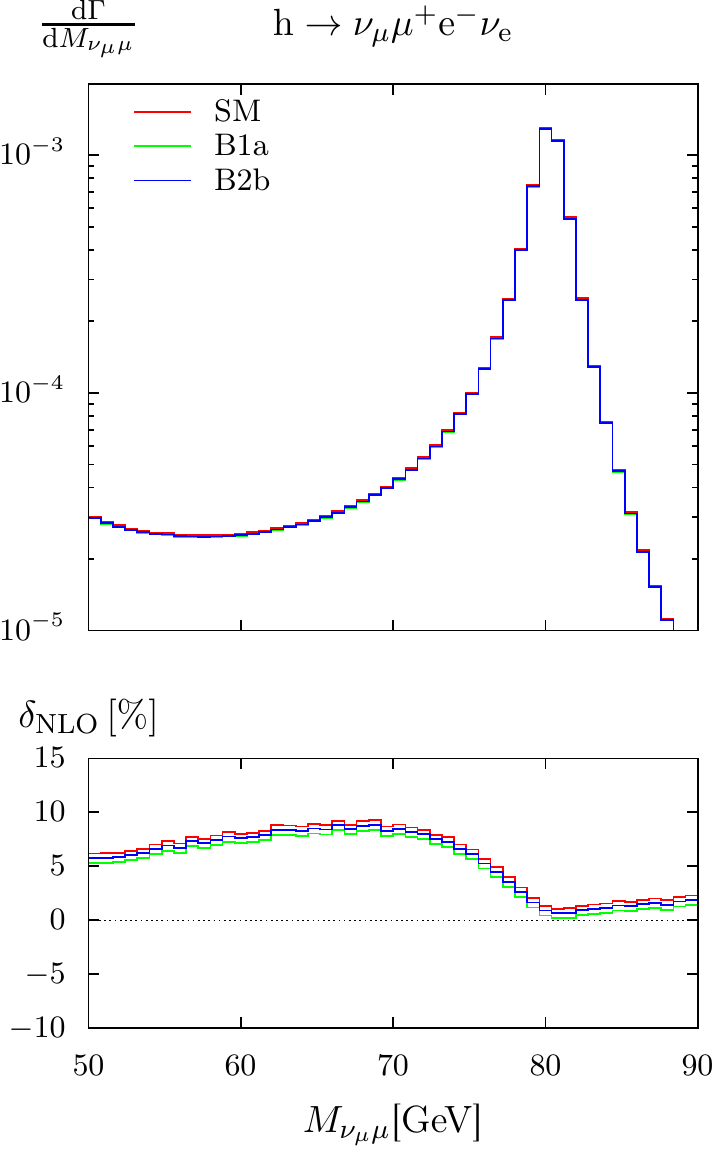}
}
\hspace{12pt}
\subfigure[]{
\label{fig:plot_HM_vmuev_phi}
\includegraphics[scale=0.8]{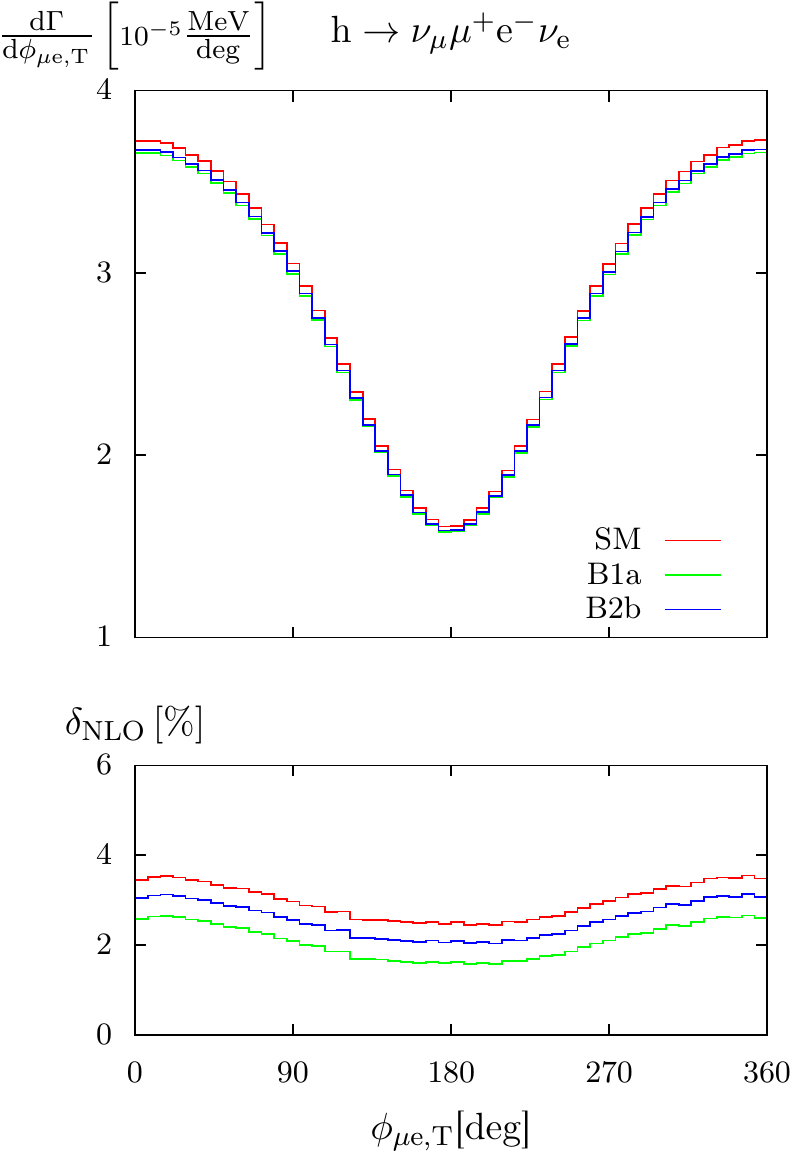}
}
\\[-.7em]
  \caption{Invariant-mass (a) and angular distributions (b) of the leptonic charged-current decay $\mr{h} \to \nu_\mu \mu^+ \Pe^- \bar\nu_\Pe$ for the SM and the THDM scenarios~B1a and B2b. The relative NLO corrections to the distributions are plotted in the lower panels.}
\label{fig:distr_HM_vmuev}
\end{figure}

\begin{figure}
  \centering
  \subfigure[]{
\label{fig:plot_HM_qqee_inv12}
\includegraphics[scale=0.8]{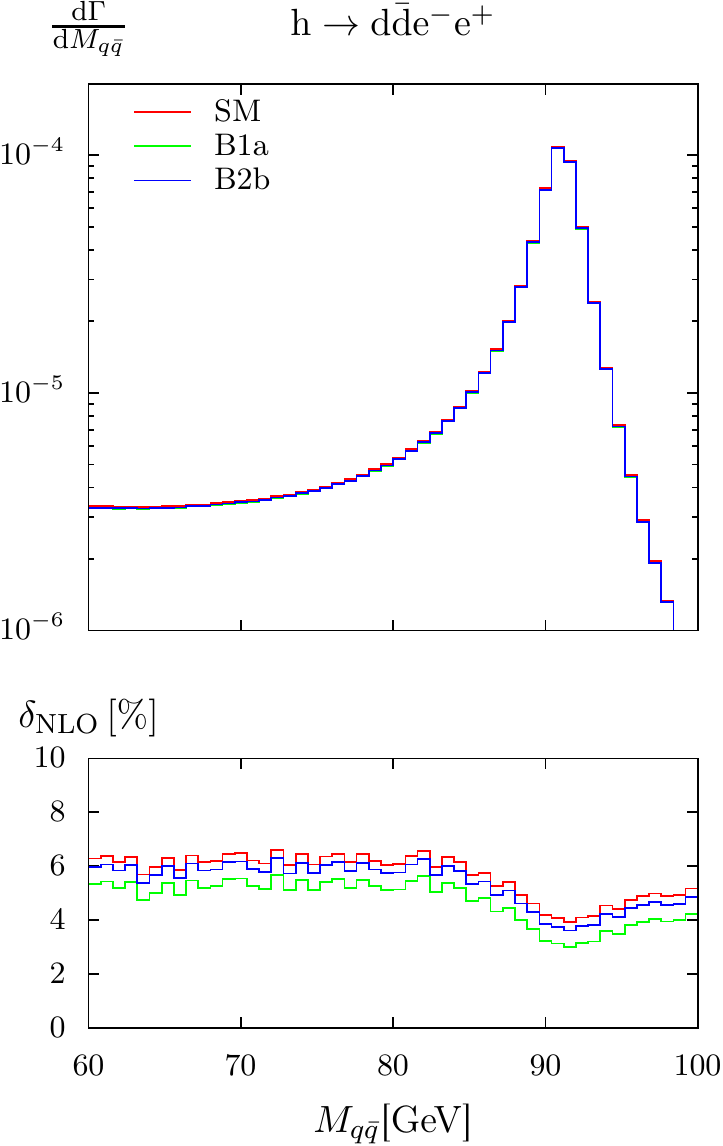}
}
\hspace{12pt}
\subfigure[]{
\label{fig:plot_HM_qqee_phi}
\includegraphics[scale=0.8]{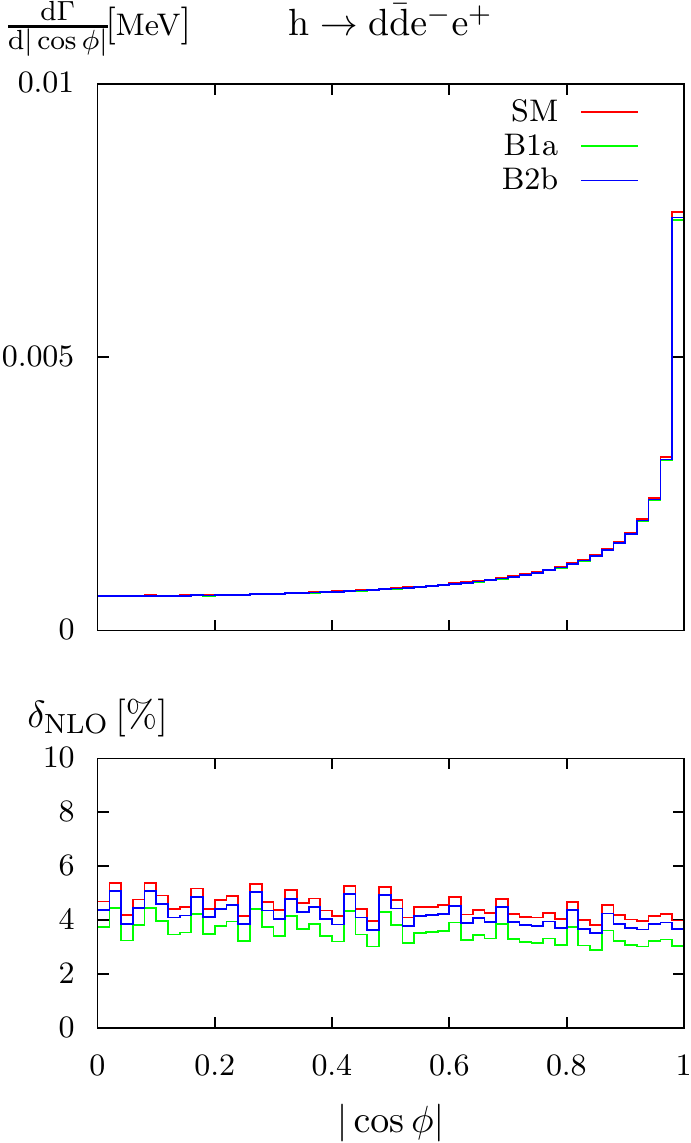}
}
\\[-.7em]
  \caption{Invariant-mass (a) and angular distributions (b) of the charged-current semi-leptonic decay $\mr{h} \to \Pd \bar{\Pd} \Pe^- \Pe^+$ for the SM and the THDM scenarios~B1a and B2b. The relative NLO corrections to the distributions are plotted in the lower panels.}
\label{fig:distr_HM_qqee}
\vspace*{1em}
%\end{figure}
%
%\begin{figure}
  \centering
  \subfigure[]{
\label{fig:plot_HM_veqq_inv12}
\includegraphics[scale=0.8]{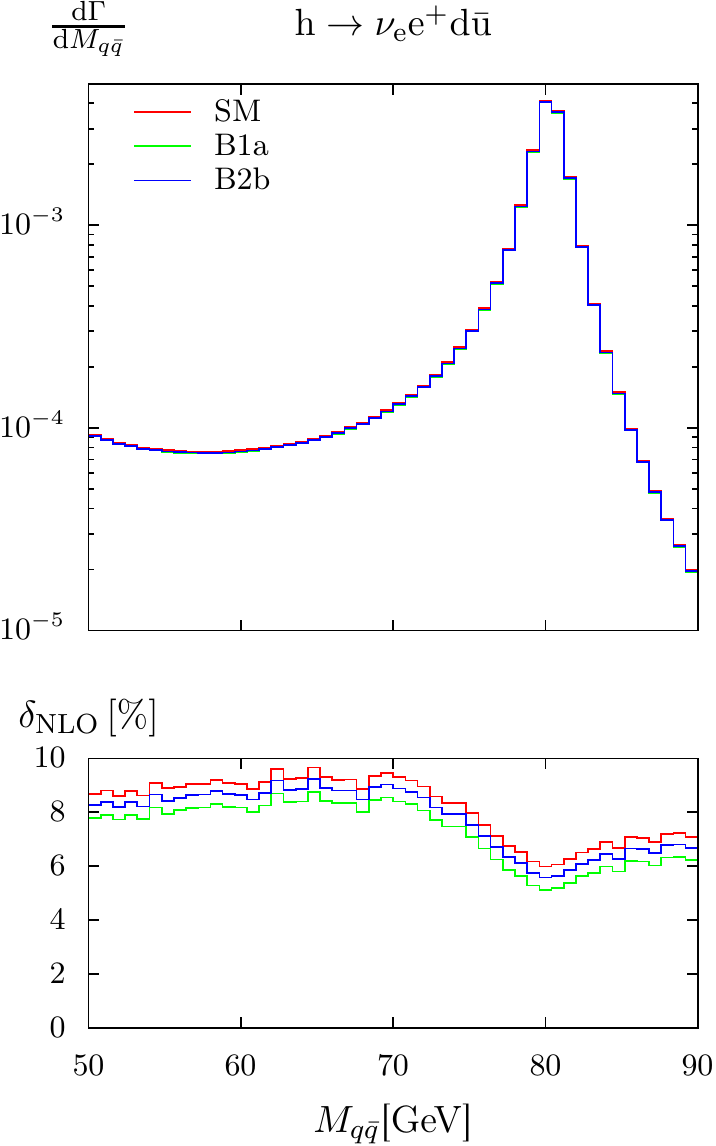}
}
\hspace{12pt}
\subfigure[]{
\label{fig:plot_HM_veqq_phi}
\includegraphics[scale=0.8]{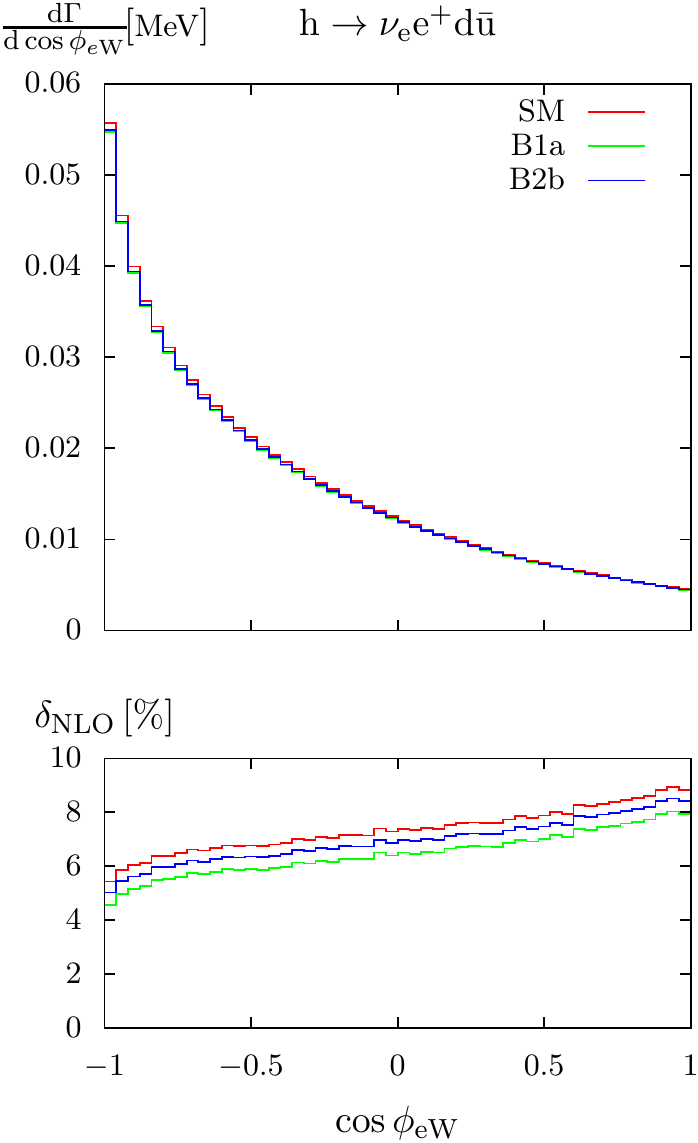}
}
\\[-.7em]
  \caption{Invariant-mass (a) and angular distributions (b) of the charged-current semi-leptonic decay $\mr{h} \to  \nu_\Pe \Pe^+ \Pd \bar{\Pu}$ for the SM and the THDM scenarios~B1a and B2b. The relative NLO corrections to the distributions are plotted in the lower panels.}
\label{fig:distr_HM_veqq}
\end{figure}

For none of the considered benchmark scenarios, 
we have observed any distortion in the shapes of differential distributions for $\Ph{\to} 4f$ decays
in the transition from the SM to the THDM.
For the low-mass scenarios~Aa and Ab this was illustrated in \refse{sec:diffdistr} for some selected 
leptonic and semileptonic final states.
Here we show the respective distributions for the scenarios~B1a and B2b 
in \reffis{fig:distr_HM_mumuee}--\ref{fig:distr_HM_veqq}.

\clearpage

\bibliographystyle{JHEPmod}            
\bibliography{bibliography}

\end{document}